\documentclass[12pt]{article}
\usepackage{lineno} % 添加这个包

\usepackage{authblk}
\DeclareUnicodeCharacter{2212}{-}
\usepackage{scicite}  % template
\usepackage{times}    % font
\usepackage{graphicx} % needed for figures
\usepackage{graphics} % needed for figures
\usepackage{float}
\usepackage[superscript, nomove]{cite} % needed for citations
\usepackage{amsmath} % need for equation
\usepackage{amssymb} % for \dashrightarrow and other math symbols

\usepackage{booktabs}
\usepackage{threeparttable}
\usepackage{fix-cm} % Any font size is allowed
\usepackage{xcolor}
\usepackage{ulem} % delete line
\newcommand{\KI}{\text{KI}}

\makeatletter
\let\NAT@parse\undefined
\makeatother
\usepackage[colorlinks=false,citecolor=blue]{hyperref}
\hypersetup{hidelinks}%hide the frame

\usepackage{booktabs} % Three line table

\usepackage[font=scriptsize]{caption}
\usepackage[left=2.15cm, right=2.15cm, top=2cm, bottom=2cm]{geometry} %page setup

\newenvironment{sciabstract}{%
\begin{quote} \bf}
{\end{quote}}

\def\scititle{Knowledge Independence Breeds Disruption but Limits Recognition}
\title{\bfseries \boldmath \scititle}

\author{
	Xiaoyao Yu$^{1,2}$,
	Talal Rahwan$^{2\ast}$,
	Tao Jia$^{1,3\ast}$\\ \vspace{0.6em}
	\small$^{1}$College of Computer and Information Science, Southwest University, Chongqing 400715, China.\\ \vspace{0.6em}
	\small$^{2}$Computer Science, Science Division, New York University Abu Dhabi, Abu Dhabi 129188, UAE.\\ \vspace{0.6em}
	\small$^{3}$College of Computer and Information Science, Chongqing Normal University, Chongqing 401331, China.\\ \vspace{0.6em}
	\small$^\ast$Corresponding author. Email: talal.rahwan@nyu.edu (T.~R.); tjia@swu.edu.cn (T.~J.)\and
}

\date{}

\begin{document}
%\begin{linenumbers}% This is the first page with line numbers.

% Double-space the manuscript.

\baselineskip24pt

% Make the title.

\maketitle
% Adjust the vertical space between affil and sciabstract
\vspace{-6em}

% Place your abstract within the special {sciabstract} environment.

\begin{sciabstract}
Despite extensive research on scientific disruption, two questions remain: why disruption has declined amid growing knowledge, and why disruptive work receives fewer and delayed citations. One way to address these questions is to identify an intrinsic, paper-level property that reliably predicts disruption and explains both patterns. Here, we propose a novel measure, knowledge independence, capturing the extent to which a paper draws on references that do not cite one another. Analyzing 114 million publications, we find that knowledge independence strongly predicts disruption and mediates the disruptive advantage of small, onsite, and fresh teams. Its long-term decline, nonreproducible by null models, provides a mechanistic explanation for the parallel decline in disruption. Causal and simulation evidence further indicates that knowledge independence drives the persistent trade-off between disruption and impact. Taken together, these findings fill a critical gap in understanding scientific innovation, revealing a universal law: Knowledge independence breeds disruption but limits recognition.

% SUBMISSION VERSION EXTENDED
% Despite extensive research on scientific disruption, two questions remain: why disruption has declined amid growing knowledge, and why disruptive work receives fewer and delayed citations. One way to address these questions is to identify an intrinsic, paper-level property that reliably predicts disruption and explains both patterns. A promising candidate is the atypicality of knowledge combinations, as recombinant growth posits that expanding knowledge enables potentially innovative pairings. Yet existing atypicality measures, based on how bibliographies combine journals or disciplines, do not reliably predict disruption. Here, we propose a novel measure, knowledge independence, capturing the extent to which a paper draws on references that do not cite one another. Analyzing 114 million publications, we find that knowledge independence strongly predicts disruption and mediates the disruptive advantage of small, onsite, and fresh teams. Its long-term decline, nonreproducible by null models, provides a mechanistic explanation for the parallel decline in disruption. Causal and simulation evidence further indicates that knowledge independence drives the persistent trade-off between disruption and impact. Taken together, these findings fill a critical gap in understanding scientific innovation, revealing a universal law: Knowledge independence breeds disruption but limits recognition.
%

% }
\end{sciabstract}

%%%% TEASER
\begin{sciabstract}
\textit{Teaser}\\
\textit{Knowledge independence, a declining intrinsic driver, mechanistically explains disruption's paradoxes: its system-wide decline and its trade-off with impact.}
\end{sciabstract}
\clearpage

% In setting up this template for *Science* papers, we've used both
% the \section* command and the \paragraph* command for topical
% divisions.  Which you use will of course depend on the type of paper
% you're writing.  Review Articles tend to have displayed headings, for
% which \section* is more appropriate; Research Articles, when they have
% formal topical divisions at all, tend to signal them with bold text
% that runs into the paragraph, for which \paragraph* is the right
% choice. Either way, use the asterisk (*) modifier, as shown, to
% suppress numbering.

\section*{Introduction}

\noindent
% \color{purple}
%%%%%%%%%%%%%%%% 1: PARADOXES IN DISRUPTION'S DECLINING AND RECOGNITION PENALTY
Scientific disruption has become a central topic in studies of scientific innovation~\cite{wu2019large,park2023papers,lin2023remote,zeng2021fresh,deng2025critical,kozlov2023disruptive,leahey2023types}. Unlike most citation-based metrics~\cite{garfield1972citation,hirsch2005index,egghe2006improvement,shen2014collective,ruiz2015field,hutchins2016relative,ke2023network,meng2024hidden}, which primarily quantify the magnitude of scientific impact, the disruption index~\cite{funk2017dynamic,wu2019large} captures the type of influence a paper exerts---specifically, whether it displaces or develops the prior work on which it builds. Despite numerous studies examining this topic, two unexplained phenomena persist. First, \textit{disruption is declining over time despite unprecedented growth in scientific output}~\cite{park2023papers} and ``no one knows why''~\cite{kozlov2023disruptive}. This puzzle is compounded by yet another unexplained phenomenon: \textit{disruption is correlated with fewer citations and delayed recognition}~\cite{bornmann2020disruption,lin2022new,liang2023bias,zeng2023disruptive,deng2025critical}. Together, these open questions constitute a fundamental gap in our understanding of scientific disruption. 

%%%%%%%%%%%%%%%% 2: THE NEED OF IDENTIFYING AN INTRINSIC PROPERTY
One way to address these questions is to identify an intrinsic property of research papers that reliably predicts their disruption. After all, the disruptive potential of a paper must stem from something it possesses \textit{ex ante}---a feature present at the moment of publication. By contrast, the disruption index, which captures how future research cites the paper, is a proxy that offers an \textit{ex post} assessment: it reflects the consequences, rather than the underlying causes, of disruptive potential. Once such an intrinsic property is identified, it can shed light on the aforementioned phenomena. More specifically, if this property has declined systematically over the past decades, it would naturally account for the parallel decline in disruption. Likewise, if it turns out to be negatively associated with impact, it would serve as a confounder underlying the observed association between disruption and impact. 

%%%%%%%%%%%%%%%% 3: ATYPICALITY METRICS YIELD THE EXPLANATORY IMPASSE
A promising candidate for such a property is the atypicality of the body, or ``combination,'' of the references on which a paper builds. This idea is grounded in recombinant growth theory, which posits as the stock of knowledge expands, so too does the combinatorial search space, creating more opportunities for unconventional pairings that can spark innovation~\cite{weitzman1998recombinant}. One of the most established measures that capture such atypicality is the novelty index~\cite{uzzi2013atypical}, which evaluates the rarity of co-cited journal pairs in a paper's bibliography. Intuitively, if a pair co-occurs less frequently than expected by random chance, it contributes to the paper's atypicality (see Methods for a formal definition). There are two main approaches to representing novelty based on the rarity of journal combinations: one based on the median rarity of distribution~\cite{lin2022new,park2023papers}, and another based on the left tail (10th percentile) of that distribution~\cite{uzzi2013atypical,lin2023sciscinet}. Notably, these two variants lead to contradictory conclusions: the median-based approach yields a positive association between novelty and disruption~\cite{lin2022new}, whereas the tail-based alternative produces a negative association~\cite{li2025innovation}. This divergence indicates that the association is highly sensitive to the operationalization of novelty, and hence this index cannot serve as the intrinsic property we seek.

An alternative way to assess a paper's atypicality focuses on the disciplines, not journals, that the paper combines in its bibliography~\cite{rafols2010diversity,zhang2016diversity,zhang2021scientometric,xiang2025evaluating}. Arguably, the most establish measure in this line of work is the Rao–Stirling interdisciplinarity index~\cite{stirling2007general,rafols2010diversity,zhang2016diversity,zhang2021scientometric,xiang2025evaluating}. Yet the empirical evidence linking interdisciplinarity to disruption is inconsistent. An analysis based on the Scopus dataset reported a positive association between interdisciplinarity and disruption~\cite{chen2024interdisciplinarity}; a study using the Microsoft Academic Graph dataset found the association to be negative~\cite{liu2024monodisciplinary}; and a third study using the Web of Science dataset observed a negative association with the disruption index but a positive association with the odds of a paper being disruptive~\cite{cheng2025impact}. Taken together, these mixed findings indicate that the relationship between interdisciplinarity and disruption is highly sensitive to the dataset and to the particular operationalization of disruption, suggesting that interdisciplinarity, like novelty, cannot serve as the robust predictor we are looking for.

%%%%%%%%%%%%%%%% 3: MOTIVATION OF KONWLEDGE INDEPENDENCE
Since journal- and discipline-based atypicality indices cannot reliably predict disruption, a natural question follows: What specific form of atypical reference combinations, if any, can do so---and might such a form also clarify the broader phenomena surrounding disruption? To address this question, we turn to a largely overlooked aspect of knowledge combination: path dependence. Here, by path dependence we do not refer to the traditional sociological concept of historical lock-in and self-reinforcement~\cite{page2006path,strambach2013reconceptualizing,puffert2024path}, but rather to the extent to which technological innovations rely on established sequences, or ``paths,'' of prior inventions~\cite{xie2016does,zhang2023knowledge}. From the latter perspective, studies have shown that reliance on established technological trajectories tends to reinforce incremental improvements by reducing search and development costs~\cite{ahuja2001entrepreneurship,katila2002something,kapoor2012firms}, and hinders radical innovation by narrowing the exploratory space~\cite{kaplan2015double,xie2016does,schillebeeckx2021knowledge}. What remains unknown, however, is whether similar dynamics extend from technological innovation to scientific disruption. Unfortunately, existing measures of this notion of knowledge path dependence are either qualitative~\cite{sydow2009organizational,jackson2010formulation,jackson2012modeling} or defined at the institutional- or regional-level~\cite{jaffe1986technological,zhang2023knowledge}, leaving us without a quantitative, paper-level metric capable of capturing path dependence in scientific research

%%%%%%%%%%%%%%%% 4: KNOWLEDGE INDEPENDENCE IN SCIENCE AND DISRUPTION
Here, we propose a measure that quantifies the extent to which a paper's references are ``independent.'' More concretely, it calculates the proportion of references that are independent from, i.e., do not cite, other references in the paper's bibliography. Consider, for instance, a pair of references $(r_1, r_2)$ in the focal paper's bibliography. If it turns out that $r_1$ cites $r_2$, then this can be thought of as a path transmitting knowledge from $r_2$ to $r_1$. As such path-dependent references become more frequent, knowledge path independence decreases. For brevity, we hereafter omit the term ``path'' and simply write ``knowledge independence,'' denoted by $\KI$. Note that our measure captures the independence, rather than dependence, of knowledge paths, since we are primarily interested in atypical, rather than typical, combinations. Conceptually, a paper built on independent knowledge acts as a ``knowledge broker'' connecting intellectual entities that were previously unlinked. This is quite different from the role played by papers that are atypical from the perspective of novelty (combining unconventional journal pairs) or interdisciplinarity (combining divergent disciplines); those papers act as ``knowledge importers'' bridging distant knowledge domains. Importantly, $\KI$ does not necessarily correlate with novelty or interdisciplinarity. For instance, a paper can have low novelty but high $\KI$, e.g., if it cites papers from two commonly co-cited journals, but its references hardly cite each other. Likewise, a paper may have high interdisciplinarity but low $\KI$, e.g., if it links divergent disciplines, but many of its references happen to cite one another.

%%%%%%%%%%%%%%%% 5: SUMMARY
Armed with our measure, we examine 53.8 million publications from the Web of Science, and find that $\KI$ positively correlates with disruption. Causal inference using standard matching techniques (CEM and PSM) along with normalized regression analyses reveal that this association is significant. Moreover, both the effect magnitude and the explanatory power of $\KI$ on disruption are substantially greater than that of numerous intrinsic properties, including the paper's discipline, the publication year, the team composition, and various reference attributes. Our analysis also reveals $\KI$'s mediating role in the well-documented association between team composition and disruption. More specifically, we show that small~\cite{wu2019large}, onsite~\cite{lin2023remote}, and fresh~\cite{zeng2021fresh} teams exhibit a systematic preference for higher $\KI$, which helps explain their disruptive potential. Taken together, these findings highlight $\KI$ as the intrinsic paper property that most strongly predicts disruption. Finally, we show that $\KI$ provides a plausible unified explanation for the aforementioned paradoxes surrounding disruption. In particular, we find that $\KI$ has undergone a decades-long decline across all disciplines---a trend that cannot be replicated by null models---signaling a genuine shift in scientists' behaviour, and explaining the parallel decline in disruption. Furthermore, we observe that papers built on higher $\KI$ experience lower and delayed citation recognition. Once $\KI$, the confounder, is controlled for, the negative correlation between disruption and citation recognition is substantially attenuated, providing a plausible explanation for the second paradox. More broadly, our findings fill a critical gap in our understanding of scientific innovation, revealing a universal law: Knowledge independence breeds disruption but limits recognition.

% \color{black}

\section*{Results}

\subsection*{Quantifying knowledge independence}

\begin{figure}[ht]
    \begin{center}
        % \resizebox{15cm}{!}{\includegraphics{figures/FigOA_1_KI_Stats.pdf}}
        \includegraphics[width=.84\textwidth]{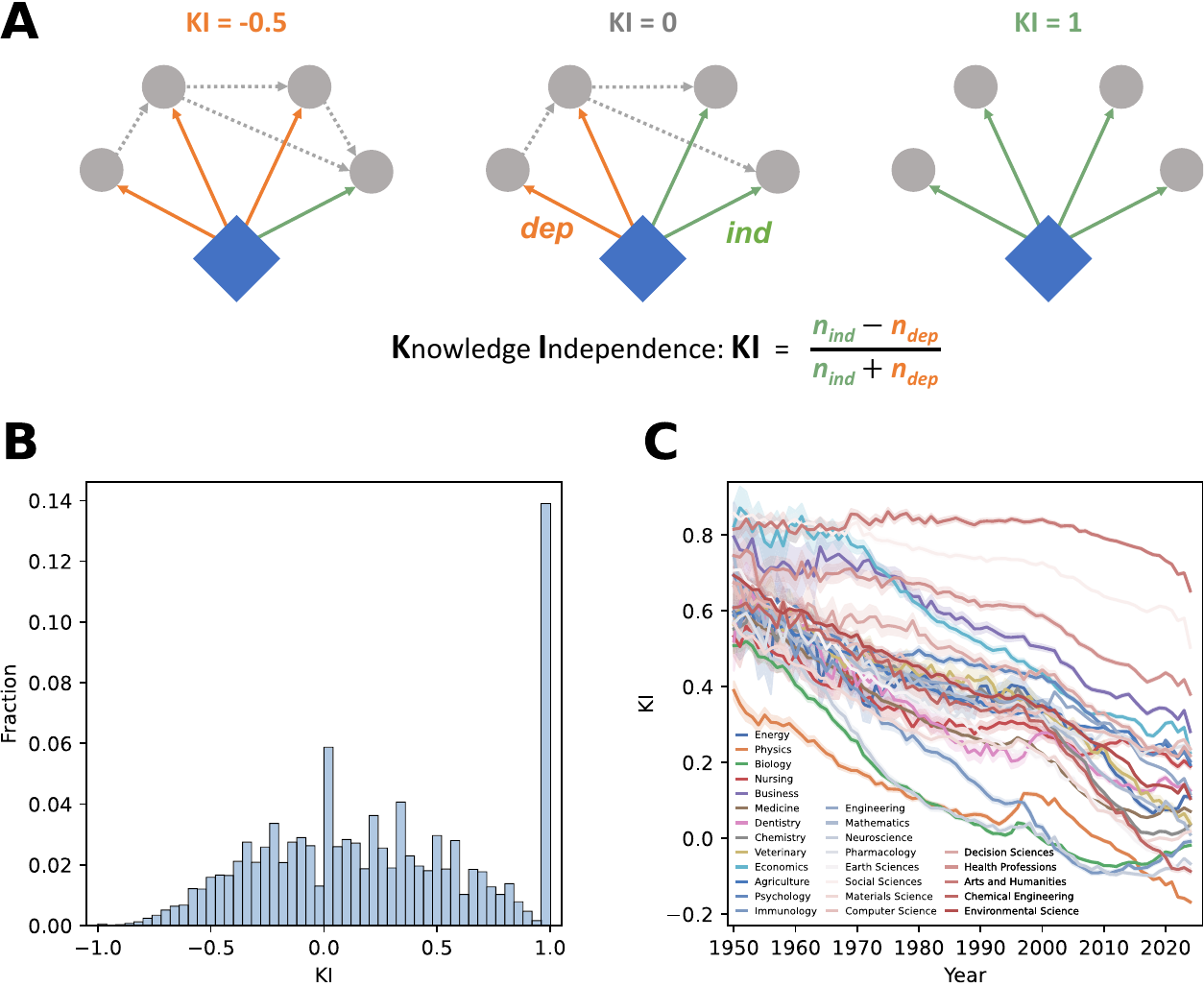}
        \caption{\textbf{$|$ Quantifying knowledge independence.}
            (\textbf{A}) Three examples of how knowledge independence ($\KI$) is calculated for a given focal paper (blue diamond) based on its references (gray circles). A reference is of $ind$-type (green) if it does not cite any other reference, or of $dep$-type (orange) if it does. $\KI$ is then the difference in the fraction of $ind$-type and $dep$-type references.
            (\textbf{B}) The $\KI$ distribution of 73,256,108 papers with at least two references published between 1950 and 2024 in the OpenAlex.
            (\textbf{C}) The pervasive downtrend of $\KI$ over time across disciplines. Bootstrapped 95\% confidence intervals are shown as shaded bands.
            \label{fig1}}
    \end{center}
\end{figure}

% THE DEFINITION AND MEASURE OF KI
To quantify knowledge independence in the context of bibliometrics, we develop a measure that relies on the citation network. This is because citations serve as tangible indicators of knowledge flow, reflecting the dependency of the citing paper on its references. Intuitively, for any given paper, the measure quantifies the degree to which its references are ``independent'' in the sense that they do not cite one another. To this end, let us introduce two types of references: An $ind$-type reference is one that does not cite any other work within the same reference list, whereas a $dep$-type reference is one that does cite at least one other work in this list. Then, knowledge independence ($\KI$) is measured as the difference between the fraction of $ind$-type and $dep$-type references. More formally:
\begin{equation}
\KI = \frac{n_{\rm ind}-n_{\rm dep}}{n_{\rm ind}+n_{\rm dep}},
\end{equation}
\noindent where $n_{\rm ind}$ and $n_{\rm dep}$ are the numbers of $ind$-type and $dep$-type references, respectively. This yields a continuous value between $-1$ and $1$. On one end of the spectrum, we have papers in which none of the references cites any of the other references, corresponding to a $\KI$ value of $1$ (since $n_{\rm dep}=0$). On the other extreme, we have papers in which every reference (apart from the oldest one) cites at least one other reference, corresponding to a $\KI$ value that approaches $-1$ ($\frac{1-n_{\rm dep}}{1+n_{\rm dep}}$, since the oldest reference cannot cite any of the other ones, which yields $n_{\rm ind}=1$).
Fig.~\ref{fig1}A illustrates three examples, each depicting a focal paper (represented as blue diamond) along with its references (gray circles), demonstrating how $\KI$ changes with different values of $n_{\rm ind}$ and $n_{\rm dep}$. 
Throughout our analyses, we exclude papers that cite only one reference, since their $\KI$ is bound to be $1$ by definition, lacking any meaningful interpretation.

% THE DISTRIBUTION AND EVOLUTION OF KI
The quantification of $\KI$ makes it possible to systematically explore the distribution and evolution of knowledge independence in scientific papers. In general, the $\KI$ distribution seems to resemble a bell-shaped curve (Fig.~\ref{fig1}B), punctuated by two outlier peaks at values $0$ and $1$. This anomaly is due to the large number of papers with relatively short bibliographies (Extended Data Fig.~\ref{figS_RC_Dis}). For every such paper, the range of $\KI$ contains relatively few (if any) values other than $0$ and $1$, e.g., for any paper with two references, $\KI\in\{0,1\}$ (i.e., $0$ and $1$ are the only possible values), and for any paper with four references, $\KI\in\{-0.5,0,0.5,1\}$ (i.e., $0$ and $1$ represent half the possible values). Additionally, $1$ is the only value that appears in the range of $\KI$ regardless of the number of references, while $0$ is the only value (other than $1$) that appears in the range whenever the number of references is even. Fig.~\ref{fig1}C depicts the evolution of $\KI$. As can be seen, there has been a notable decline in papers' $\KI$ over the past six decades, regardless of the discipline under consideration.

To validate the $\KI$ metric and its observed patterns in distribution and temporal trend, we conducted a two-stage robustness analysis. First, we show through direct statistical controls that our findings are robust to variations in reference count, novelty index, and interdisciplinarity index (see Section~\ref{SI_KI_Robustness} in Supplementary Materials). 
% This addresses a critical concern, as $\KI$ and the established atypicality metrics (novelty index and interdisciplinarity) all relate to bibliography structure. We confirm that $\KI$ is a distinct metric from atypicality metrics, exhibiting a pronounced inverted U-shaped relationship stratifying by time periods and reference count, respectively (Extended Data Fig.~\ref{figS_RNA-B, Extended Data Fig.~\ref{figS_RM}A-B), which confirms this relationship is robust and not an artifact of temporal evolution or reference count.} 
More fundamentally, to ensure these patterns are not artifacts of network evolution, we employed a random rewiring Monte Carlo simulation. The results reveal that the empirically observed $\KI$ has a distribution and historical decline that are starkly different from the random expectation (see Section~\ref{SI_Monte_Carlo_Rewire} in Supplementary Materials). This provides strong evidence that our findings reflect a substantive, non-random shift in scientific behavior.

\subsection*{Knowledge independence is associated with scientific disruption}
% MAIN RESULTS
% In principle, building on independent knowledge creates more opportunities for unconventional combinations, thereby fostering innovation with a greater chance. To test this hypothesis, we examined the correlation between $\KI$ and the disruption index~\cite{funk2017dynamic,wu2019large}, which quantifies the extent to which a paper's contributions overshadow the prior work it builds upon (see Methods). 
In principle, building on independent knowledge creates more opportunities for unconventional combinations, thereby fostering innovation with a greater chance. To test this hypothesis, we examined the correlation between $\KI$ and the disruption index~\cite{funk2017dynamic,wu2019large}. The disruption index quantifies the extent to which a paper introduces novel contributions that overshadow the prior work it builds upon, leading subsequent scholars to cite the paper itself without citing its references. This index belongs to $[-1, 1]$, representing a spectrum of papers ranging from fully developmental ($-1$) to fully disruptive ($1$) (see the Disruption section in Methods).

\begin{figure}[ht]
    \begin{center}
        % \resizebox{17cm}{!}{\includegraphics{figures/FigOA_2_KI_D.pdf}}
        \includegraphics[width=1\textwidth]{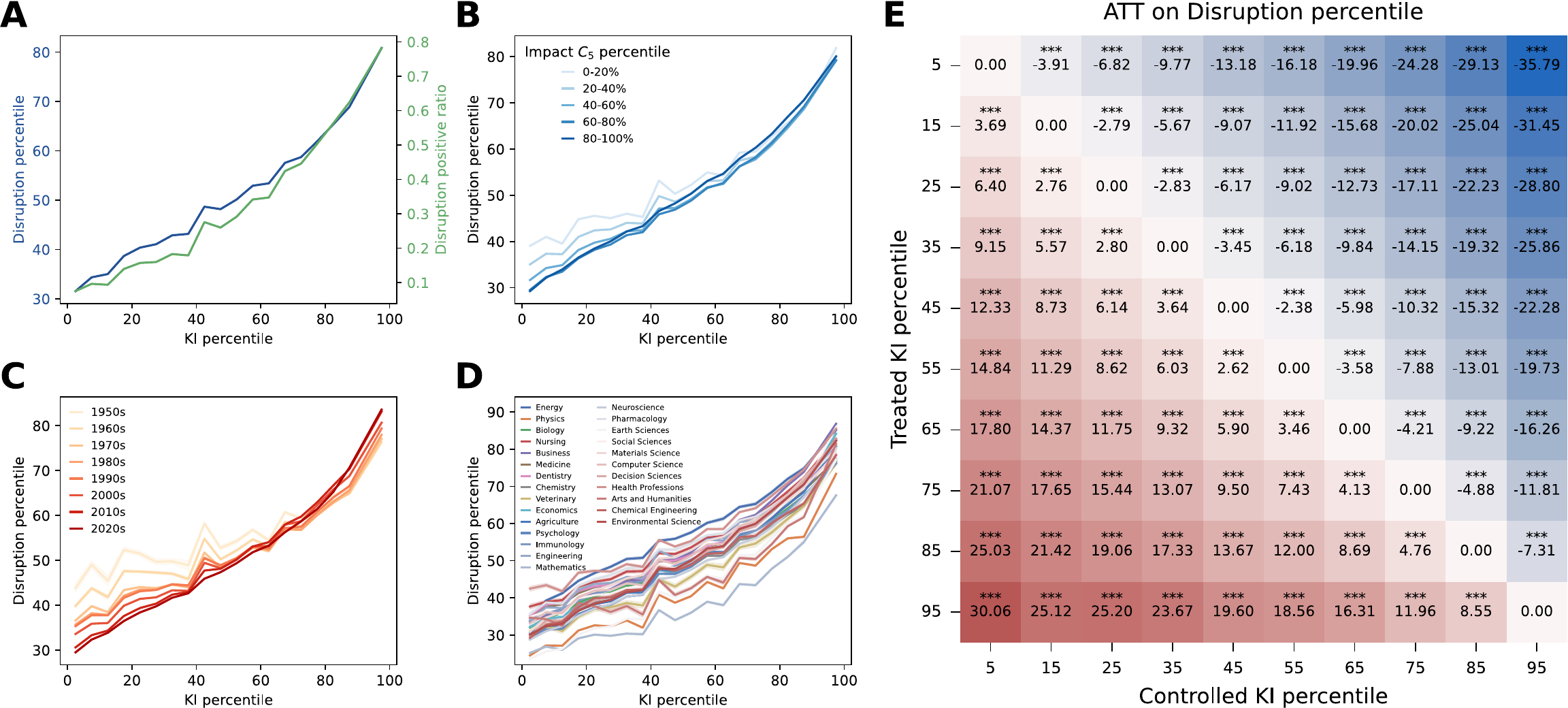}
        \caption{\textbf{$|$ Knowledge independence is associated with scientific disruption.}
            (\textbf{A}) For the 55,576,192 papers published before 2023 in the OpenAlex that are cited at least once, the average disruption percentile (blue curve, left y-axis) and the disruption positive ratio (green curve, right y-axis) increase with the $\KI$ percentile.
            (\textbf{B}) The association between $\KI$ and disruption persists, regardless of the impact of the focal paper. Here, impact is measured as the number of citations received within the first five years after publication, denoted by $C_5$.
            (\textbf{C}) The association between $\KI$ and disruption persists across decades.
            (\textbf{D}) The association between $\KI$ and disruption persists, regardless of discipline.
            Bootstrapped 95\% confidence intervals are shown as shaded bands in panels~\textbf{A-D}.
            (\textbf{E}) The ATT (average treatment effect on the treated) matrix of $\KI$ on disruption via coarsened exact matching (CEM). Each controlled group is set as a baseline, and ATTs are calculated for comparisons between the baseline and each of the treated groups. Blue cells represent negative ATTs, while red ones represent positive ATTs, with color intensity proportional to the absolute value.
            Within each controlled group, the ATT increases as the treated size grows, transitioning from negative to positive when the treated size approaches the controlled size.
            Each ATT is tested against the null hypothesis (ATT equals 0) using a two-sided \textit{t}-test. (*$p < 0.05$, **$p < 0.01$, ***$p < 0.001$).
            \label{fig2}}
    \end{center}
\end{figure}

% MAIN RESULTS
$\KI$ is an individual measure of the extent to which a paper connects originally isolated knowledge within the internal reference network, whereas the disruption index is a collective measure of future peer recognition for a paper's contribution relative to its predecessors. Put it differently, the former measure is solely rooted in the cited bibliography at the publication moment, but the latter index integrates the past context (cited bibliography) with the future recognition patterns (received citations). Seemingly, a paper's intrinsic structural property ($\KI$) and its future extrinsic recognition (disruption) are not necessarily associated with each other.
But our findings strongly support the hypothesis of innovation: papers with higher $\KI$ are more disruptive (Fig.~\ref{fig2}A, blue). This relationship remains robust even when focusing exclusively on disruptive papers (disruption index $>0$), suggesting that increasing $\KI$ not only correlates with higher disruption scores but also raises the odds of a paper being disruptive (Fig.~\ref{fig2}A, green). This pattern is not an artifact of reference count, as it holds when controlling for the reference count within the focal paper (Extended Data Fig.~\ref{figS_RC}); Our observations are also robust to the variations of $\KI$ and disruption indices (definitions in Methods, and results in Extended Data Fig.~\ref{figS_AI}).

% CONTROLLING FOR IMPACT, TIME PERIOD, DISCIPLINE, AND NOVELTY
Next, we control for three properties of the focal paper: impact, publication date, scientific discipline. Starting with impact, it is measured as the number of citations the paper accumulates within the first five years post-publication, denoted by $C_5$. As impact increases, the association between disruption and $\KI$ becomes more pronounced, especially when the $\KI$ value is small (Fig.~\ref{fig2}B). Moving on to publication date, we find that the relationship between $\KI$ and disruption has strengthened in recent decades (Fig.~\ref{fig2}C). We also find a systematic decline in disruption over time, consistent with prior findings~\cite{park2023papers}. Next, we consider the discipline of the focal paper. As shown in Fig.~\ref{fig2}D, the relationship between $\KI$ and disruption holds universally, from natural to social sciences, suggesting that it is a fundamental feature of the interplay between knowledge recombination and scientific disruption.

% CONTROLLING FOR REFERENCE PROPERTIES
Having controlled for various properties of the focal paper, let us now control for the properties of the references within that paper. 
One aspect lies in atypical combinations of references at the journal and disciplinary level, as reflected in the novelty index and the interdisciplinarity index, respectively (details in Methods). The relationship between $\KI$ and disruption remains robust after controlling for both indices (Extended Data Fig.~\ref{figS_RN}B-C, Extended Data Fig.~\ref{figS_RM}B-C). 
This addresses a critical concern: while established atypicality metrics (novelty and interdisciplinarity) primarily measure the conceptual distance based on meso-level co-occurrence (journals or disciplines), $\KI$ captures a distinct and more fine-grained micro-level structural facet: the direct independence of the relationship among papers. $\KI$'s robust and independent predictive power---even after controlling for these rivals---confirms that this structural dimension is a unique and non-redundant component.
Another aspect of the focal paper's references lies in their statistical patterns, such as average age and impact. In terms of reference age, both papers with high and low $\KI$ tend to cite relatively younger references compared to those cited by papers with intermediate $\KI$ values, i.e., those balancing dependent and independent knowledge (Extended Data Fig.~\ref{figS_RA}A). As for reference impact, papers with higher $\KI$ exhibit a preference for less impactful (Extended Data Fig.~\ref{figS_RI}A). Importantly, the association between $\KI$ and disruption remains robust after controlling for reference age (Extended Data Fig.~\ref{figS_RA}B-C) and reference impact (Extended Data Fig.~\ref{figS_RI}B-C).
% Another aspect of the focal paper's references lies in their statistical patterns, such as average age and average impact, and average disruption. In terms of reference age, both papers with high and low $\KI$ tend to cite relatively younger references compared to those cited by papers with intermediate $\KI$ values, i.e., those balancing dependent and independent knowledge (Extended Data Fig.~\ref{figS_RA}A). Furthermore, papers with higher $\KI$ exhibit a preference for less impactful (Extended Data Fig.~\ref{figS_RI}A), but more disruptive references (Extended Data Fig.~\ref{figS_RD}A). Importantly, the association between $\KI$ and disruption remains robust after controlling for reference age (Extended Data Fig.~\ref{figS_RA}B-C), reference impact (Extended Data Fig.~\ref{figS_RI}B-C), and reference disruption (Extended Data Fig.~\ref{figS_RD}B-C).

% CONTROLLING FOR TEAM COMPOSITIONS
We further investigate whether team composition influences the relationship between $\KI$ and disruption. We measure team size~\cite{wu2019large}, geographic distance~\cite{lin2023remote}, and collaborative freshness~\cite{zeng2021fresh} of the focal paper (definitions in Methods) and then control for these variables. Our results reveal that the positive correlation between disruption and $\KI$ persists across varying team size (Extended Data Fig.~\ref{figS_TS}). Similarly, the inherent advantages of onsite and fresh teams do not alter the association between $\KI$ and disruption (Extended Data Fig.~\ref{figS_TD}-\ref{figS_TF}). These findings well dispel the concerns that team composition might confound the observed relationship, affirming the robustness of the link between $\KI$ and disruption.

% QUANTIFYING THE IMPACT EFFECTS VIA CEM
Motivated by the strong association between $\KI$ and disruption, we perform a matching experiment to estimate the effect size. To this end, we conduct randomized controlled trials via Coarsened Exact Matching~\cite{blackwell2009cem,iacus2012causal} (CEM, details in Methods). In particular, each ex-ante covariate is coarsened into bins with appropriate scheme, allowing for the exact matching within these coarsened strata; comprehensive robustness analyses on different coarsening schemes are displayed in Supplementary Materials (Section~\ref{SI_CEM}). The effect size, characterized as the average treatment effect on the treated (ATT), is then estimated by a weighted regression model on the matched sample using the CEM-generated weights. Fig.~\ref{fig2}E shows the pairwise ATTs across varying treated and control group sizes. The results reveal that for each controlled size, the ATT increases as the treated size grows, transiting from negative to positive when the treated size approaches the controlled size. This robustly confirms a positive effect of increasing $\KI$ on disruption. To ensure our causal estimates are robust to the matching algorithm, we supplemented our primary CEM analysis with Propensity Score Matching (PSM). This additional analysis yields fully consistent results, reinforcing the significant positive impact of $\KI$ on disruption (see Section~\ref{SI_PSM} in Supplementary Materials).

\subsection*{Knowledge independence dominantly predicts disruption and explains its decline}
The above analyses establish $\KI$ as a robust predictor of disruption. What remains unknown is whether $\KI$ can resolve the explanatory gap of atypicality metrics? To find out, we employed normalized ordinary least-squares (OLS) regression to competitively test $\KI$ against all other covariates.

\begin{table}[ht] % Do NOT use \begin{table*}
    \centering
    % Captions go above tables
    \caption{\textbf{$|$ Normalized regression models of disruption and the influencing factors.}}
    \label{table1} % give each table a logical label name
    \renewcommand\tabcolsep{0.16cm} % column spacing
    \renewcommand{\arraystretch}{0.8} % line spacing
    \begin{threeparttable}
    \fontsize{7.5}{14}\selectfont % 设置表格字体大小为 8.75pt，行距为 12pt
    \begin{tabular}{lcccccccccc}
        \hline  % top line
        & Model 1 & Model 2 & Model 3 & Model 4 & Model 5 & Model 6 & Model 7 & Model 8 & Model 9 & $R^2$\% \\ 
        \hline  % middle line 1
        $\KI$ & \textbf{0.4389}*** & \textbf{0.4389}*** & \textbf{0.4409}*** & \textbf{0.4178}*** & \textbf{0.4399}*** & \textbf{0.4382}*** & \textbf{0.4364}*** & \textbf{0.4335}*** & \textbf{0.4134}*** & \textbf{71.87} \\
         & (0.000) & (0.000) & (0.000) & (0.000) & (0.000) & (0.000) & (0.000) & (0.000) & (0.000) \\
        % Log $C_5$ &  & -0.0416*** &  &  &  &  &  &  &  &  &  & -0.0322*** & 3.91 \\
         % &  & (0.000) &  &  &  &  &  &  &  &  &  & (0.000) \\
        Novelty &  & 0.0001 &  &  &  &  &  &  & 0.0002 & 0.00 \\
         &  & (0.000) &  &  &  &  &  &  & (0.000) \\
        Interdisciplinarity &  &  & -0.0120*** &  &  &  &  &  & -0.0061*** & 0.61 \\
         &  &  & (0.000) &  &  &  &  &  & (0.000) \\
        Reference count &  &  &  & -0.0623*** &  &  &  &  & -0.0579*** & 10.82 \\
         &  &  &  & (0.000) &  &  &  &  & (0.000) \\
        Reference age &  &  &  &  & -0.0148*** &  &  &  & -0.0148*** & 0.12 \\
         &  &  &  &  & (0.000) &  &  &  & (0.000) \\
        % Reference $C_5$ &  &  &  &  &  &  & -0.0197*** &  &  &  &  & -0.0168*** & 0.39 \\
        %  &  &  &  &  &  &  & (0.000) &  &  &  &  & (0.000) \\
        % Reference disruption &  &  &  &  &  &  &  & 0.2293*** &  &  &  & 0.2231*** & 37.17 \\
        %  &  &  &  &  &  &  &  & (0.000) &  &  &  & (0.000) \\
        Team size &  &  &  &  &  & -0.0147*** &  &  & -0.0159*** & 0.68 \\
         &  &  &  &  &  & (0.000) &  &  & (0.000) \\
        Team distance &  &  &  &  &  &  & -0.0307*** &  & -0.0256*** & 1.21 \\
         &  &  &  &  &  &  & (0.000) &  & (0.000) \\
        Team freshness &  &  &  &  &  &  &  & 0.0470*** & 0.0469*** & 2.45 \\
         &  &  &  &  &  &  &  & (0.000) & (0.000) \\
        \hline % middle line 2
        Discipline fixed effects  & Yes & Yes & Yes & Yes & Yes & Yes & Yes & Yes & Yes & 10.60 \\
        Year fixed effects & Yes & Yes & Yes & Yes & Yes & Yes & Yes & Yes & Yes & 1.64 \\
        \hline % middle line 3
        N & 33304580 & 33304580 & 33304580 & 33304580 & 33304580 & 33304580 & 33304580 & 33304580 & 33304580 \\
        $R^2$ & 0.225 & 0.225 & 0.225 & 0.228 & 0.225 & 0.225 & 0.226 & 0.227  & 0.231 \\
        \hline  % bottom line
    \end{tabular}
    \begin{tablenotes}
        \small
        \item[$\dagger$] We perform normalized ordinary-least-squares (OLS) regression analyses to examine the effect of $\KI$ on disruption, while controlling for various covariates. Each regression coefficient is tested against the null hypothesis (coefficient equals 0) using a two-sided \textit{t}-test. Standard errors are provided in parentheses for each coefficient. Note that we do not apply adjustments for multiple hypothesis testing in this analysis. (*$p < 0.05$, **$p < 0.01$, ***$p < 0.001$).
    \end{tablenotes}
    \end{threeparttable}
\end{table}

% RESOLVE THE IMPASSE ON NOVELTY INDEX
The results are conclusive. In the full model (Table~\ref{table1}, Model 9) accounting for all ex-ante variables, $\KI$ maintains its large, positive, and significant association with disruption ($0.4134$), with a dominant share in accounting for $R_2$ ($71.87\%$). In contrast, the atypicality metrics, both novelty index and interdisciplinarity display a minor association ($0.0002$ and $-0.0061$) with low share in $R_2$ ($0.00\%$ and $0.61\%$). This finding persists when applied to alternative disruption metrics (Table~\ref{table_SI_ALL_OLS} in Section~\ref{SI_ALTER_DISRUPTION}), suggesting that the key to disruptive innovation is not the combination of conceptually distant knowledge (novelty or interdisciplinarity), but the integration of independent knowledge ($\KI$).

\textcolor{black}{
$\KI$'s dominant predicting power provides an immediate and direct mechanistic explanation for the first paradox, i.e., the systematic decline in disruption~\cite{park2023papers}. As shown in Fig.~\ref{fig1}C, $\KI$ itself has undergone a steady, decades-long decline across all disciplines. Meanwhile, we find that the relationship between $\KI$ and disruption has strengthened in recent decades and accompanied with a systematic decline in disruption over time (Fig.~\ref{fig2}C), consistent with prior findings~\cite{park2023papers}.
One might argue that this parallel decline could be caused by mere chance or by the structural evolution of the citation network itself. We rule out this possibility with our randomized rewiring Monte Carlo analysis (see Section~\ref{SI_Monte_Carlo_Rewire} in Supplementary Materials): the randomized network fails to reproduce the decline trend of $\KI$, signaling that it is a consequence of a behavioural shift in the way knowledge is combined. The randomized network also fails to replicate the $\KI$-disruption association, further indicating that the phenomenon we observe is derived from non-trivial scientists' behaviour.
This temporal alignment---where the intrinsic property ($\KI$) and the extrinsic outcome (disruption) decline in parallel---provides a plausible explanation for why disruption is decreasing.}
\color{black}

% REPRODUCE RESULTS ON PATENT DATA
Finally, we extended this analyzing framework to the technological domain. Results from the OECD patent dataset consistently show that a higher $\KI$ is also a key driver of technological disruption (see Section~\ref{SI_OECD_results} in Supplementary Materials). The consistency of these results across the scientific and technological domains strongly reinforces the predictive power of the $\KI$ metric on disruption.

\subsection*{Small, onsite, and fresh teams prefer independent knowledge}
% EMPHASIZE THAT $\KI$ IS AN INHERENT PROPERTY OF PAPERS
Despite the extensive utilization and discussion on scientific disruption, one critical question remains unexplored: what type of papers tend to be disruptive? The revealed dominant effect of $\KI$ on disruption offers a quantitative answer, bridging the collective measure of peer recognition with an inherent property of the paper. This insight prompts us to check if previously identified correlates of disruption are, in fact, driven or mediated by $\KI$.
\begin{figure}[ht]
    \begin{center}
        % \resizebox{17cm}{!}{\includegraphics{figures/FigOA_4_team_preference.pdf}}
        \includegraphics[width=1\textwidth]{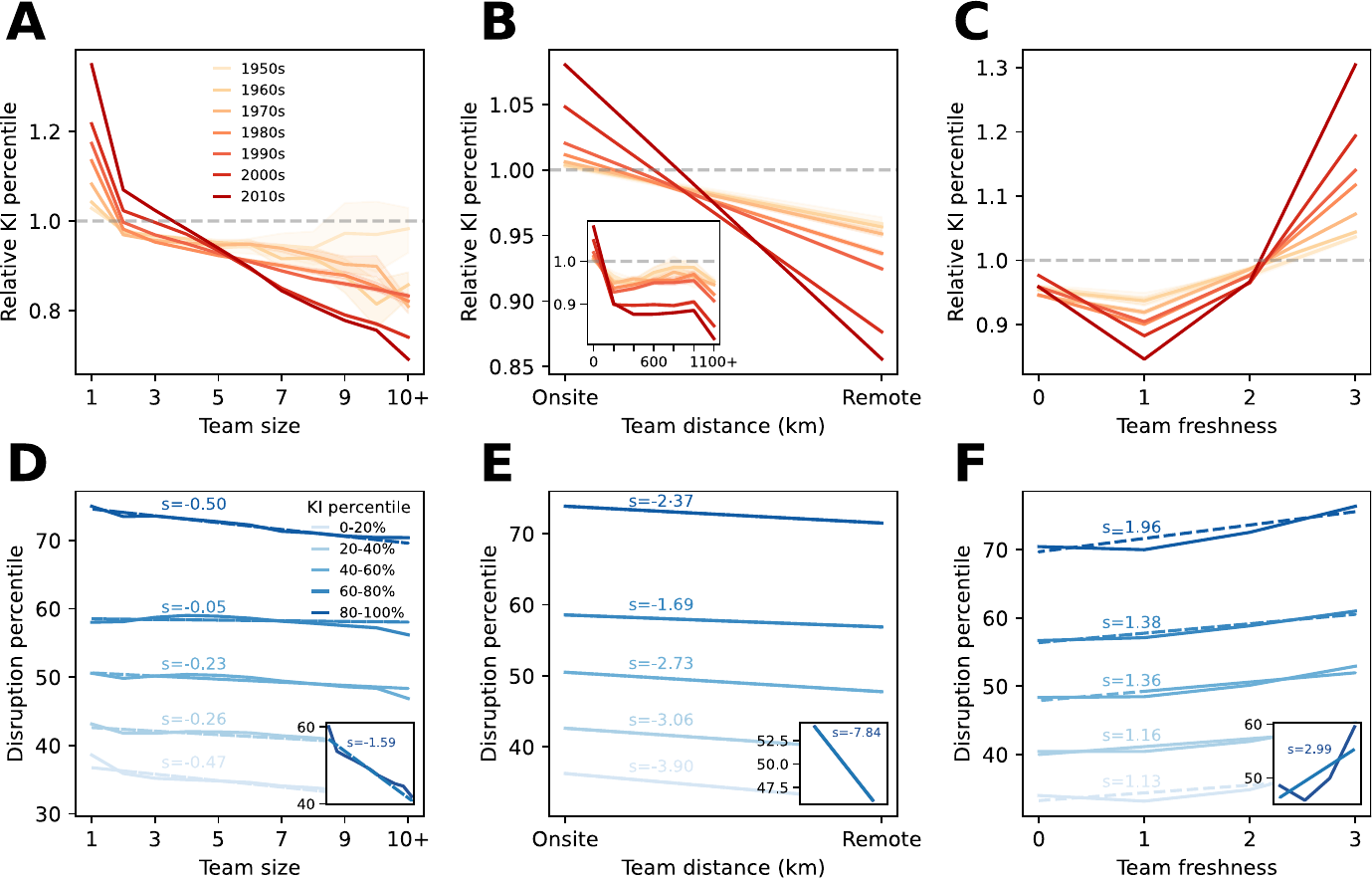}
        \caption{\textbf{$|$ Team's preference for knowledge independence.}
            (\textbf{A-C}) The $\KI$ percentiles in each bin are rescaled by the average value of the respective time period to highlight the trends. Curves with larger bounds are displayed in the inset to improve visualization. Over time, the relative $\KI$ percentile decreases with team size (panel~\textbf{A}) and team distance (panel~\textbf{B}), and increases with team freshness (panel~\textbf{C}).
            Here we calculate the geographic team distance by the coordinate information of affiliations, and categorize teams into two types---onsite ($\leq100$ km) and remote ($>100$ km) teams (panel~\textbf{B}). The relationship between $\KI$ and average collaboration distance is displayed in inset of panel~\textbf{B}.
            (\textbf{D-F}) Solid lines depict the relationships between disruption and team properties with fixed $\KI$ values, alongside the uncontrolled cases (insets). Dashed lines represent linear fitted curves. Teams with higher $\KI$ are consistently more disruptive. Moreover, the slopes of the fitted curves with fixed $\KI$ are flatter compared to the uncontrolled curves (insets).
            Bootstrapped 95\% confidence intervals are shown as shaded bands.
            \label{fig3_team_preference}}
    \end{center}
\end{figure}

% TEAM COMPOSITIONS RELATED TO DISRUPTION
The most well-known research on disruption stems from team science. Prior studies have highlighted the impact of team composition on disruption, emphasizing factors such as team size~\cite{wu2019large}, geographic distance~\cite{lin2023remote}, and collaboration freshness~\cite{zeng2021fresh}. These studies suggest that small, onsite, and fresh teams are more likely to produce disruptive research. Based on these insights and our discovery of the link between $\KI$ and disruption, we investigate whether such teams tend to build on independent knowledge. Indeed, Fig.~\ref{fig3_team_preference}A-C confirms that this has been the case for decades, indicating that $\KI$ may serve as a mediator in producing disruptive research. 
To explore this possibility, we analyze the relationship between disruption and team composition while controlling for $\KI$. Our investigations (Fig.~\ref{fig3_team_preference}D-F) reveal that teams favoring higher $\KI$ systematically produce more disruptive research. However, once $\KI$ is controlled for, the slope of the disruption–team relationship becomes notably weaker compared to the uncontrolled case (inset, Fig.~\ref{fig3_team_preference}D-F), indicating the existence of the mediating role of $\KI$. Mediation analyses~\cite{nguyen2021clarifying} (details in Methods) further confirm the non-trivial mediating effect from $\KI$, particularly when compared to other variables (Extended Data Table~\ref{tableS_mediation_TS}-\ref{tableS_mediation_TF}). These findings persist when applied to alternative disruption metrics (Table~\ref{table_SI_ALL_mediation_TS}-\ref{table_SI_ALL_mediation_TF} in Section~\ref{SI_ALTER_DISRUPTION}), suggesting that a team's preference for independent knowledge amplifies its disruptive potential, regardless of its size, geographic distance, or collaboration freshness.

\subsection*{Knowledge independence draws limited and delayed recognition}
\begin{figure}[htbp]
    \begin{center}
        % \resizebox{17cm}{!}{\includegraphics{figures/FigOA_3_paradox.pdf}}
        \includegraphics[width=1\textwidth]{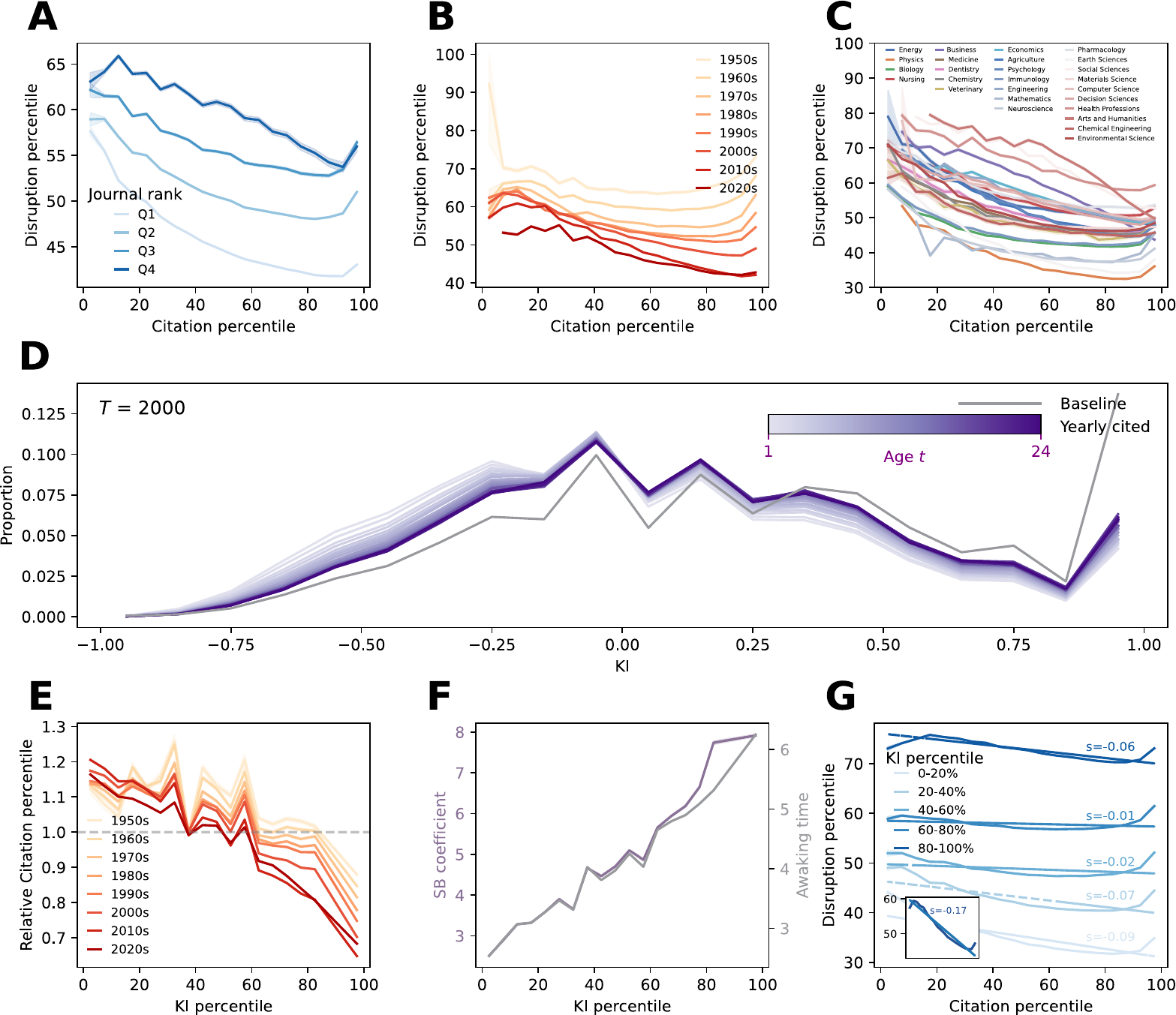}
        \caption{\textbf{$|$ Knowledge independence explains the negative relationship between disruption and impact.}
            We observe a universal negative relationship between papers' disruption and impact, regardless of the journal rank (panel~\textbf{A}), time period (panel~\textbf{B}), and discipline (panel~\textbf{C}). Here, a paper's impact is measured by the percentile of its total citations within the same publication year and discipline, and the journal rank is based on Scimago Journal Rank.
            (\textbf{D}) For papers published in the year $T=2000$, we measure the $\KI$ distribution among these papers as the baseline distribution (gray curve). Then, for all papers published in a subsequent year, $T+t$ ($t\geq 1$), we examine their bibliographies, extract the references that were published in $T$, and generate the $\KI$ distribution of those references (purple curve). Compared to the baseline, papers with negative $\KI$ are overrepresented while those with positive $\KI$ are underrepresented. Moreover, as the cited age $t$ increases, the corresponding distribution converges towards the baseline (see Extended Data Fig.~\ref{figS_PS} for different values of $T$).
            (\textbf{E}) The citation percentiles in each bin are rescaled by the average value of the respective time period to highlight the trends. Over time, the relative citation percentile decreases with the $\KI$ percentile.
            (\textbf{F}) A paper with a higher $\KI$ is more likely to become a Sleeping Beauty (SB coefficient, blue curve, left y-axis) and takes a longer silence to usher in the citation burst (Awaking time, green curve, right y-axis). The SB coefficient and Awaking time~\cite{ke2015defining} capture a delayed scientific recognition by quantifying the convexity and the age of abrupt change in the distribution of annual citations of a paper.
            (\textbf{G}) Solid lines depict the relationships between disruption and impact with fixed $\KI$ values, alongside the uncontrolled case (inset). Dashed lines represent linear fitted curves. For papers of comparable impact, those with higher $\KI$ are consistently more disruptive. Moreover, the slopes of the fitted curves with fixed $\KI$ are significantly flatter compared to the uncontrolled curve (inset).
            Bootstrapped 95\% confidence intervals are shown as shaded bands.
            \label{fig3_paradox}}
    \end{center}
\end{figure}

% THE NEGATIVE RELATIONSHIP BETWEEN DISRUPTION AND IMPACT
A well-documented phenomenon in the study of scientific impact is the negative correlation between a paper's disruption index and its citation-based impact~\cite{bornmann2020disruption,liang2023bias,zeng2023disruptive}. We provide evidence indicating that this pattern persists across journal ranks, publication years, and disciplines (Fig.~\ref{fig3_paradox}A-C). However, the mechanism underlying this correlation remains to be explored. Considering that both the disruption and citation are the extrinsic measures of the collective peer recognition, it indicates that there might be some symbiosis among these citation-based extrinsic metrics, and the correlation between them is likely to be a spurious or biased association stemming from a more intrinsic attribute.

% THE CONFUNDING ROLE OF $\KI$ BENEATH THE NEGATIVE RELATIONSHIP
Given the advantages of interconnected knowledge in facilitating knowledge dissemination and adoption~\cite{nooteboom2009cognitive,deichmann2020ideas}, it is reasonable to hypothesize that higher $\KI$ may hinder citation-based impact. To test this, for any given year $T$, we measure how $\KI$ is distributed across all papers published in that year, thereby forming a baseline distribution. Then, for each subsequent year, $T+t$ ($t\geq 1$), we examine the bibliographies of all papers published in that year, and extract only the references that were published in year $T$. Finally, we compute the $\KI$ distribution of the extracted references---where a paper may be counted multiple times if cited more than once---and compare it against the baseline distribution. The results show a systematic shift towards lower $\KI$ values, i.e., papers with negative $\KI$ values are consistently overrepresented among cited references compared to the baseline distribution, while those with positive $\KI$ values are consistently underrepresented (Fig.~\ref{fig3_paradox}D). This pattern supports our hypothesis, indicating a negative correlation between $\KI$ and citation-based impact (Fig.~\ref{fig3_paradox}E). In addition, as cited age $t$ increases, the $\KI$ distribution of yearly cited papers gradually converges towards the baseline distribution, suggesting that the citation disadvantage of higher-$\KI$ papers diminishes over time. This delayed recognition indicates that higher-$\KI$ papers are more likely to be ``sleeping beauties''~\cite{ke2015defining}, remaining dormant for longer periods before experiencing citation bursts (Fig.~\ref{fig3_paradox}F).

The analyses above verify that $\KI$ is associated with both limited and delayed academic recognition, suggesting that the negative relationship between disruption and impact may be influenced by $\KI$.
Given that both disruption and recognition are ex-post variables relative to $\KI$, their relationship is more potentially and logically depicted as being confounded by $\KI$, an ex-ante variable that precedes and affects both outcomes.
To explore this possibility, we analyze the relationship between disruption and impact while controlling for $\KI$. Our results (Fig.~\ref{fig3_paradox}G) show that among papers with comparable impact, disruption systematically increases with $\KI$. Notably, once $\KI$ is controlled for, the negative slope of the disruption-impact relationship becomes significantly weaker compared to the uncontrolled case (inset, Fig.~\ref{fig3_paradox}G), highlighting the confounding potential of $\KI$. Further confounding analyses reveal that $\KI$ accounts for a substantial portion of this observed negative relationship, surpassing the explanatory power of other factors (Extended Data Table~\ref{tableS_confounding_impact}), which provides additional analytical evidence for $\KI$'s confounding role. These findings persist when applied to alternative metrics of disruption (Table~\ref{table_SI_ALL_confounding_impact} in Section~\ref{SI_ALTER_DISRUPTION}), supporting our hypothesis that $\KI$ increases the challenge of knowledge diffusion while simultaneously enhancing the disruptive potential of scientific work.

\section*{Discussion}
% SUMMARY OF FINDINGS
% This paper offers new insights into two open questions: Why has disruption declined systematically despite unprecedented knowledge growth? And why is disruption associated with fewer citations and delayed recognition? To answer these questions, we identify an intrinsic property of scientific papers, knowledge independence ($\KI$), that reliably predicts disruption. We also show that this property mediates the disruptive advantage of small, onsite, and fresh teams, confirming its role as a core driver of disruptive processes. Moreover, the decades-long decline of $\KI$ provides a mechanistic explanation for the parallel decline in disruption. Finally, causal inference methods, coupled with Monte Carlo simulations, suggest that $\KI$ is the underlying driver behind the persistent trade-off between disruption and impact, as it simultaneously fosters the former while constraining the latter.
To explain why disruption declined systematically and associated with fewer citations and delayed recognition, we identify an intrinsic property of scientific papers, knowledge independence ($\KI$), that reliably predicts disruption. We also show that this property mediates the disruptive advantage of small, onsite, and fresh teams, confirming its role as a core driver of disruptive processes. Moreover, the decades-long decline of $\KI$ provides a mechanistic explanation for the parallel decline in disruption. Finally, causal inference methods coupled with Monte Carlo simulations, suggest that $\KI$ is the underlying driver behind the persistent trade-off between disruption and impact.

Our work complements prior studies examining the relationship between team composition and scientific disruption, which highlight the disruptive potential of small~\cite{wu2019large}, onsite~\cite{lin2023remote}, and fresh teams~\cite{zeng2021fresh}. In particular, we find that such teams exhibit a stronger preference to build on independent knowledge. Moreover, we demonstrate that the relationship between team composition and disruption weakens when controlling for $\KI$. These findings highlight the role $\KI$ as a mediator, offering an additional boost to disruption. Consequently, our work suggests a practical strategy for scientists aiming to achieve disruptive discoveries: researchers should prioritize ideas composed of previously unlinked but implicitly related knowledge in their collaborative efforts, regardless of team size, location, or collaboration experience.

The decades-long decline of $\KI$ provides a mechanistic explanation for the decline in disruption. One may argue that this parallel decline might be caused by mere chance, i.e., by the structural evolution of the citation network itself. We rule out this possibility with our randomized rewiring Monte Carlo analysis (see Section~\ref{SI_Monte_Carlo_Rewire} in Supplementary Materials): the randomized network fails to replicate the decline of $\KI$, signaling that it is a consequence of a behavioural shift in the way knowledge is combined. The randomized network also fails to reproduce the $\KI$-disruption association, further indicating that the phenomenon we observe is derived from non-trivial scientists' behaviour. This temporal alignment---where the intrinsic property ($\KI$) and the extrinsic outcome (disruption) decline in parallel---provides a plausible explanation for why disruption is decreasing.

% EXPLAIN THE NEGATIVE RELATIONSHIP BETWEEN DISRUPTION AND IMPACT
Moreover, our work extends previous observations~\cite{bornmann2020disruption,liang2023bias,zeng2023disruptive} on the negative relationship between disruption and impact, demonstrating that this pattern persists across journal rank, publication year, and discipline. More importantly, we offer a novel mechanistic explanation for this phenomenon. Our analysis reveals that the seemingly negative relationship between disruption and impact is confounded by an intrinsic property of the paper: its knowledge independence ($\KI$). When controlling for $\KI$, this relationship is substantially attenuated. Further confounding analysis confirms that $\KI$ plays a dominant role, suggesting that it drives the observed relationship between these two extrinsic, citation-based metrics. Intuitively, $\KI$ plays a dual role: when a paper is grounded in independent knowledge, the non-redundant information it introduces allows it to function as a ``knowledge broker'', attracting dedicated niche attention and thus increasing its disruptive potential. However, the very lack of explicit connections between the combined knowledge simultaneously impedes its broader diffusion, leading to lower and more delayed recognition.

To rigorously test our mechanism against the null hypothesis of trivial structural determinism---the idea that our finding is a mere artifact of random network evolution---we conducted a suite of Monte Carlo simulations. First, our falsification test (see Section~\ref{SI_Monte_Carlo_Rewire} in Supplementary Materials), which randomizes the empirical network while preserving its core structure, finds that the $\KI$-Disruption and $\KI$-Impact correlations completely disappear. This provides evidence that our findings are not driven by a trivial, universal network property but a substantive, but rather by non-random pattern of scientific behavior. 
Second, our generative Monte Carlo simulations (see Section~\ref{SI_Monte_Carlo_Generate} in Supplementary Materials) serves to validate the explanatory sufficiency of our proposed mechanism, rather than to identify predictors. We find that standard network models cannot reproduce this phenomenon. By embedding the specific cognitive preference for independent knowledge---which was empirically isolated as the dominant driver in our earlier analyses---into the model, we demonstrate that this mechanism is sufficient to reproduce the work that is highly disruptive but less recognized.
This comprehensive evidence, from large-scale statistical observations to targeted mechanistic simulations, supports our conclusion that a paper's intrinsic knowledge independence is the non-trivial, human-driven missing link that explains the trade-off between disruption and recognition in science.

\section*{Methods}

\subsection*{Data}
Our data source is from the OpenAlex~\cite{priem2022openalex}, which includes over 114 million publications from 1950 to 2024, with 
92,659,091 articles, 3,411,018 reviews, 2,693,331 preprints, and 1,391,755 letters. Other types of records that fall out of these ``traditional'' forms of articles are excluded from our analyses. The OpenAlex consolidates 244 subject categories into 26 major fields of discipline: Energy, Physics, Biology, Nursing, Business, Medicine, Dentistry, Economics, Chemistry, Psychology, Veterinary, Immunology, Mathematics, Engineering, Agriculture, Neuroscience, Pharmacology, Earth Sciences, Social Sciences, Computer Science, Materials Science, Decision Sciences, Health Professions, Arts and Humanities, Chemical Engineering, and Environmental Science. We also replicate our main findings on two additional datasets, SciSciNet~\cite{lin2023sciscinet} and Web of Science (WOS). The SciSciNet dataset comprises approximately 130 million publications from 1950 to 2022, and the WOS dataset comprises approximately 53.8 million publications from 1955 to 2017. Similarly, we consider the ``traditional'' forms of articles in our analyses, such as journal and conference papers. The results based on SciSciNet and WOS datasets are provided in the Supplementary Materials (Section~\ref{SI_SciSciNet_results}, \ref{SI_WebofScience_results}). We further extend our analysis on the relationship between $\KI$ and disruption to technical area based on OECD patent dataset (latest version: Feb-2025), which comprises approximately 30 million patents filed at the European Patent Office (EPO) and the United States Patent and Trademark Office (USPTO), or via the Patent Co-operation Treaty (PCT) from 1978 to 2024. The results based on OECD patent dataset are provided in the Supplementary Materials (Section~\ref{SI_OECD_results}).

\subsection*{Knowledge independence}
Four alternative measures of knowledge independence are proposed for comparison. $\KI$ is the principal definition used in the main text. $\KI^{'}$ simplifies the equation by only measuring the fraction of $ind$-type references. $\KI_{\rm{adj}}$ subtracts 1 from the count of $ind$-type references, to account for the oldest reference in the bibliography, which cannot cite any other reference in that list. Finally, $\KI^{'}_{\rm{adj}}$ simplifies the previous equation by only measuring the fraction of $ind$-type reference. More formally, the four alternative measures are defined as follows:
\begin{itemize}
    \item $\KI=\dfrac{n_{\rm ind}-n_{\rm dep}}{n_{\rm ind}+n_{\rm dep}}$
    \item $\KI^{'}=\dfrac{n_{\rm ind}}{n_{\rm ind}+n_{\rm dep}}$
    \item $\KI_{\rm{adj}}=\dfrac{(n_i-1)-n_{\rm dep}}{(n_i-1)+n_{\rm dep}}$
    \item $\KI^{'}_{\rm{adj}}=\dfrac{n_{\rm ind}-1}{(n_i-1)+n_{\rm dep}}$
\end{itemize}

\subsection*{Disruption}
The groundbreaking work of Funk and Owen-Smith~\cite{funk2017dynamic} proposed an index that analyzes patent citations to quantify technological change. This index was then advocated by the highly influential work of Wu et al.~\cite{wu2019large}, who coined the term ``disruption index'' and were the first to apply it in the analysis of scientific papers. The basic idea behind this index is that a disruptive work shifts the focus of subsequent research from its prior knowledge to itself. Therefore, the disruption index is calculated as $D=\dfrac{c_i-c_j}{c_i+c_j+c_k}$, where $c_i$ represents the number of works that only cite the focal work, $c_j$ represents the number of works that cite both the focal work and its references, and $c_k$ represents the number of subsequent works that cite the references of the focal work but not the focal work itself. There has been ongoing debate on the rationale behind the inclusion of the term $c_k$ in the formula~\cite{wu2019confusing,bornmann2020disruption,leibel2024we}. Specifically, $c_k$ quantifies the attention directed towards the focal work's references while bypassing the focal work itself. Under fixed $c_i$ and $c_j$, focal works with higher $c_k$ values would be considered less disruptive. However, a contradiction arises in cases of negative $D$, where an increase in $c_k$ paradoxically increases $D$ rather than decreasing it---contrary to the conceptual definition of disruption. To address this inconsistency, we adopt a modified definition of disruption that excludes the term $c_k$. This modified measure serves as the principal metric of disruption in our study, while the original definition is retained as an alternative for robustness check. Additionally, we implement an open time window to count the citations a focal work receives ($c^{o}$) and use a 5-year time window ($c^{5}$) as an alternative measure.

To summarize, the disruption measures that we consider are defined as follows: $D_0$ is the principal definition used in the main text. $D_1$ takes the form of original definition with an open time window $c^{o}$. $D_2$ is defined the same way as $D_0$, but with a different time window $c^{5}$. $D_3$ takes the form of original definition with a 5-year time window $c^{5}$. More formally, these measures are defined as follows:
\begin{itemize}
    \item $D_0=\dfrac{c^{o}_i-c^{o}_j}{c^{o}_i+c^{o}_j}$
    \item $D_1=\dfrac{c^{o}_i-c^{o}_j}{c^{o}_i+c^{o}_j+c^{o}_k}$
    \item $D_2=\dfrac{c^{5}_i-c^{5}_j}{c^{5}_i+c^{5}_j}$
    \item $D_3=\dfrac{c^{5}_i-c^{5}_j}{c^{5}_i+c^{5}_j+c^{5}_k}$
\end{itemize}

\subsection*{Novelty}
The seminal work of Uzzi et al.~\cite{uzzi2013atypical} demonstrated that papers built on atypical combinations of prior work are more likely to be highly influential. This finding is based on a novelty index, which quantifies the degree of atypicality in the combinations of references at the journal level. More specifically, this index is calculated by comparing the observed co-occurrence of journal pairs in actual citation networks with their randomized co-occurrence under randomized citation networks. This comparison allows for computing the z-score of each possible journal pair cited by a given focal paper. A high and positive z-score indicates a conventional combination of journal pairs, while a low and negative z-score reflects a novel combination that deviates significantly from randomized co-occurrence patterns. Following the way that novelty is computed in the seminal work of Uzzi et al.\cite{uzzi2013atypical} and SciSciNet~\cite{lin2023sciscinet} dateset, we calculate the 10th percentile of the z-score distribution of the journal pairs cited by the focal paper, and apply its negative value to represent the 90th percentile of novelty. This approach underscores the relatively top atypical combinations of references within the paper, capturing its degree of novelty.

\color{black}
\subsection*{Interdisciplinarity}
To assess the impact of a paper's interdisciplinary nature, we apply a interdisciplinarity index for each paper. This index is calculated as the Rao–Stirling indicator~\cite{stirling2007general,rafols2010diversity,zhang2016diversity,zhang2021scientometric,xiang2025evaluating} by combining three components: variety, balance, and disparity among the categories (e.g., scientific fields) to which a focal paper's bibliography belongs. Specifically, for each field $i$, compute its share $p_i$ in the bibliography with whole counting method, which captures variety and balance. Then, for each pair of fields ($i,j$), calculate a distance measure 
$d_{ij}$ based on its co-citation frequency annually, which represents how intellectually far apart the fields are, capturing disparity in each year. Finally, the Rao–Stirling interdisciplinarity index is computed as:
\begin{align}
    RS &= \sum\nolimits_{i,j(i\ne j)}^{n}p_ip_jd_{ij} \label{(eq_RaoStirling)}.
\end{align}
This formula increases when: more categories are represented, the distribution is more even, and the fields are cognitively more distant (less being co-cited).
\color{black}

\subsection*{Team distance}
Lin et al.~\cite{lin2023remote} reported the rise of remote collaborations that span locations and time zones. These collaborations are less likely to produce disruptive works as they tend to focus more on technical tasks rather than conceptual ones. The original calculation of team distance relies on the latitude and longitude of the affiliations listed in a paper's byline. 
We follow this methodology to calculate the geographic team distance by the coordinate information of affiliations, which can further categorize teams into two types---onsite ($\leq100$ km) and remote ($>100$ km) teams as defined by Lin et al~\cite{lin2023remote}.
% Due to the absence of coordinate information in the Web of Science dataset, we instead use the following hierarchical categorization of the affiliations:
% \begin{itemize}
% \item $0$: Sub-organization / Department level; 
% \item $1$: Organization / University level; 
% \item $2$: City level; 
% \item $3$: State / Province level; 
% \item $4$: National level; 
% \item $5$: International level.
% \end{itemize}
% Using this hierarchy, the distance of any given team is categorized based on the highest level within the team. For instance, if there are three authors, two of which are in the same university, while the third is in another university situated at the same city, the distance will be categorized as $2$, implying that this is a city-level collaboration.

\subsection*{Team freshness}
Zeng et al.~\cite{zeng2021fresh} revealed a positive correlation between team freshness and scientific disruption, emphasizing the importance of new team dynamics in fostering innovative research. To quantify team freshness, we analyze the prior collaboration network of the authors involved in a given focal paper. In this network, the undirected edges represent prior collaborations between pairs of authors before coauthoring the focal paper. We classify team freshness into mutually exclusive categories based on the topological structure of the collaboration network, where $V$, $k$, and $n$ represent the node set, node degree, and network size, respectively. More specifically, the classification is as follows:
\begin{itemize}
\item $0$: All authors have previously collaborated with each other, $\{\forall v_i \in V, \ k_i = n-1\}$; 
\item $1$: All authors have at least one prior collaboration link, $\{\forall v_i \in V, 1 \leq \ k_i \leq n-1\}$; 
\item $2$: Some authors have at least one prior collaboration link, $\{\exists v_i \in V, \ k_i = 0\}$; 
\item $3$: No author has any prior collaboration link, $\{\forall v_i \in V, \ k_i = 0\}$.
\end{itemize}

\subsection*{Coarsened exact matching (CEM)}
Coarsened Exact Matching (CEM) is a non-parametric matching method that directly coarsens the covariates into bins and then performs exact matching within these coarsened bins. This ensures excellent covariate balance within the matched samples by guaranteeing that all matched units have identical values on the coarsened covariates. The methodological procedure follows:

\begin{itemize}
    \item \textbf{Treatment and Control Groups}: We categorize papers into treated and control groups based on their $\KI$ values.
    \item \textbf{Covariates:} We utilize the set of ex-ante covariates: focal paper properties (discipline and publication year), reference properties (novelty, interdisciplinarity, reference count, and average reference age), and team properties (size, geographic distance, and collaboration freshness).
    \item \textbf{Coarsening and matching:} For categorical and numerical covariates with limited distinct values, like discipline and team freshness, we directly use their natural strata for binning. For other numerical covariates, we choose appropriate coarsening schemes by comparing the weighted $L_1$ imbalance and standardized mean difference (SMD); comprehensive analyses of coarsening schemes are displayed in Section~\ref{SI_CEM}. CEM then identified strata where exact matches on these coarsened covariates could be found. Only units within these matched strata were included in the subsequent analysis.
    \item \textbf{Weighting:} To equalize the distribution of covariates between the groups, CEM weights are generated. Matched treated units are assigned a weight of $1$. Matched control units are weighted such that their distribution across strata mirrors that of the treated group. Formally, the weight for a control unit $i$ in stratum $s$ is defined as: $$w_i = \frac{m_C}{m_T} \cdot \frac{m_{T}^s}{m_{C}^s}$$ where $m_C$ and $m_T$ are the total matched sample sizes in controlled and treated group, and $m_{C}^s$ and $m_{T}^s$ are the number of controlled and treated units in stratum $s$.
    \item \textbf{Effect estimation:} To estimate the Average Treatment Effect on the Treated (ATT), we fit a Weighted Least Squares (WLS) regression model on the matched sample using the CEM-generated weights, which is specified as follows: 
    $$D = b + a_{ATT} \cdot T + \delta \cdot \rm Covariates + \epsilon$$
    Where: $D$ is the outcome variable of disruption. $T$ is the binary treatment indicator of $\KI$. Regarding doubly robust estimation, we control for the original Covariates to exploit within-stratum variation. The regression is weighted by the CEM weights ($w_i$). The coefficient of interest, $a_{ATT}$, captures the marginal effect of the treatment on the outcome after balancing ex-ante characteristics.
\end{itemize}

\subsection*{Normalized ordinary-least-squares (OLS) regression}
To analyze the relationships between scientific disruption and various ex-ante covariates, we employ normalized ordinary least squares (OLS) regression models. This approach allows us to evaluate whether $\KI$ has a dominant effect on disruption. The regression models include the following covariates: focal paper properties (discipline and publication year), reference properties (novelty, interdisciplinarity, reference count, and average age), as well as team properties (size, geographic distance, and collaboration freshness). All numerical variables are standardized by z-score normalization to ensure comparability of effect coefficients across models. We start with a model that does not control for any covariates other than fixing discipline and publication year (\ref{(eq_single)}). Then, we control for each of the following covariates separately (\ref{(eq_mix)}): focal paper's novelty, interdisciplinarity, reference count, reference age, team size, team distance, and team freshness. Finally, we consider a model that accounts for all those covariates (\ref{(eq_all)}). The results are displayed in Table~\ref{table1}.
\begin{align}
    D &= b_0 + a_0\KI + c_{0}F \label{(eq_single)},\\
    D &= b_0 + a_0\KI + a_{i}C_i + c_{0}F \label{(eq_mix)},\\
    D &= b_0 + a_0\KI + \sum\nolimits_{i=1}^{7}a_{i}C_i + c_{0}F \label{(eq_all)}.
\end{align}

\color{black}
To further quantify the relative importance of individual predictors within the full model (Model~9 in Table~1) and to systematically decompose the explained variance ($R^2$), we performed a dominance analysis~\cite{azen2003dominance}. This method systematically assesses each predictor's marginal contribution to $R^2$ across all possible sub-models, thereby attributing its unique and shared contributions to the total $R^2$. It establishes three types of dominance: conditional dominance (contribution in specific subsets), partial dominance (average contribution across specific subset sizes), and general dominance (overall average contribution across all subsets). The general dominance value, which inherently accounts for shared variance and avoids order dependence, serves as our primary measure of relative importance for each predictor. This approach allows for a robust and comprehensive comparison of the relative influence of all independent variables in the model.
\color{black}

\color{black}
\subsection*{Confounding Analysis}
Confounding is a critical challenge in causal inference, where an ex-ante variable $C$ can distort the true effect of a target exposure $X$ on an outcome $Y$~($C \rightarrow X, C \rightarrow Y$), leading to spurious associations or biased estimates~($X \dashrightarrow Y$). This analysis aims to rigorously estimate the causal effect of a target variable $X$ on an outcome $Y$ by identifying and adjusting for potential confounding factors that are common causes of both $X$ and $Y$, and are not causally affected by $X$. Here, we employ a model-based confounding analysis framework, comparing two parametric regression models. The first model, the Unadjusted Model (\ref{eq_unadjusted}), formulates the baseline association between the outcome $Y$ and the target variable $X$, while controlling for other relevant covariates $F$. The coefficient $\alpha_1$ in this model represents the unadjusted (or crude) effect of $X$ on $Y$. The second model, the Adjusted Model (\ref{eq_adjusted}), expands upon the first by explicitly incorporating the identified confounder $C$ alongside $X$ and other covariates $F$. The coefficient $\delta_1$ in this model captures the adjusted effect of $X$ on $Y$, after accounting for the confounding effect of $C$. The presence and magnitude of confounding are primarily assessed by observing the change in the estimated coefficient of $X$ when the confounder $C$ is introduced into the model. The relative change in estimate (RCE) is defined as $|\frac{\alpha_1-\delta_1}{\alpha_1}|$. A substantial RCE indicates that $C$ indeed played a confounding role, distorting the unadjusted association between $X$ and $Y$. This adjustment helps to isolate $X$'s independent contribution to $Y$. 
In all conducted confounding analyses~(i.e., citation impact as the target variable, and each of the ex-ante covariates as the potential confounder), we standardize the numerical variables by z-score normalization to ensure comparability of effect coefficients across models, with discipline and publication year controlled.
\begin{align}
Y &= \alpha_{0} + \alpha_{1}X + \alpha_{2}F \label{eq_unadjusted},\\
Y &= \delta_{0} + \delta_{1}X + \delta_{2}C + \delta_{3}F \label{eq_adjusted}.
\end{align}
\color{black}

\subsection*{Mediation analysis}
Over the past decades, mediation analysis has seen rapid methodological development~\cite{nguyen2021clarifying}, becoming particularly prevalent in biomedical and social sciences. This approach is often applied to data from randomized controlled trials, aiming to investigate whether and how the effect of a target variable $X$ on an outcome $Y$ operates through a mediating variable $M$~($X\rightarrow M \rightarrow Y$), where $M$ is an ex-post variable relative to $X$ and an ex-ante variable relative to $Y$. Here we utilize the model-based mediation framework, comprising two parametric regression models: Outcome Model (\ref{(eq_output)}) formulates the relation between the outcome $Y$ and the target variable $X$, as well as the potential mediator $M$, with covariates $F$ controlled. On the other hand, the Mediator Model (\ref{(eq_mediator)}) formulates the relation between the mediator $M$ and the target variable $X$, also controlling for covariates $F$. The indirect effect (ACME) is defined as the product of the coefficient $\beta_1$ of $X$ in model (\ref{(eq_mediator)}) and the coefficient $\theta_2$ of $M$ in model (\ref{(eq_output)}). This represents the portion of $X$'s effect on $Y$ mediated through $M$. The direct effect (ADE) is defined as the coefficient $\theta_1$ of $X$ in model (\ref{(eq_output)}), capturing the direct relationship between $X$ and $Y$, independent of $M$. The total effect (TE) is defined as the sum of the indirect effect (ACME) and direct effect (ADE). 
In all conducted mediation analyses~(i.e., team composition as the target variable, and each of the ex-ante covariates as the potential mediator), we standardize the numerical variables by z-score normalization to ensure comparability of effect coefficients across models, with discipline and publication year controlled. 
\begin{align}
    Y &= \theta_{0} + \theta_{1}X + \theta_{2}M + \theta_{3}F \label{(eq_output)},\\
    M &= \beta_{0} + \beta_{1}X + \beta_{2}F \label{(eq_mediator)}.
\end{align}

%%%%%%%%%%%%%%%% Reference %%%%%%%%%%%%%%%
\bibliographystyle{nature}
\bibliography{Knowledge_Independence_Disruption}

%%%%%%%%%%%%%%%% ACKNOWLEDGEMENTS %%%%%%%%%%%%%%%

\section*{Acknowledgements}
We thank P.~Holme and M.-Y.~Luo for helpful discussions. We thank A.-L.~Barabási for giving access to the Web of Science data. The support and resources of the High Performance Computing Center from Southwest University and New York University Abu Dhabi are gratefully acknowledged.

\subsection*{Funding:}
National Natural Science Foundation of China (NSFC), No.72374173 (T.~J.)\\
University Innovation Research Group of Chongqing, No.CXQT21005 (T.~J.)\\
Fundamental Research Funds for the Central Universities, No.SWU-XDJH202303 (T.~J.)\\
China Scholarship Council (CSC) program, No.202306990091 (X.-Y.~Y.)\\ 
Chongqing Graduate Research Innovation Project, No.CYB22129 (X.-Y.~Y.)

\subsection*{Author contributions:}
Conceptualization: X.-Y.~Y., T.~R. and T.~J.\\
Methodology: X.-Y.~Y. and T.~J.\\
Investigation: X.-Y.~Y.\\
Visualization: X.-Y.~Y.\\
Supervision: T.~R. and T.~J.\\
Writing---original draft, review \& editing: X.-Y.~Y., T.~R. and T.~J.

\subsection*{Competing interests:}
The authors declare no competing interests.

% \subsection*{Inclusion \& ethics:}
% All authors have agreed to all manuscript contents, the author list and its order, and the author contribution statements. Any changes to the author list after submission will be subject to approval by all authors.

\subsection*{Data and materials availability:}
Correspondence and requests for materials should be addressed to T.~R. or T.~J.

\noindent \textbf{Data availability:}
Our study draws on data from three publication datasets (OpenAlex~\cite{priem2022openalex}, SciSciNet~\cite{lin2023sciscinet}, and Web of Science) and one patent dataset (OECD-Patent). For OpenAlex dataset, you can access directly from https://docs.openalex.org/. For SciSciNet dataset\footnote{We use version 1 of SciSciNet, as version 2 updates its publication records based on OpenAlex, the same source as our main dataset, which would result in data redundancy.}, you can access directly from https://doi.org/10.6084/m9.figshare.c.6076908.v1. The Web of Science dataset is not publicly available, and was used under license from their respective publishers. If you are interested, you may request access to the API through Clarivate, which requires an additional subscription or permission (https://clarivate.com/academia-government/scientific-and-academic-research/research-discovery-and-\\referencing/web-of-science/web-of-science-core-collection/). For OECD-Patent dataset, you can access directly from https://www.oecd.org/en/data/datasets/intellectual-property-statistics.html.

\noindent \textbf{Code availability:}
% Computational codes for data processing, analysis, and model simulation are available upon request.
The computational codes for data processing, analysis, and visualization are available on a public repository (https://github.com/XiaoyaoYu95/Knowledge-Independence-Disruption).

\section*{Supplementary Materials}
\textbf{This PDF file includes:}\\
Supplementary Text~\ref{SI_KI_Robustness} to~\ref{SI_OECD_results}\\
Figures~\ref{FigOAS_Reference_count_KI} to~\ref{FigOECD_2_PSM_CEM_binary}\\
Tables~\ref{table_SI_ALL_OLS} to~\ref{table_OECD_OLS_binary}

\clearpage

%%%%%%%%%%%%%%%% EXTENDED FIGURES %%%%%%%%%%%%%%%

\setcounter{figure}{0} % figures below are counted start 1
\captionsetup[figure]{labelfont={bf},labelformat={default},name={Extended Data Fig.},labelsep=space,font=small}
\begin{figure}[htbp]
    \begin{center}
        % \resizebox{12cm}{!}{\includegraphics{figures/FigOAE_Reference_length_distribution.pdf}}
        \includegraphics[width=0.6\textwidth]{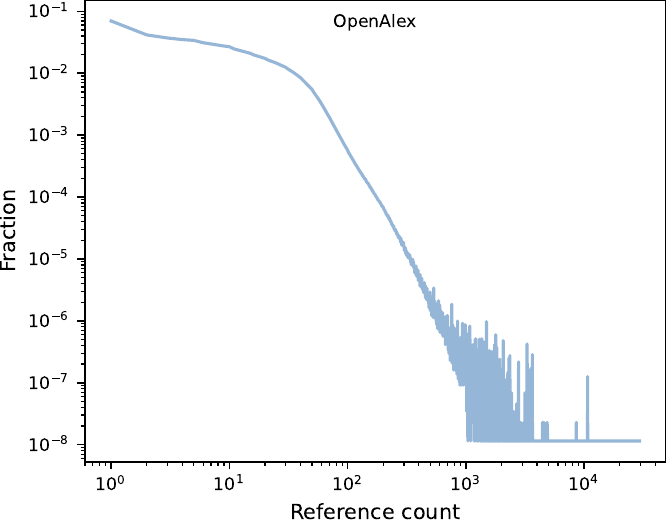}
        \caption{\textbf{$|$ Distribution of reference count.}
        The distribution of reference count for 87,419,190 articles recorded in OpenAlex follows a stretched exponential pattern, with a large number of papers containing relatively short reference lists.
        \label{figS_RC_Dis}}
    \end{center}
\end{figure}
\clearpage

\begin{figure}[htbp]
    \begin{center}
        % \resizebox{17cm}{!}{\includegraphics{figures/FigOAE_Reference_length.pdf}}
        \includegraphics[width=1.0\textwidth]{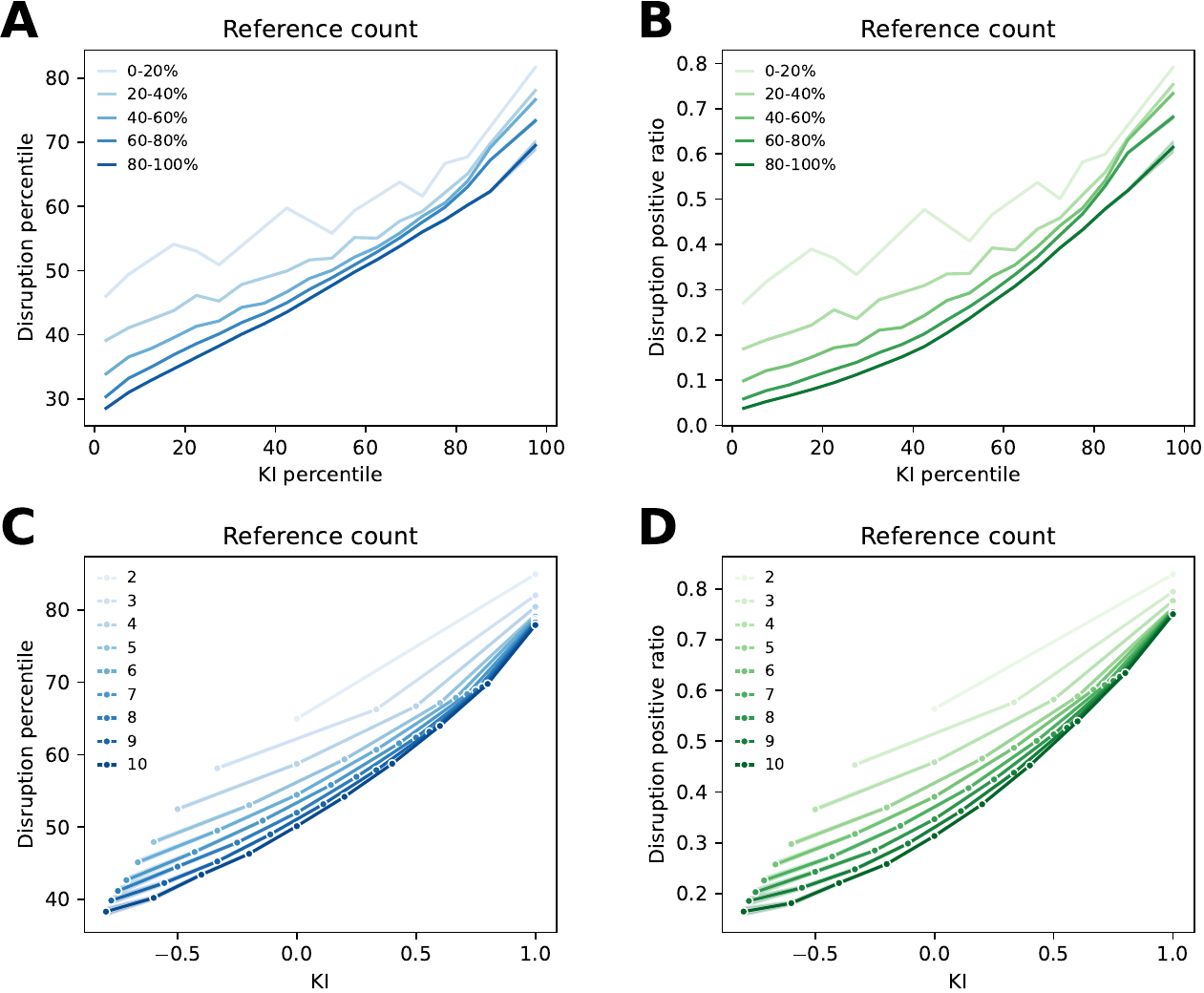}
        \caption{\textbf{$|$ The association between disruption and $\KI$ is robust across different reference counts.}
        (\textbf{A-B}) When controlling for the percentile of reference count, both the percentile and positive ratio of disruption continue to increase with $\KI$ across all levels of reference count.
        (\textbf{C-D}) When controlling for the reference count within the bottom group ($<=10$), where each reference count induces a limited range of $\KI$ values, both the percentile and positive ratio of disruption continue to increase with $\KI$ across all groups of reference count.
        \label{figS_RC}}
    \end{center}
\end{figure}
\clearpage

\begin{figure}[htbp]
    \begin{center}
        % \resizebox{17cm}{!}{\includegraphics{figures/FigOAE_alter_index.pdf}}
        \includegraphics[width=1.0\textwidth]{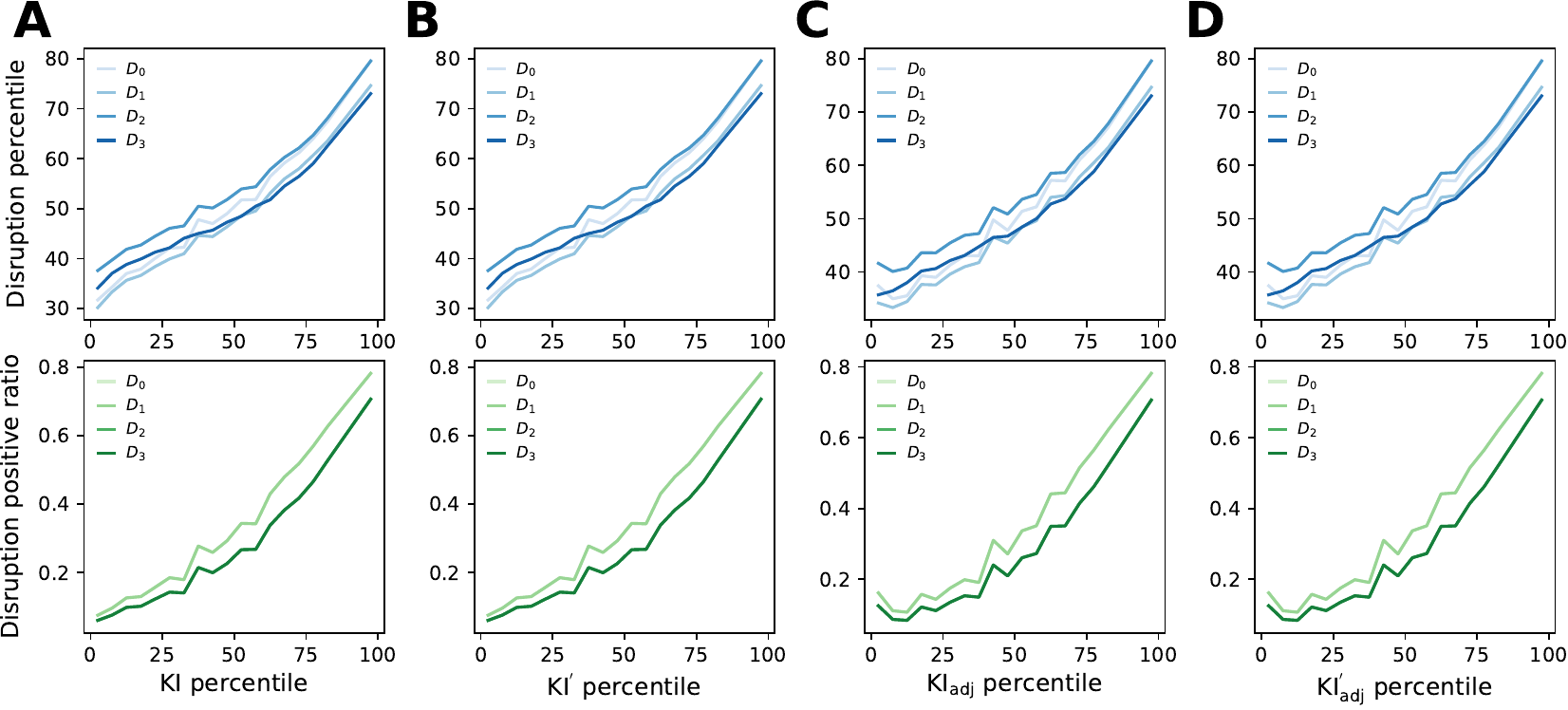}
        \caption{\textbf{$|$ The association between disruption and $\KI$ is robust across alternative measures.}
        (\textbf{A-D})
        \textbf{Upper panel}: from the perspective of percentile, all disruption measures (Methods) show a consistent increasing trend with all $\KI$ measures.
        \textbf{Lower panel}: from the perspective of positive ratio, all disruption measures similarly increase with all $\KI$ measures. The positive ratio represents the likelihood of having a higher proportion of $ind$-type references compared to $dep$-type references. Consequently, the variation curves for $D_1$ and $D_3$ align with those for $D_0$ and $D_2$, respectively.
        \label{figS_AI}}
    \end{center}
\end{figure}
\clearpage

\begin{figure}[htbp]
    \begin{center}
        % \resizebox{17cm}{!}{\includegraphics{figures/FigOAE_Novelty.pdf}}
        \includegraphics[width=1.0\textwidth]{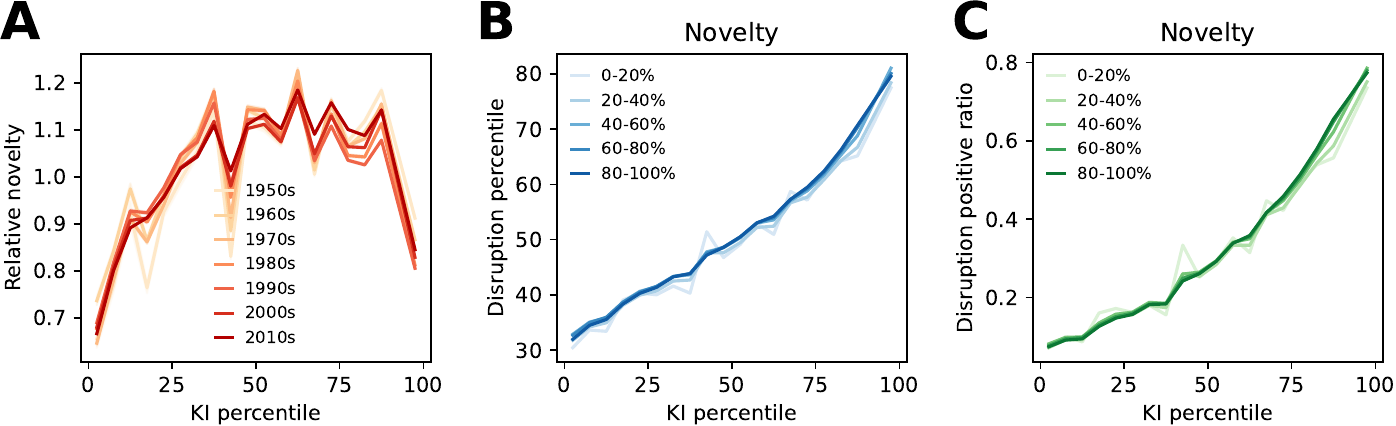}
        \caption{\textbf{$|$ The association between disruption and $\KI$ is robust when controlling for novelty.}
        (\textbf{A}) The percentile of novelty index (Methods) within each bin is re-scaled by the average value of the corresponding period to highlight the trends. Papers utilizing dependent or independent knowledge tend to rely on less novel combinations of references, whereas neutral papers---those maintaining a balance between dependent and independent knowledge---show a preference for more novel combinations of references. This pattern remains consistent across time periods.
        (\textbf{B-C}) When controlling for focal papers' novelty, both the percentile and positive ratio of disruption continue to increase with $\KI$ across all levels of novelty.
        \label{figS_RN}}
    \end{center}
\end{figure}
\clearpage

\begin{figure}[htbp]
    \begin{center}
        % \resizebox{17cm}{!}{\includegraphics{figures/FigOAE_Interdisciplinarity.pdf}}
        \includegraphics[width=1.0\textwidth]{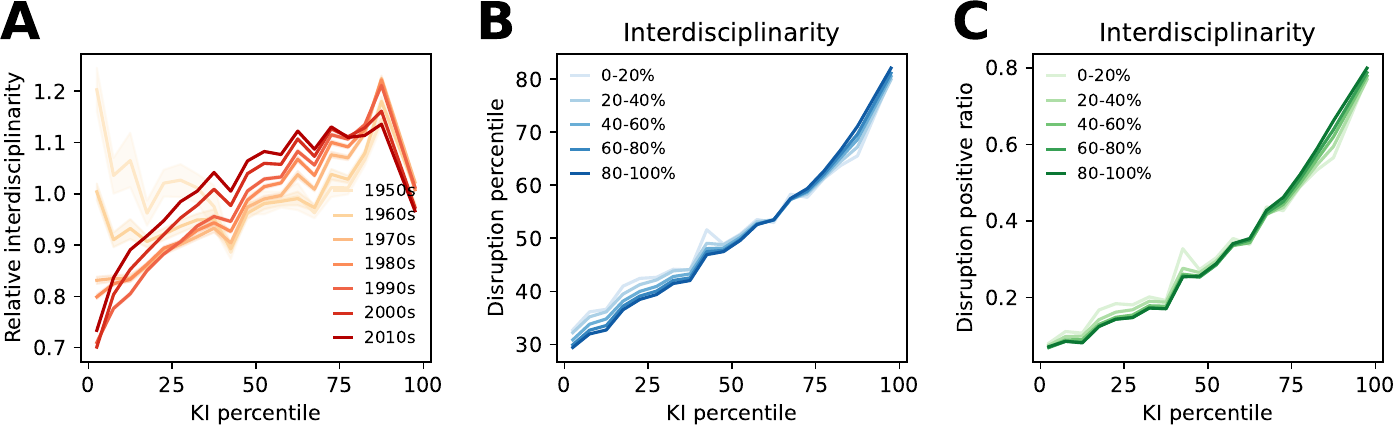}
        \caption{\textbf{$|$ The association between disruption and $\KI$ is robust when controlling for novelty.
        (\textbf{A}) The percentile of interdisciplinarity index (Methods) within each bin is re-scaled by the average value of the corresponding period to highlight the trends. Papers utilizing dependent or independent knowledge tend to rely on slightly less interdisciplinary combinations of references, which remains consistent across time periods.
        (\textbf{B-C}) When controlling for focal papers' interdisciplinarity, both the percentile and positive ratio of disruption continue to increase with $\KI$ across all levels of interdisciplinarity.}
        \label{figS_RM}}
    \end{center}
\end{figure}
\clearpage

\begin{figure}[htbp]
    \begin{center}
        % \resizebox{17cm}{!}{\includegraphics{figures/FigOAE_Reference_age.pdf}}
        \includegraphics[width=1.0\textwidth]{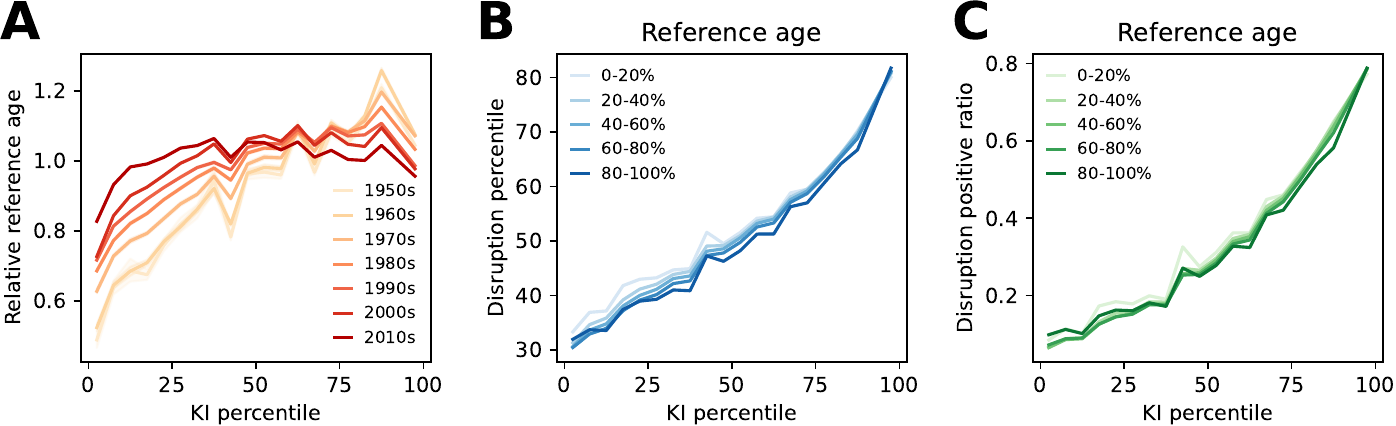}
        \caption{\textbf{$|$ The association between disruption and $\KI$ is robust when controlling for reference age.}
        (\textbf{A}) We represent the reference age of a given focal paper by calculating the the average age of cited references. The percentile of reference age within each bin is re-scaled by the average value of the corresponding period to highlight the trends. Papers utilizing dependent or independent knowledge tend to rely on relatively younger references, whereas neutral papers---those maintaining a balance between dependent and independent knowledge---show a preference for older references. This pattern remains consistent across time periods.
        (\textbf{B-C}) When controlling for reference age, both the percentile and positive ratio of disruption continue to increase with $\KI$ across all levels of reference age.
        \label{figS_RA}}
    \end{center}
\end{figure}
\clearpage

\begin{figure}[htbp]
    \begin{center}
        % \resizebox{17cm}{!}{\includegraphics{figures/FigOAE_Reference_impact_C5.pdf}}
        \includegraphics[width=1.0\textwidth]{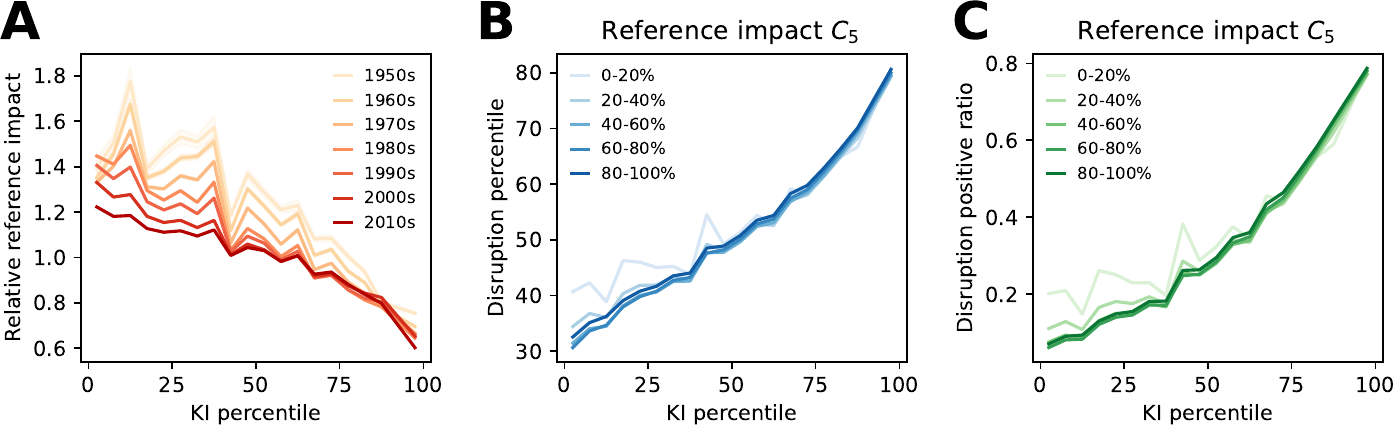}
        \caption{\textbf{$|$ The association between disruption and $\KI$ is robust when controlling for reference impact.}
        (\textbf{A}) We represent the reference impact of a given focal paper by calculating the average $C_5$ of cited references. The percentile of reference impact within each bin is re-scaled by the average value of the corresponding period to highlight the trends. Papers that draw on more independent knowledge tend to cite less impactful references. This pattern remains consistent across time periods.
        (\textbf{B-C}) When controlling for reference impact, both the percentile and positive ratio of disruption continue to increase with $\KI$ across all levels of reference impact. Additionally, papers citing less impactful references tend to exhibit slightly higher disruption.
        \label{figS_RI}}
    \end{center}
\end{figure}
\clearpage

\begin{figure}[htbp]
    \begin{center}
        % \resizebox{17cm}{!}{\includegraphics{figures/FigOAE_Reference_disruption.pdf}}
        \includegraphics[width=1.0\textwidth]{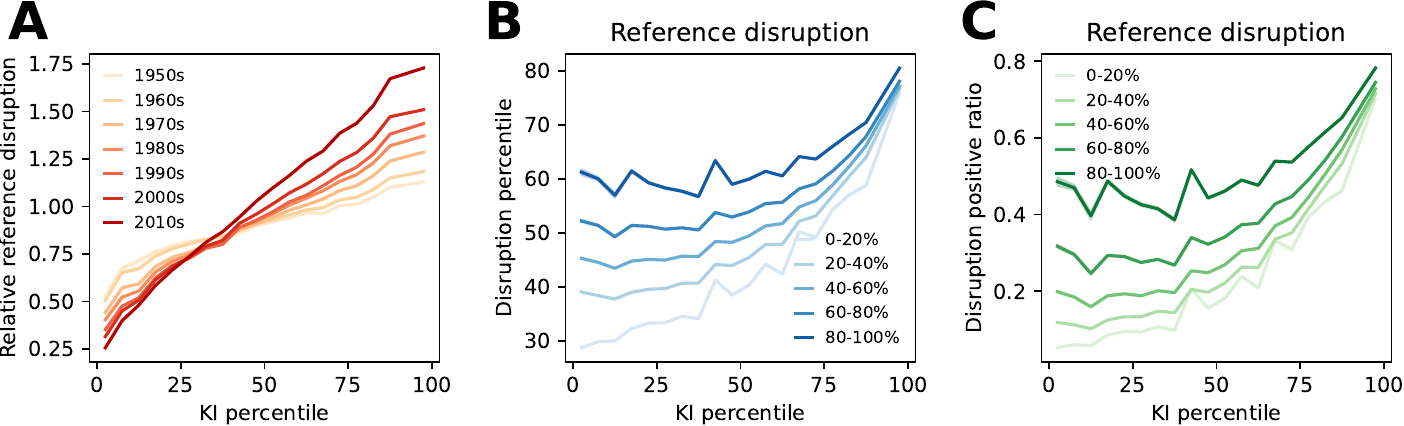}
        \caption{\textbf{$|$ The association between disruption and $\KI$ is robust when controlling for reference disruption.}
        (\textbf{A}) We represent the reference disruption of a given focal paper by calculating the average disruption of cited references. The percentile of reference disruption within each bin is re-scaled by the average value of the corresponding period to highlight the trends. Papers that draw on more independent knowledge tend to cite more disruptive references. This pattern remains consistent across time periods.
        (\textbf{B-C}) When controlling for reference disruption, both the percentile and positive ratio of disruption continue to increase with $\KI$ across all levels of reference disruption. Additionally, papers citing more disruptive references tend to exhibit higher disruption.
        \label{figS_RD}}
    \end{center}
\end{figure}
\clearpage

\begin{figure}[htbp]
    \begin{center}
        % \resizebox{17cm}{!}{\includegraphics{figures/FigOAE_Team_size.pdf}}
        \includegraphics[width=1.0\textwidth]{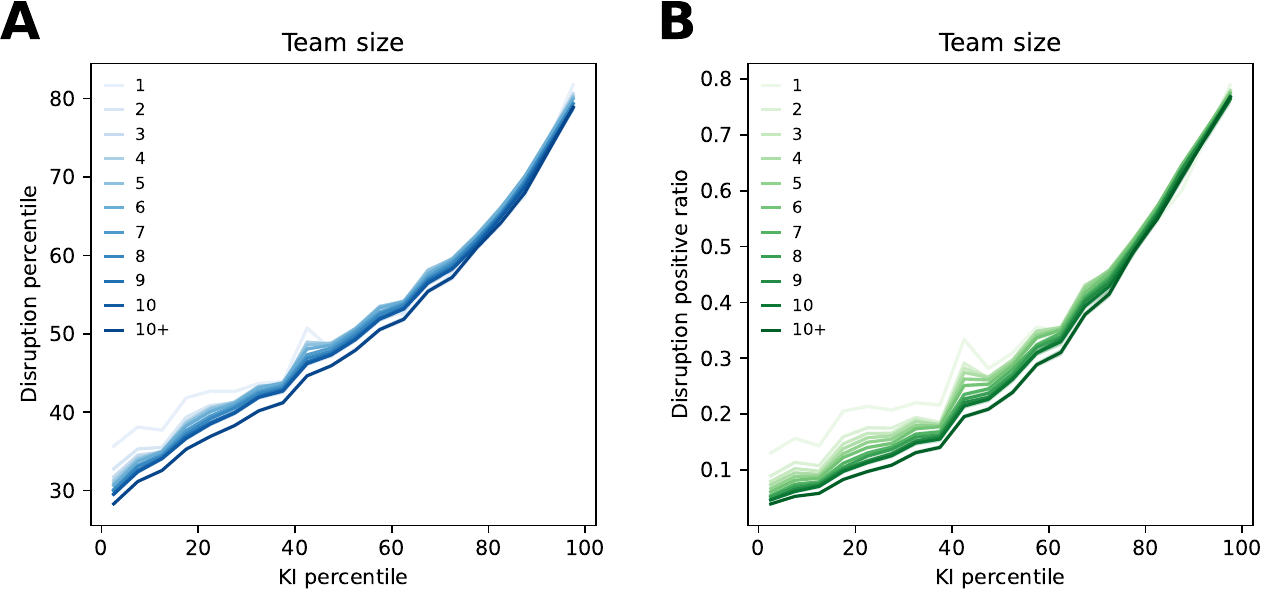}
        \caption{\textbf{$|$ The association between disruption and $\KI$ is robust when controlling for team size.}
        (\textbf{A-B}) When controlling for team size, both the percentile and positive ratio of disruption continue to increase with $\KI$ across different team sizes. Additionally, papers produced by smaller teams tend to exhibit slightly higher disruption.
        \label{figS_TS}}
    \end{center}
\end{figure}
\clearpage

\begin{figure}[htbp]
    \begin{center}
        % \resizebox{17cm}{!}{\includegraphics{figures/FigOAE_Team_distance.pdf}}
        \includegraphics[width=1.0\textwidth]{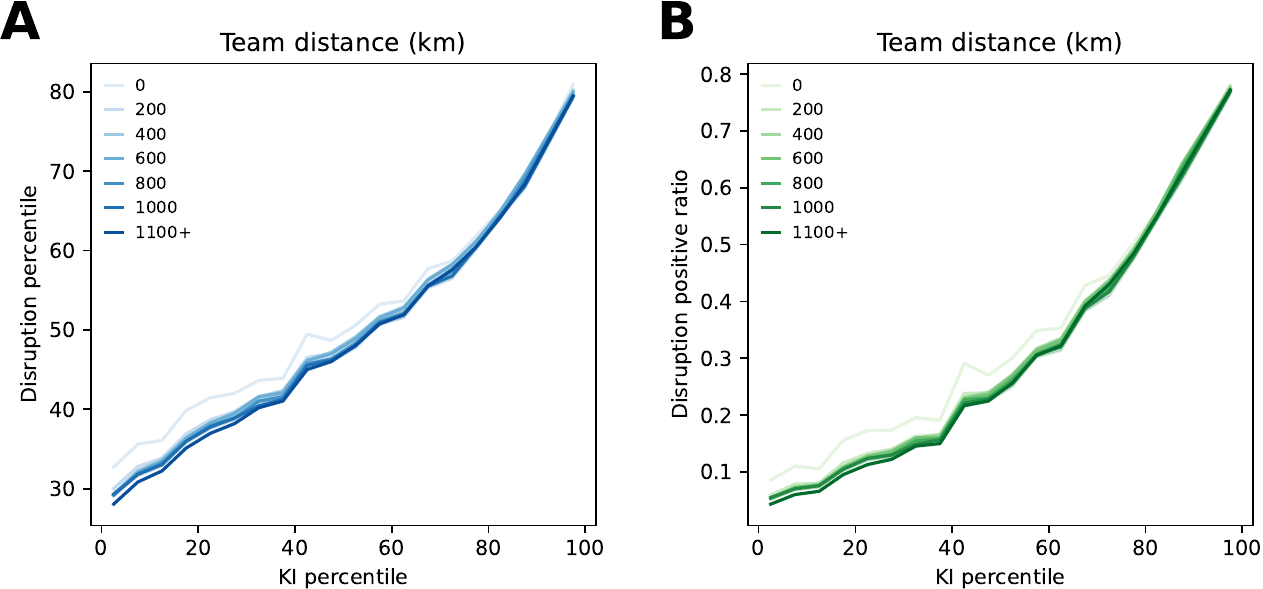}
        \caption{\textbf{$|$ The association between disruption and $\KI$ is robust when controlling for team distance.}
        (\textbf{A-B}) 
        We calculate the geographic team distance by the coordinate information of affiliations (Methods). When controlling for team distance, both the percentile and positive ratio of disruption continue to increase with $\KI$ across different team distances. Additionally, papers produced by geographically closer teams tend to exhibit slightly higher disruption.
        \label{figS_TD}}
    \end{center}
\end{figure}
\clearpage

\begin{figure}[htbp]
    \begin{center}
        % \resizebox{17cm}{!}{\includegraphics{figures/FigOAE_Team_freshness.pdf}}	
        \includegraphics[width=1.0\textwidth]{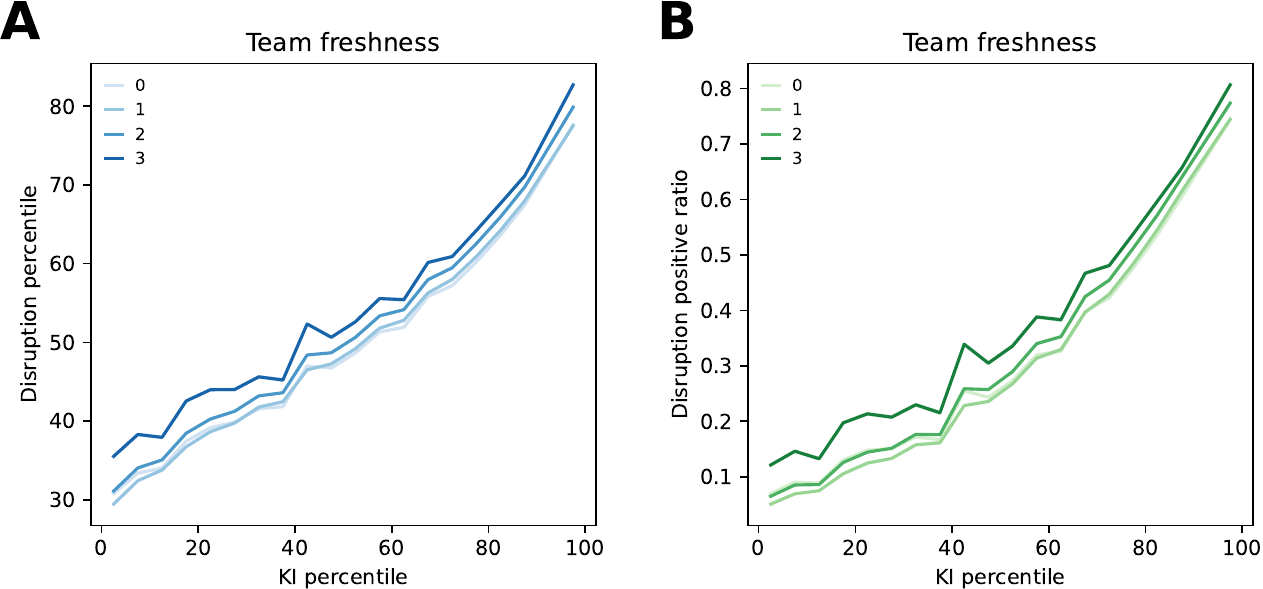}
        \caption{\textbf{$|$ The association between disruption and $\KI$ is robust when controlling for team freshness.}
        (\textbf{A-B}) We classify the team freshness of a given focal paper by the topological structure of the prior collaboration network, where the value from $0$ to $3$ refers to network topology from a full-connected network to an isolated network (Methods). When controlling for team freshness, both the percentile and positive ratio of disruption continue to increase with $\KI$ across different team freshnesses. Additionally, papers produced by fresher teams tend to exhibit slightly higher disruption.
        \label{figS_TF}}
    \end{center}
\end{figure}
\clearpage

\begin{figure}[htbp]
    \begin{center}
        \includegraphics[width=1.0\textwidth]{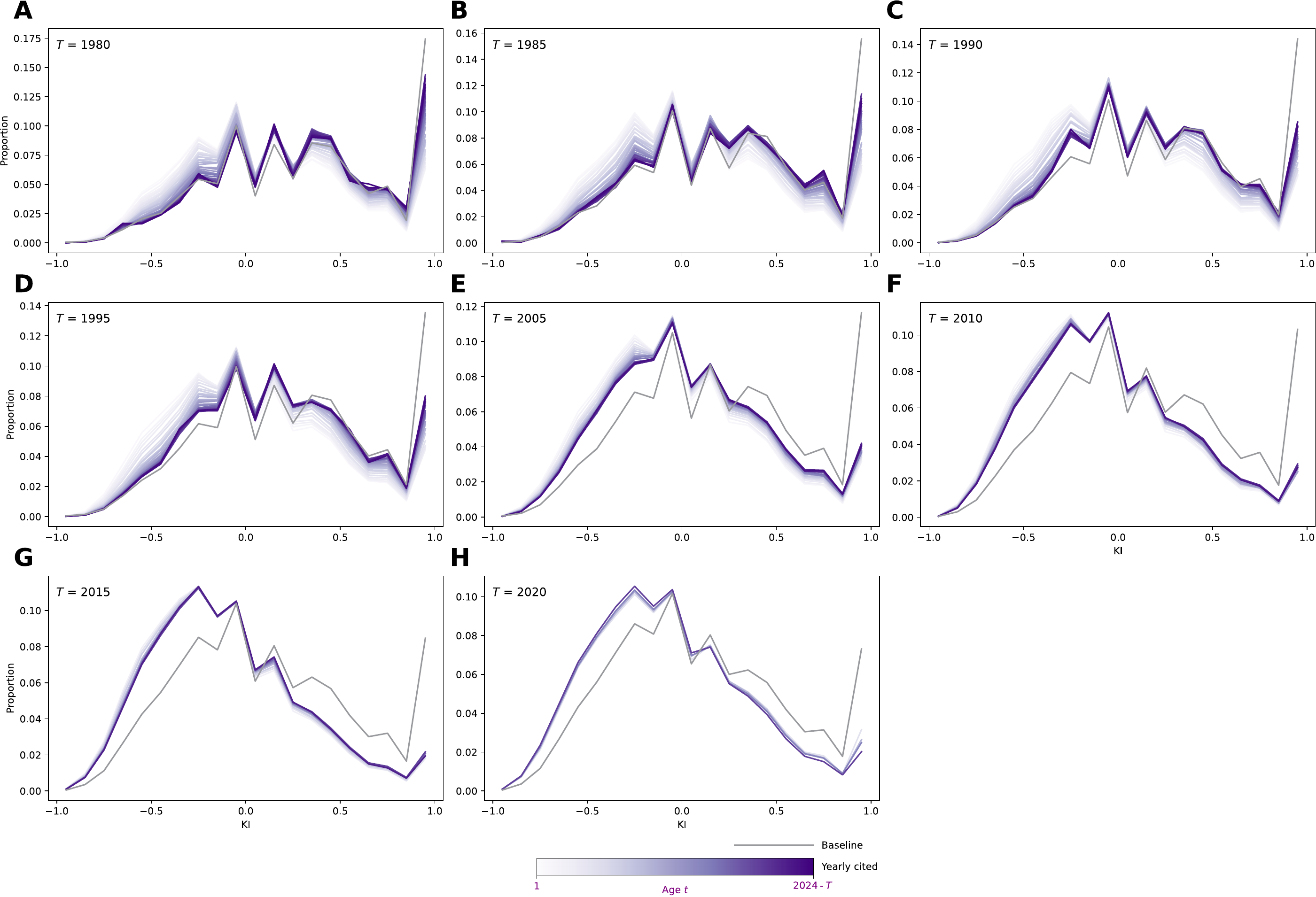}
        \caption{\textbf{$|$ Papers with higher $\KI$ are underrepresented among cited references and receiving delayed citations.}
        (\textbf{A-H}) 
        For papers published in the year $T$, we measure the $\KI$ distribution among these papers as the baseline distribution (gray curve). Then, for all papers published in a subsequent year, $T+t$ ($t\geq 1$), we examine their bibliographies, extract the references that were published in $T$, and generate the $\KI$ distribution of those references (purple curve). Compared to the baseline, papers with negative $\KI$ are overrepresented while those with positive $\KI$ are underrepresented. Moreover, as the cited age $t$ increases, the corresponding distribution converges towards the baseline.
        \label{figS_PS}}
    \end{center}
\end{figure}
\clearpage

% \begin{figure}[htbp] % Do not use \begin{figure*}
% 	\centering
% 	\includegraphics[width=1.0\textwidth]{figures/FigSSN_award_fund.pdf}
%     \caption{\textbf{$|$ Knowledge independence is awarded but underfunded.}
%         (\textbf{A}) Based on the dataset for Nobel laureates~\cite{li2019dataset}, We match 425 Nobel Prize-winning papers that cover the time period 1902–2016 in SciSciNet dataset, including 78 papers in Physics, 133 papers in Chemistry, and 214 papers in Medicine. The probability of winning the Nobel prize increases with the $\KI$ value of papers.
%         (\textbf{B}) Based on the funding information recorded in SciSciNet dataset, we extract 263,5812 papers funded by NIH and 92,9177 papers funded by NSF. Here we repeat counting when a paper is funded by both agencies. The probability of being funded decreases with the $\KI$ value of papers.
%         \label{figSSN_award_fund}}
% \end{figure}
% \clearpage

%%%%%%%%%%%%%%%% EXTENDED TABLES %%%%%%%%%%%%%%%

\setcounter{table}{0} % table below are counted start 1
\captionsetup[table]{labelfont={bf},labelformat={default},name={Extended Data Table},labelsep=space,font=small}

\clearpage
\begin{table} % Do not use \begin{table*}
\centering
% Captions go above tables
\caption{\textbf{$|$ Normalized mediation regression on team size.}}
\label{tableS_mediation_TS} % give each table a logical label name
\renewcommand\tabcolsep{0.2cm} % column spacing
\renewcommand{\arraystretch}{1.5} % line spacing
\begin{threeparttable}
\fontsize{12}{10}\selectfont % 设置表格字体大小为 8.75pt，行距为 12pt
\begin{tabular}{llcccc} % four columns, alignment for each
    \hline
    Independent variable & Mediator & ACME ($\beta_1\theta_2$) & ADE ($\theta_1$) & TE ($\theta_1 + \beta_1\theta_2$) & $|$ACME / TE$|$ \\
    \hline
    Team size & $\KI$ & \textbf{-0.0196}***  & \textbf{-0.0149}*** & -0.0345*** & \textbf{0.5680} \\
     &  & (0.0011) & (0.0009) & (0.0020) \\\\
    % Team size & \sout{Log $C_5$} & -0.0065***  & -0.0058*** & -0.0345*** \\
    %  &  & (0.0006) & (0.0007) & (0.0020) \\\\
    Team size & Novelty & -0.0000***  & -0.0345*** & -0.0345*** & 0.0000 \\
     &  & (0.0000) & (0.0020) & (0.0020) \\\\
    Team size & Interdisciplinarity & -0.0020***  & -0.0325*** & -0.0345*** & 0.0575 \\
     &  & (0.0001) & (0.0018) & (0.0020) \\\\
    Team size & Reference count & -0.0049***  & -0.0296*** & -0.0345*** & 0.1413 \\
     &  & (0.0003) & (0.0017) & (0.0020) \\\\
    Team size & Reference age & -0.0011***  & -0.0333*** & -0.0345*** & 0.0330 \\
     &  & (0.0001) & (0.0019) & (0.0020) \\\\
    % Team size & Reference $C_5$ & -0.0004***  & -0.0120*** & -0.0345*** & 0.0359 \\
    %  &  & (0.0000) & (0.0020) & (0.0020) \\\\
    % Team size & \sout{Reference disruption} & -0.0064***  & -0.0059*** & -0.0345*** \\
    %  &  & (0.0007) & (0.0006) & (0.0020) \\\\
    Team size & Team distance & -0.0092***  & -0.0253*** & -0.0345*** & 0.2678 \\
     &  & (0.0005) & (0.0015) & (0.0020) \\\\
    Team size & Team freshness & 0.0082***  & -0.0426*** & -0.0345*** & 0.2368 \\
     &  & (0.0005) & (0.0024) & (0.0020) \\
    \hline
\end{tabular}
\begin{tablenotes}
    \small
    \item[$\dagger$] We perform normalized mediation analysis to assess the effect of team size on disruption. The candidate mediating variables include: $\KI$, novelty, interdisciplinarity, reference count, reference age, team distance, and team freshness, with discipline and publication year controlled. ACME refers to average causal mediation effects driven from the mediating variable, ADE stands for average direct effects caused by team size, and TE is defined as the sum of the ACME and ADE. Each regression coefficient is tested against the null hypothesis (coefficient equals 0) using a two-sided \textit{t}-test. Standard errors are provided in parentheses for each coefficient. Note that we do not apply adjustments for multiple hypothesis testing in this analysis. (*$p < 0.05$, **$p < 0.01$, ***$p < 0.001$).
\end{tablenotes}
\end{threeparttable}
\end{table}

\clearpage
\begin{table} % Do not use \begin{table*}
\centering
% Captions go above tables
\caption{\textbf{$|$ Normalized mediation regression on team distance.}}
\label{tableS_mediation_TD} % give each table a logical label name
\renewcommand\tabcolsep{0.2cm} % column spacing
\renewcommand{\arraystretch}{1.5} % line spacing
\begin{threeparttable}
\fontsize{12}{10}\selectfont % 设置表格字体大小为 8.75pt，行距为 12pt
\begin{tabular}{llcccc} % four columns, alignment for each
    \hline
    Independent variable & Mediator & ACME ($\beta_1\theta_2$) & ADE ($\theta_1$) & TE ($\theta_1 + \beta_1\theta_2$) & $|$ACME / TE$|$ \\
    \hline
    Team distance & $\KI$ & \textbf{-0.0330}***  & \textbf{-0.0310}*** & -0.0640*** & \textbf{0.5151} \\
     &  & (0.0001) & (0.0003) & (0.0003) \\\\
    % Team distance & \sout{Log $C_5$} & -0.0174***  & -0.0399*** & -0.0640*** \\
    %  &  & (0.0001) & (0.0003) & (0.0003) \\\\
    Team distance & Novelty & 0.0000***  & -0.0640*** & -0.0640*** & 0.0000 \\
     &  & (0.0000) & (0.0003) & (0.0003) \\\\
    Team distance & Interdisciplinarity & 0.0008***  & -0.0648*** & -0.0640*** & 0.0121 \\
     &  & (0.0000) & (0.0003) & (0.0003) \\\\
    Team distance & Reference count & -0.0181***  & -0.0459*** & -0.0640*** & 0.2824 \\
     &  & (0.0001) & (0.0003) & (0.0003) \\\\
    Team distance & Reference age & -0.0002***  & -0.0638*** & -0.0640*** & 0.0036 \\
     &  & (0.0000) & (0.0003) & (0.0003) \\\\
    % Team distance & Reference $C_5$ & -0.0003***  & -0.0547*** & -0.0640*** & 0.0056 \\
    %  &  & (0.0000) & (0.0003) & (0.0003) \\\\
    % Team distance & \sout{Reference disruption} & -0.0244***  & -0.0330*** & -0.0640*** \\
    %  &  & (0.0003) & (0.0003) & (0.0003) \\\\
    Team distance & Team size & -0.0039***  & -0.0601*** & -0.0640*** & 0.0605 \\
     &  & (0.0002) & (0.0004) & (0.0003) \\\\
    Team distance & Team freshness & 0.0005***  & -0.0645*** & -0.0640*** & 0.0079 \\
     &  & (0.0000) & (0.0003) & (0.0003) \\
    \hline
\end{tabular}
\begin{tablenotes}
    \small
    \item[$\dagger$] We perform normalized mediation analysis to assess the effect of team distance on disruption. The candidate mediating variables include: $\KI$, novelty, interdisciplinarity, reference count, reference age, team size, and team freshness, with discipline and publication year controlled. ACME refers to average causal mediation effects driven from the mediating variable, ADE stands for average direct effects caused by team distance, and TE is defined as the sum of the ACME and ADE. Each regression coefficient is tested against the null hypothesis (coefficient equals 0) using a two-sided \textit{t}-test. Standard errors are provided in parentheses for each coefficient. Note that we do not apply adjustments for multiple hypothesis testing in this analysis. (*$p < 0.05$, **$p < 0.01$, ***$p < 0.001$).
\end{tablenotes}
\end{threeparttable}
\end{table}

\clearpage
\begin{table} % Do not use \begin{table*}
\centering
% Captions go above tables
\caption{\textbf{$|$ Normalized mediation regression on team freshness.}}
\label{tableS_mediation_TF} % give each table a logical label name
\renewcommand\tabcolsep{0.2cm} % column spacing
\renewcommand{\arraystretch}{1.5} % line spacing
\begin{threeparttable}
\fontsize{12}{10}\selectfont % 设置表格字体大小为 8.75pt，行距为 12pt
\begin{tabular}{llcccc} % four columns, alignment for each
    \hline
    Independent variable & Mediator & ACME ($\beta_1\theta_2$) & ADE ($\theta_1$) & TE ($\theta_1 + \beta_1\theta_2$) & $|$ACME / TE$|$ \\
    \hline
    Team freshness & $\KI$ & \textbf{0.0465***}  & \textbf{0.0480***} & 0.0946*** & \textbf{0.4922} \\
     &  & (0.0002) & (0.0003) & (0.0003) \\\\
    % Team freshness & \sout{Log $C_5$} & -0.0001***  & -0.0003*** & 0.0946*** \\
    %  &  & (0.0001) & (0.0006) & (0.0003) \\\\
    Team freshness & Novelty & 0.0000***  & 0.0946*** & 0.0946*** & 0.0000 \\
     &  & (0.0000) & (0.0003) & (0.0003) \\\\
    Team freshness & Interdisciplinarity & 0.0001***  & 0.0945*** & 0.0946*** & 0.0013 \\
     &  & (0.0000) & (0.0003) & (0.0003) \\\\
    Team freshness & Reference count & 0.0136***  & 0.0810*** & 0.0946*** & 0.1439 \\
     &  & (0.0001) & (0.0003) & (0.0003) \\\\
    Team freshness & Reference age & 0.0000***  & 0.0946*** & 0.0946*** & 0.0002 \\
     &  & (0.0000) & (0.0004) & (0.0003) \\\\
    % Team freshness & Reference $C_5$ & 0.0000***  & 0.0946*** & 0.0946*** & 0.0118 \\
    %  &  & (0.0000) & (0.0006) & (0.0003) \\\\
    % Team freshness & \sout{Reference disruption} & 0.0003***  & -0.0007*** & 0.0946*** \\
    %  &  & (0.0003) & (0.0006) & (0.0003) \\\\
    Team freshness & Team size & -0.0035***  & 0.0981*** & 0.0946*** & 0.0375 \\
     &  & (0.0002) & (0.0004) & (0.0003) \\\\
    Team freshness & Team distance & -0.0003***  & 0.0949*** & 0.0005*** & 0.0036 \\
     &  & (0.0000) & (0.0003) & (0.0003) \\
    \hline
\end{tabular}
\begin{tablenotes}
    \small
    \item[$\dagger$] We perform normalized mediation analysis to assess the effect of team freshness on disruption. The candidate mediating variables include: $\KI$, novelty, interdisciplinarity, reference count, reference age, team size, and team distance, with discipline and publication year controlled. ACME refers to average causal mediation effects driven from the mediating variable, ADE stands for average direct effects caused by team freshness, and TE is defined as the sum of the ACME and ADE. Each regression coefficient is tested against the null hypothesis (coefficient equals 0) using a two-sided \textit{t}-test. Standard errors are provided in parentheses for each coefficient. Note that we do not apply adjustments for multiple hypothesis testing in this analysis. (*$p < 0.05$, **$p < 0.01$, ***$p < 0.001$).
\end{tablenotes}
\end{threeparttable}
\end{table}

\clearpage
\begin{table} % Do not use \begin{table*}
\centering
% Captions go above tables
\caption{\textbf{$|$ Normalized confounding regression on paper impact.}}
\label{tableS_confounding_impact} % give each table a logical label name
\renewcommand\tabcolsep{0.2cm} % column spacing
\renewcommand{\arraystretch}{1.5} % line spacing
\begin{threeparttable}
\fontsize{12}{10}\selectfont % 设置表格字体大小为 8.75pt，行距为 12pt
\begin{tabular}{llccc} % four columns, alignment for each
    \hline
    Independent variable & Confounder & Unadjusted ($\alpha_1$) & Adjusted ($\delta_1$) & RCE ($|\alpha_1 - \delta_1|/|\alpha_1|$) \\
    \hline
    Citation percentile & $\KI$ & -0.1349***  & \textbf{-0.0248}***  & \textbf{0.8164} \\
     &  & (0.0004) & (0.0004) \\\\
    Citation percentile & Novelty & -0.1349***  & -0.1349*** & 0.0000 \\
     &  & (0.0004) & (0.0004) \\\\
    Citation percentile & Interdisciplinarity & -0.1349***  & -0.1355*** & 0.0044 \\
     &  & (0.0004) & (0.0004) \\\\
    Citation percentile & Reference count & -0.1349***  & -0.0762*** & 0.4356 \\
     &  & (0.0004) & (0.0005) \\\\
    Citation percentile & Reference age & -0.1349***  & -0.1355*** & 0.0045 \\
     &  & (0.0004) & (0.0004) \\\\
    % Citation percentile & Reference $C_5$ & -0.1349***  & -0.1001*** & 0.0073 \\
    %  &  & (0.0004) & (0.0004) \\\\
    % Citation percentile & \sout{Reference disruption} & -0.1349***  & -0.0534*** \\
    %  &  & (0.0004) & (0.0006) \\\\
    Citation percentile & Team size & -0.1349***  & -0.1328*** & 0.0157 \\
     &  & (0.0004) & (0.0004) \\\\
    Citation percentile & Team distance & -0.1349***  & -0.1295*** & 0.0402 \\
     &  & (0.0004) & (0.0004) \\\\
    Citation percentile & Team freshness & -0.1349***  & -0.1298*** & 0.0380 \\
     &  & (0.0004) & (0.0004) \\
    \hline
\end{tabular}
\begin{tablenotes}
    \small
    \item[$\dagger$] We perform normalized confounding analysis to assess the effect of paper impact on disruption. The candidate confounding variables include: $\KI$, novelty, interdisciplinarity, reference count, reference age, team size, team distance, and team freshness, with discipline and publication year controlled. The relative change in estimate (RCE) refers to the relative change of effect after controlling for the confounder. Each regression coefficient is tested against the null hypothesis (coefficient equals 0) using a two-sided \textit{t}-test. Standard errors are provided in parentheses for each coefficient. Note that we do not apply adjustments for multiple hypothesis testing in this analysis. (*$p < 0.05$, **$p < 0.01$, ***$p < 0.001$).
\end{tablenotes}
\end{threeparttable}
\end{table}
\clearpage

%\end{linenumbers}% This is the last page with line numbers.

%%%%%%%%%%%%%%%% START OF SUPPLEMENT %%%%%%%%%%%%%%%
% Figures, tables, equations and pages in the supplement are numbered S1, S2 etc.

\renewcommand{\thesubsection}{S\arabic{subsection}}
\renewcommand{\thefigure}{S\arabic{figure}}
\captionsetup[figure]{labelfont={bf},labelformat=default,name={Fig.},labelsep=space,font=small}
\renewcommand{\thetable}{S\arabic{table}}
\captionsetup[table]{labelfont={bf},labelformat=default,name={Table},labelsep=space,font=small}
\renewcommand{\theequation}{S\arabic{equation}}
\renewcommand{\thepage}{S\arabic{page}}
\setcounter{figure}{0}
\setcounter{table}{0}
\setcounter{equation}{0}
\setcounter{page}{1} % not 0 as \newpage already started a supplementary page
% References continue the numbering from the main text.

%%%%%%%%%%%%%%%% SUPPLEMENT TITLE PAGE %%%%%%%%%%%%%%%

\begin{center}
\section*{Supplementary Materials for\\ \scititle}
% Author list for the supplement
% Indicate the corresponding authors, but do NOT include institutions here
% It would be nice if the template auto-generated this, but doing so is complicated.
Xiaoyao Yu,
Talal Rahwan$^{\ast}$,
Tao Jia$^{\ast}$\\ % we're not in a \author{} environment this time, so use \\ for a new line
\small$^\ast$Corresponding author. Email: talal.rahwan@nyu.edu (T.~R.); tjia@swu.edu.cn (T.~J.)
%\small$^\dagger$These authors contributed equally to this work.
\end{center}

% Fill out the numbers for each type of supplementary information,
% and delete any lines that aren't applicable.
% These are just example numbers that don't match the rest of this template.

% \begin{sciabstract}
% \begin{center}
% \textbf{Supplementary results are based on publication data from SciSciNet\footnote{https://doi.org/10.6084/m9.figshare.c.6076908.v1}.}
% \end{center}
% \end{sciabstract}

\subsubsection*{This PDF file includes:}
% Materials and Methods\\
Supplementary Text~\ref{SI_KI_Robustness} to~\ref{SI_OECD_results}\\
Figures~\ref{FigOAS_Reference_count_KI} to~\ref{FigOECD_2_PSM_CEM_binary}\\
Tables~\ref{table_SI_ALL_OLS} to~\ref{table_OECD_OLS_binary}
% Captions for Movies S1 to S2\\
% Captions for Data S1 to S2

% \subsubsection*{Other Supplementary Materials for this manuscript:}
% Movies S1 to S2\\
% Data S1 to S2

\newpage

%%%%%%%%%%%%%%%% SUPPLEMENTARY TEXT %%%%%%%%%%%%%%%
\subsection{Robustness and validity checks for knowledge independence}\label{SI_KI_Robustness}

\subsubsection{Validating $\KI$'s robustness to reference count effects}
{\fontsize{12pt}{14pt}\selectfont % {fontsize}{linespace}
We systematically investigated the influence of reference count on our $\KI$ metric and its associated findings. We observed a piecewise nonlinear relationship: $\KI$ is sensitive to reference count in shorter reference lists (2-10 references), but this effect becomes flat and moderate in longer ones (Fig.~\ref{FigOAS_Reference_count_KI}A). Furthermore, when stratifying by reference count, the $\KI$ distribution slightly shifts toward lower values as reference count increases (Fig.~\ref{FigOAS_Reference_count_KI}B).

Most critically, this allowed us to quantitatively partition the causes of the historical decline of $\KI$. While a portion of the decline is attributable to the structural effect of bibliography length, as the trend becomes steeper in strata with higher reference count, a significant decline still remains in all strata from 1950 (Fig.~\ref{FigOAS_Reference_count_KI}C). This suggests that the trend is not fully explained away as an artifact of bibliography growth and that a genuine shift in authorial behavior is a key component.

Finally, our core hypothesis remains robust even after controlling for reference count. Within each stratum, $\KI$ maintains a significant positive relationship with scientific disruption (Fig.~\ref{FigOAS_Reference_count_KI}D). The negative relationship between $\KI$ and citation impact remains evident in most strata, following a piecewise nonlinear decline (Fig.~\ref{FigOAS_Reference_count_KI}E). Together, these results confirm that $\KI$'s predictive power is not simply a byproduct of reference‐list length but reflects deeper structural properties of knowledge integration.

\begin{figure}[htbp]
    \begin{center}
        \includegraphics[width=.875\textwidth]{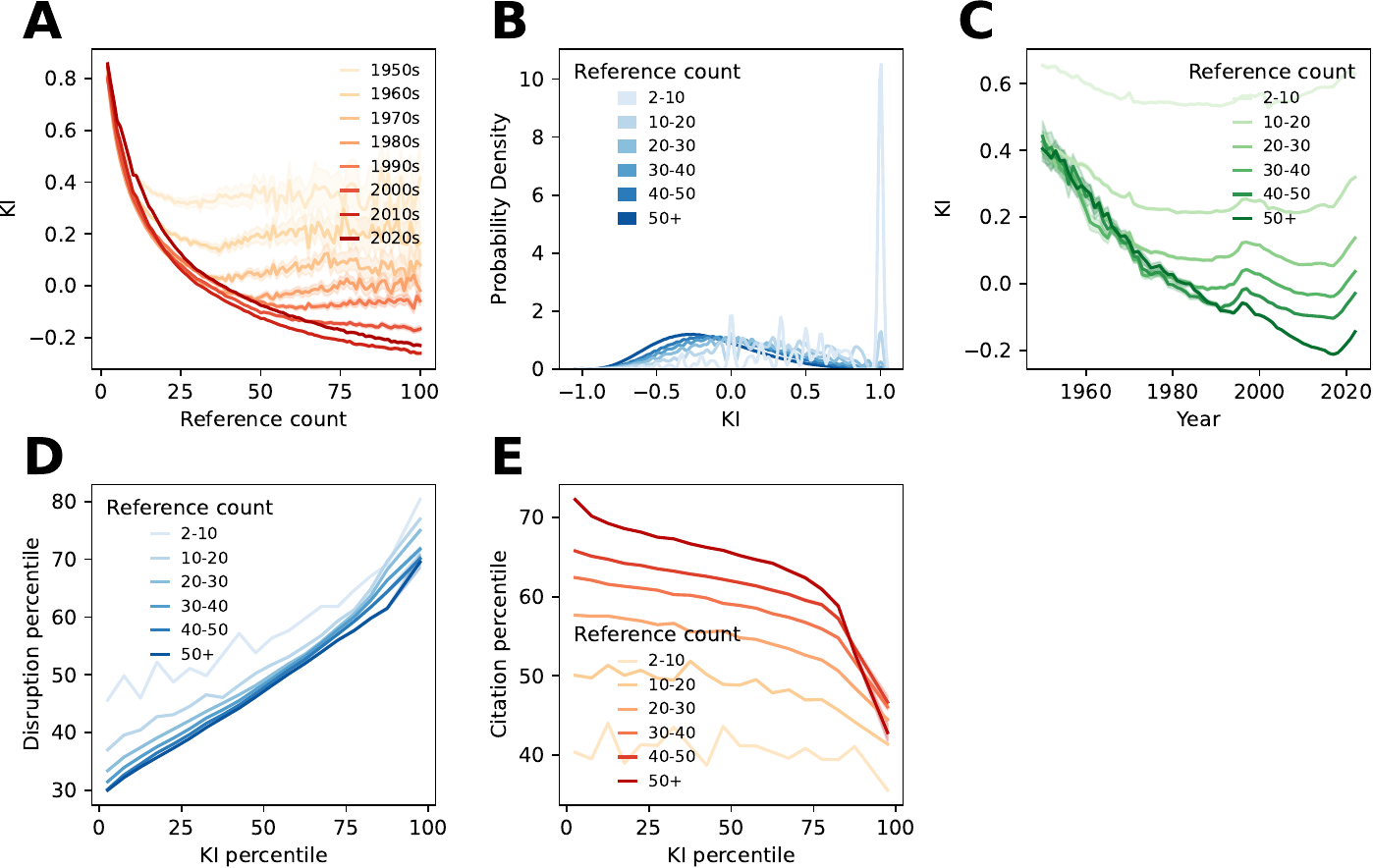}
        \caption{\textbf{$|$ Robustness of $\KI$ patterns after controlling for reference count.}
        (\textbf{A}) $\KI$ declines in papers with few references (2-10), then the trend becomes flat. The slope of this gradual decline has steepened in more recent decades.
        (\textbf{B}) Each curve represents the $\KI$ distribution for a specific reference count bin. The distribution slightly shifts toward lower values as reference count increases.
        (\textbf{C}) Each line represents the average $\KI$ over time for a specific reference count bin. A significant declining trend persists within all strata, which is slightly flatter among papers with lower reference count.
        (\textbf{D}) The positive association between $\KI$ and disruption remains robust and significant across all reference count groups.
        (\textbf{E}) The negative association between $\KI$ and citation impact remains robust and significant across all reference count groups.
        \label{FigOAS_Reference_count_KI}}
    \end{center}
\end{figure}}

\newpage
\subsubsection{Validating $\KI$'s robustness to novelty effects}\label{SI_novelty}
{\fontsize{12pt}{14pt}\selectfont % {fontsize}{linespace}
We performed an analogous systematic analysis to assess the impact of a paper's novelty in combining knowledge at journal level (Methods). Empirically, we found that $\KI$ exhibits an inverted U-shaped relationship with novelty, peaking around the 50th percentile, but has undergone a systematic and universal decline across all novelty levels from the 1950s to the 2020s (Fig.~\ref{FigOAS_Reference_Novelty_KI}A). Moreover, the overall $\KI$ distribution remains remarkably stable across different levels of novelty (Fig.~\ref{FigOAS_Reference_Novelty_KI}B). The most striking result comes from the temporal analysis: we observe a consistent, parallel decline in $\KI$ across all novelty levels~(Fig.~\ref{FigOAS_Reference_Novelty_KI}C). This strongly suggests that the historical decline of $\KI$ is independent of novelty characteristic, effectively ruling out novelty as a significant driver of the observed trend.

Finally, both the positive relationship between $\KI$ and scientific disruption, and the negative relationship with citation impact, hold across all novelty strata (Fig.~\ref{FigOAS_Reference_Novelty_KI}D-E). This demonstrates that our core findings are not artifacts of novelty effects.

\begin{figure}[htbp]
    \begin{center}
        \includegraphics[width=.875\textwidth]{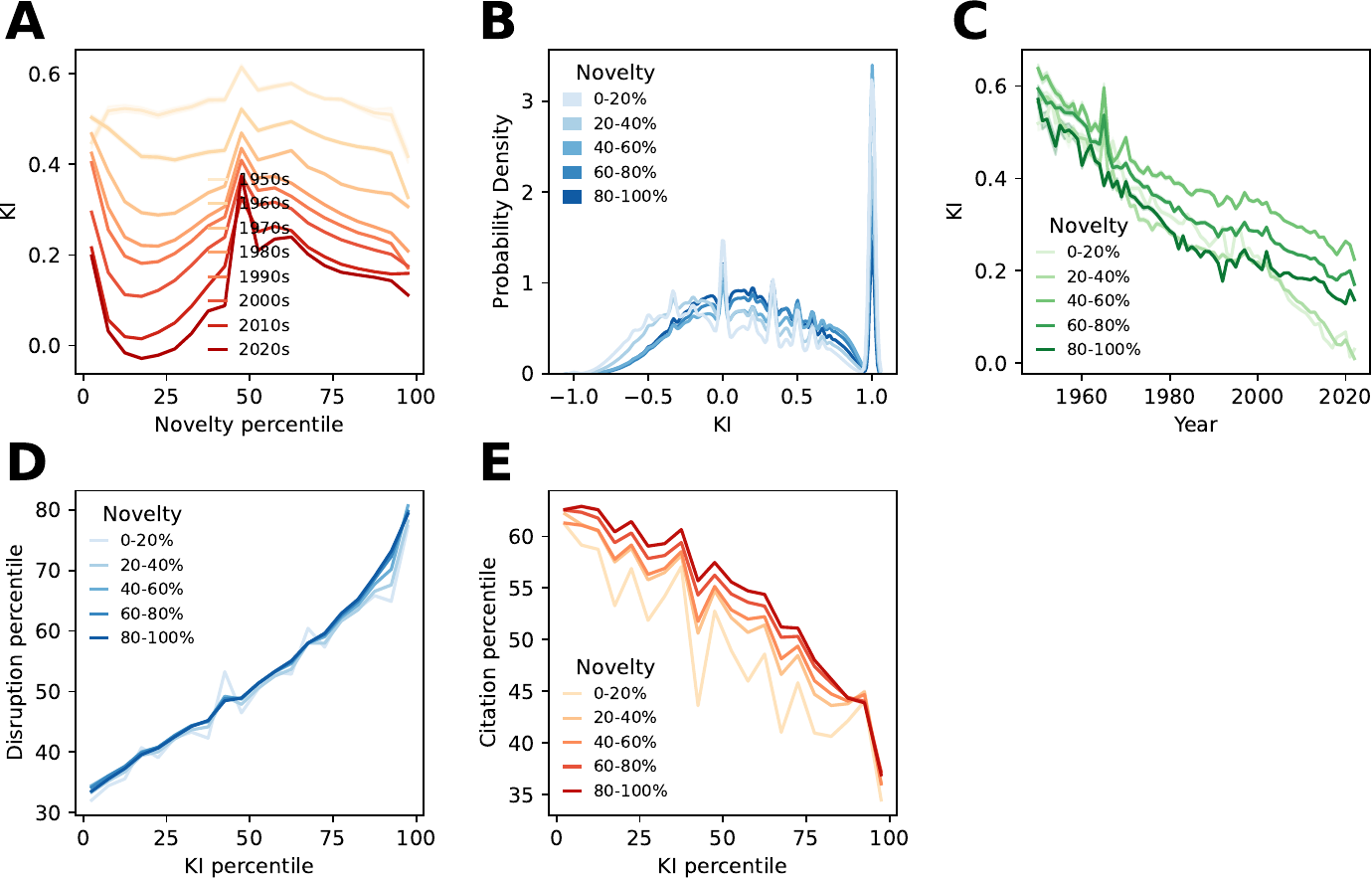}
        \caption{\textbf{$|$ Robustness of $\KI$ patterns after controlling for novelty.}
        (\textbf{A}) There is a slight negative correlation between our novelty index and $\KI$. This relationship has become flatter and less pronounced in more recent decades.
        (\textbf{B}) Each curve represents the $\KI$ distribution for a different percentile bin of novelty. The distribution remains a relatively stable bell-shaped curve across all groups.
        (\textbf{C}) Each line represents the average $\KI$ over time for a specific novelty bin. All groups exhibit a consistent and parallel declining trend.
        (\textbf{D}) The positive association between $\KI$ and disruption remains robust and significant across all levels of novelty.
        (\textbf{E}) The negative association between $\KI$ and citation impact remains robust and significant across all levels of novelty.
        \label{FigOAS_Reference_Novelty_KI}}
    \end{center}
\end{figure}}

\newpage
\subsubsection{Validating $\KI$'s robustness to interdisciplinarity effects}\label{SI_interdisciplinarity}
{\fontsize{12pt}{14pt}\selectfont % {fontsize}{linespace}
We performed an analogous systematic analysis to assess the impact of a paper's interdisciplinary nature. To do this, we apply the Rao–Stirling interdisciplinarity index (Methods). Empirically, we found a slight negative correlation between interdisciplinarity index and $\KI$, with a systematic and universal decline across all interdisciplinarity levels from the 1950s to the 2020s  (Fig.~\ref{FigOAS_Reference_Interdisciplinarity_KI}A). Besides, and the overall $\KI$ distribution remains remarkably stable across different levels of interdisciplinarity (Fig.~\ref{FigOAS_Reference_Interdisciplinarity_KI}B). As for the temporal pattern, we observe a consistent, parallel decline in $\KI$ across all interdisciplinarity levels~(Fig.~\ref{FigOAS_Reference_Interdisciplinarity_KI}C). This strongly suggests that the historical decline of $\KI$ is independent of disciplinary characteristic, effectively ruling out interdisciplinarity as a significant driver of the observed trend. 

Finally, both the positive relationship between $\KI$ and scientific disruption, and the negative relationship with citation impact, hold across all interdisciplinarity strata (Fig.~\ref{FigOAS_Reference_Interdisciplinarity_KI}D-E). This demonstrates that our core findings are not artifacts of interdisciplinarity effects.

\begin{figure}[htbp]
    \begin{center}
        \includegraphics[width=.875\textwidth]{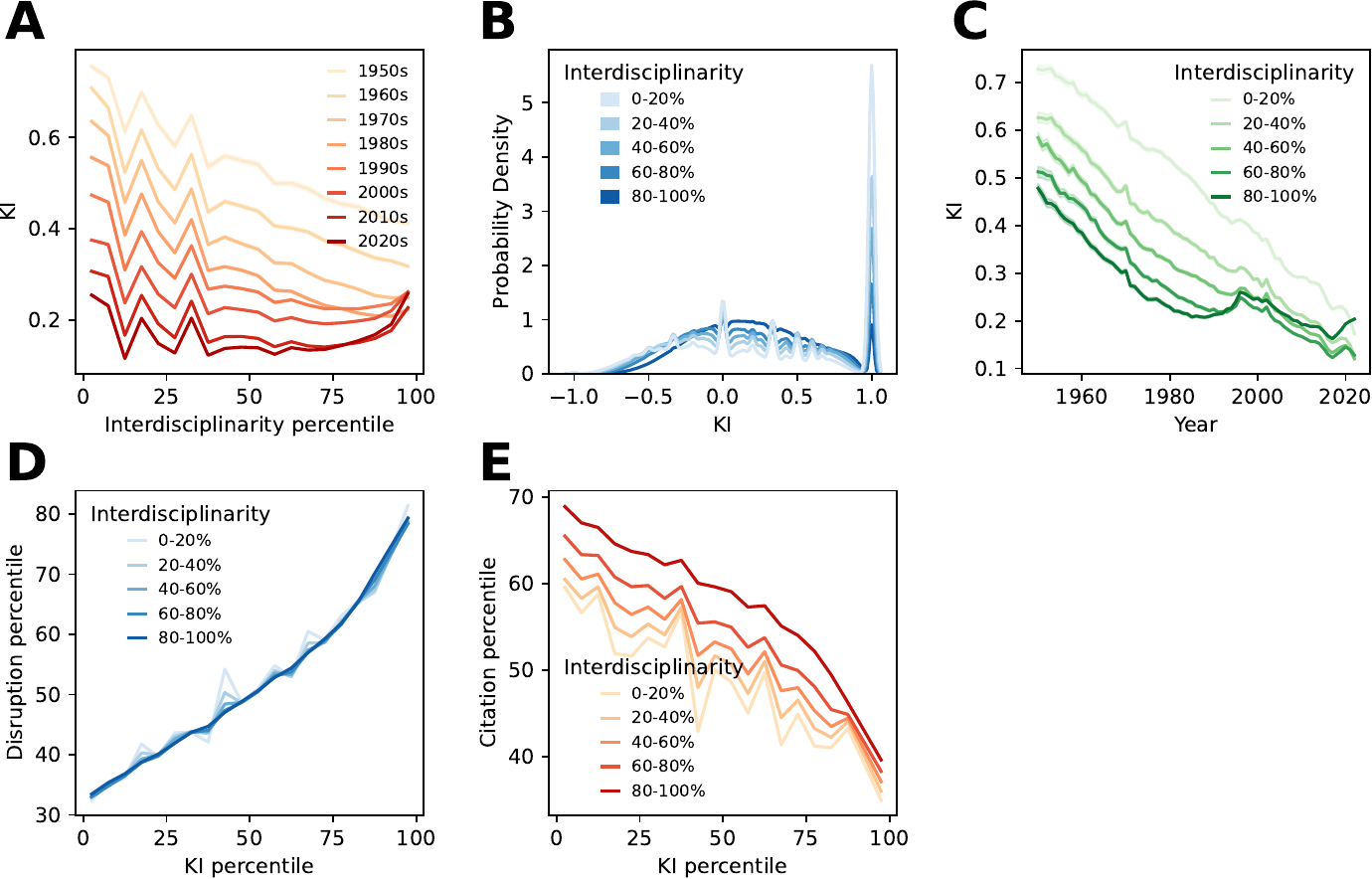}
        \caption{\textbf{$|$ Robustness of $\KI$ patterns after controlling for interdisciplinarity.}
        (\textbf{A}) There is a slight negative correlation between our interdisciplinarity index and $\KI$. This relationship has become flatter and less pronounced in more recent decades.
        (\textbf{B}) Each curve represents the $\KI$ distribution for a different percentile bin of interdisciplinarity. The distribution remains a relatively stable bell-shaped curve across all groups.
        (\textbf{C}) Each line represents the average $\KI$ over time for a specific interdisciplinarity bin. All groups exhibit a consistent and parallel declining trend.
        (\textbf{D}) The positive association between $\KI$ and disruption remains robust and significant across all levels of interdisciplinarity.
        (\textbf{E}) The negative association between $\KI$ and citation impact remains robust and significant across all levels of interdisciplinarity.
        \label{FigOAS_Reference_Interdisciplinarity_KI}}
    \end{center}
\end{figure}}

\clearpage
\subsection{A Dual Monte Carlo Simulation Framework}\label{SI_Monte_Carlo}
{\fontsize{12pt}{14pt}\selectfont % {fontsize}{linespace}
To rigorously evaluate whether observed $\KI$ values and their temporal trends deviate significantly from random expectations, and check the predictive power of $\KI$ on scientific disruption and citation impact, we developed two distinct but complementary Monte Carlo simulation strategies. The first, a Random Rewiring Method, tests whether our observed patterns are non-random artifacts of the existing network structure. The second, a Network Generative Method, goes a step further to validate the explanatory sufficiency of mechanisms linking an author's citation behavior to the disruptive potential and citation impact of their work.

\subsubsection{Random rewiring Monte Carlo method to test for non-randomness}\label{SI_Monte_Carlo_Rewire}
First, to rigorously test whether the observed patterns in $\KI$, particularly its historical decline and its correlation with disruption and citation impact, represent a substantive shift in scientific practice rather than an artifact of the evolving structure of citation networks, we implemented a random rewiring simulation. The objective is to construct a null model that characterizes the component of $\KI$ attributable to the baseline expectation generated by structural factors alone.

\noindent \textbf{Method}: To construct our null model, we generated randomized versions of the empirical citation network using a degree- and age-preserving randomization algorithm that follows established protocols. This procedure systematically rewires the network through an ``edge swapping'' process: two citation links are randomly selected, and their endpoints are swapped. Crucially, a swap is accepted only if it maintains the exact in-degree (citations received), out-degree (references made), and citation-age profile for all papers involved, ensuring our randomized networks serve as robust counterfactuals that exhibit the same degree distribution and temporal dynamics as the empirical network.

To ensure thorough randomization while managing the computational load, we performed $2E*log(E)$ swaps, where $E$ is the total number of edges in the network, to create each of the 10 independent, rewired network replicas.

After generating this ensemble of randomized networks, we calculated the $\KI$ score for every paper within each replica. We then averaged these 10 scores for each paper to derive its randomized $\KI$ under our constrained null model, a metric we denote as $\KI_{\rm ran}$. This value represents the Knowledge Independence score a paper would likely have based purely on its structural position and the random chance inherent in network evolution, and serves as our null-model baseline for comparison against the empirically observed $\KI$.

\noindent \textbf{Findings and Implications}: 
To ensure our findings are not just artifacts of random network evolution, we compared our observed $\KI$ with a Monte Carlo null model, denoted as $\KI_{\rm ran}$. This parallel analysis provided powerful, unequivocal evidence that the patterns we observed are genuine and not random.
\begin{itemize}
\item \textbf{Divergent Distributions}: The distributions of the two metrics diverge dramatically (Fig.~\ref{FigOAS_Monte_Carlo_Reshuffle}A). The observed $\KI$ follows a bell-shaped density distribution, whereas the $\KI_{\rm ran}$ peaks sharply near 1, indicating that the empirical pattern is highly non-random and unlikely to to occur by chance.
\item \textbf{Divergent Historical Trends}: The historical trends also differ markedly (Fig.~\ref{FigOAS_Monte_Carlo_Reshuffle}B). The observed $\KI$ has undergone a steep, sustained decline over six decades. In contrast, the $\KI_{\rm ran}$ has remained flat, consistent with its values concentrating near 1. This confirms that the observed decline reflects a genuine historical shift, not merely an artifact of evolving network structure.
\item \textbf{Divergent Relationship with Disruption and Impact}: Most critically, the observed $\KI$ correlates positively with disruption and negatively with citation impact, while the null-model $\KI_{\rm ran}$ significantly deviates from the aforementioned correlations~(Fig.~\ref{FigOAS_Monte_Carlo_Reshuffle}C–D), proving that the predictive power of $\KI$ is a unique property, not a trivial consequence of random structure.
\end{itemize}

Taken together, these comparative analyses provide powerful evidence for our central claim. The empirically observed $\KI$ exhibits distinctive statistical properties, whose distribution and temporal decline starkly deviate from what we would expect under random network evolution. Crucially, the predictive power on disruption and citation impact are unique properties of our empirically observed $\KI$. In contrast, the null-model $\KI_{\exp}$ shows no such predictive power. Hence, this confirms that our findings reflect a substantive shift in scientific behavior, and are not artifacts of the inherent, random evolution of the underlying network structure. 

\begin{figure}[ht]
    \begin{center}
        \includegraphics[width=1\textwidth]{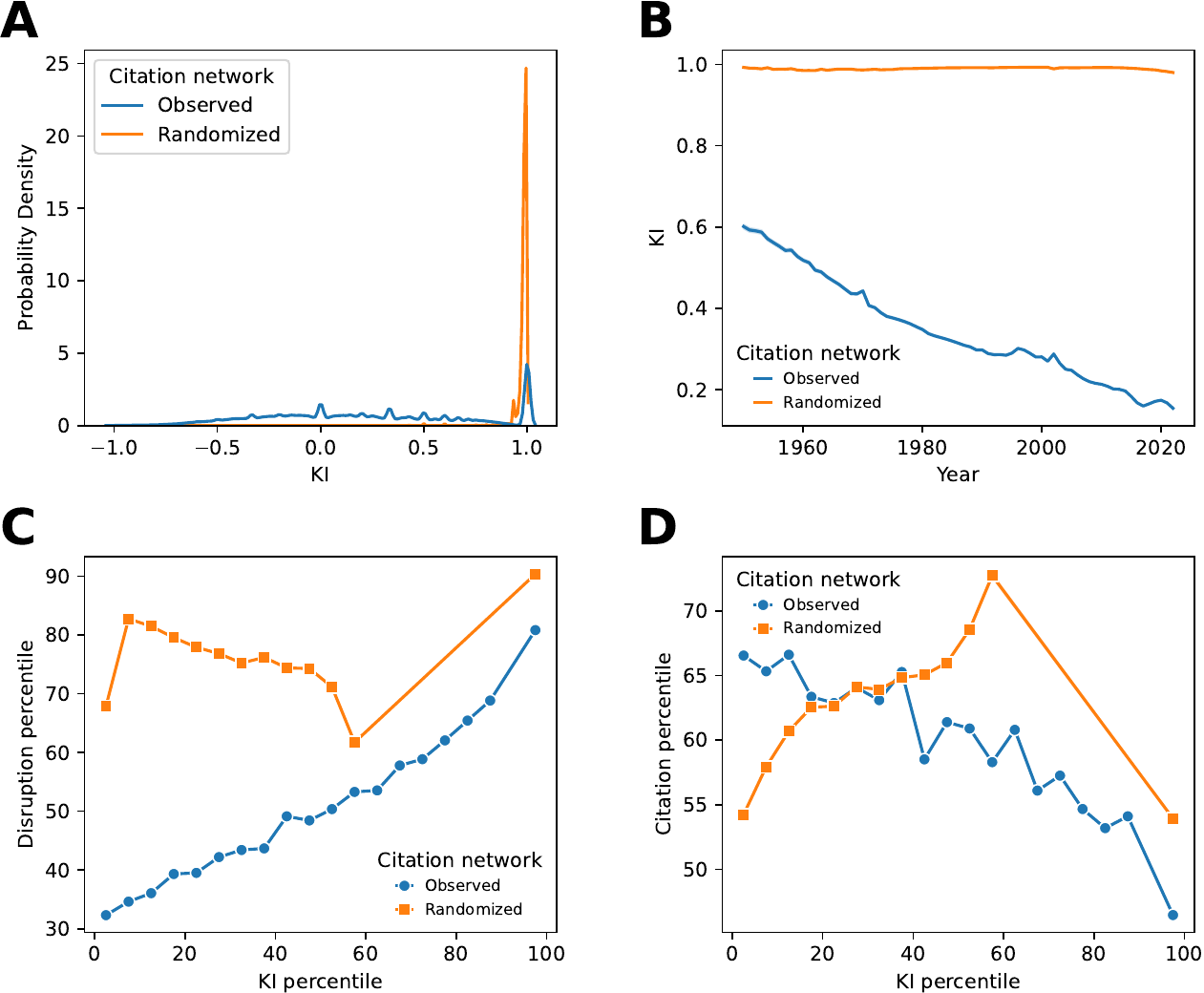}
        \caption{\textbf{$|$ Random rewiring Monte Carlo validation of $\KI$ metric.}
        (\textbf{A}) Observed $\KI$ follows a bell-shaped density curve, whereas the null-model $\KI_{\rm ran}$ peaks sharply near 1.
        (\textbf{B}) Over six decades, observed $\KI$ declines steeply, but the null-model $\KI_{\rm ran}$ remains flat.
        (\textbf{C-D}) Observed $\KI$ shows a clear positive association with disruption~(panel \textbf{C}) and a negative association with citation impact~(panel \textbf{D}). While the null-model $\KI_{\rm ran}$ appears to show slight rises or falls at the final two data points (22.5 and 97.5), these are unreliable due to sparse data and do not reflect genuine trends.
        \label{FigOAS_Monte_Carlo_Reshuffle}}
    \end{center}
\end{figure}

\subsubsection{Network generative Monte Carlo method to probe causal mechanisms}\label{SI_Monte_Carlo_Generate}
To further verify the explanatory sufficiency of mechanisms linking ``author behavior'' to our key findings, we developed a Network Generative Monte Carlo Method. This approach simulates the growth of a citation network from the ground up, allowing us to directly test how specific citation practices influence a paper's trajectory in terms of both disruption and citation impact.

\textbf{Method}: Our generative model is built upon a realistic foundation, incorporating three established mechanisms of citation dynamics: (1) preferential attachment~\cite{adamic2000power}, where highly-cited papers are more likely to be cited again; (2) node fitness~\cite{bianconi2001competition}, acknowledging that papers have an intrinsic, heterogeneous quality or appeal; and (3) an aging effect~\cite{wang2013quantifying,wang2009effect,medo2011temporal}, where a paper's citability decays over time. Crucially, we extend this baseline by incorporating two $\KI$-dependent mechanisms that we empirically observed, which directly explain the relationships between $\KI$, disruption, and citation impact:
\begin{itemize}
\item \textbf{Reference Selection (The $\KI$-Impact Mechanism)}: Our statistical analysis reveals that the probability of a paper being selected as a reference ($p_s$) is negatively correlated with its $\KI$ percentile (Fig.~\ref{FigS_model_prediction}A). That is, authors systematically prefer to cite more dependent (lower-$\KI$) works, which are often part of an established, mainstream research. Our model incorporates this behavior by making a paper's $\KI$ a factor in its general citability. This mechanism provides a direct, micro-level explanation for our finding that higher-$\KI$ papers tend to accumulate fewer citations over time.

\item \textbf{Path Following / Co-citation (The $\KI$-Disruption Mechanism)}: We also find that the probability of a paper co-citing a reference's own bibliography ($p_c$) is negatively correlated with the reference's $\KI$ percentile (Fig.~\ref{FigS_model_prediction}B). That is, a low-$\KI$ reference, being part of a dense knowledge cluster, tends to have a high probability of inducing co-citations of its own predecessors. Conversely, a high-$\KI$ reference is likely to be treated as a terminal node, discouraging further exploration of its predecessors. Our model simulates this ``path following'' behavior by making the $\KI$ of a reference, $r$, a factor in determining whether to co-cite $r$'s bibliography whenever $r$ is cited. This mechanism is the direct generative process for disruption; by discouraging co-citation, a high-$\KI$ reference is far more likely to be disruptive to the work upon which it builds.

\item  \textbf{Knowledge Independence ($\KI$) Model (Fig.~\ref{FigS_model_prediction}C)}: To ensure the model reflects real-world citation dynamics, we calibrate it using empirical distributions of $L$ (reference count), $F$ (fitness), and the temporal relaxation function $A = \phi(\tau)\propto(\tau)^{-\alpha}$ (aging effect, where $\tau$ denotes time since publication, and $\alpha$ is a decay parameter measurable from empirical data~\cite{bao2017dynamic,hu2021aging}). These distributions and function are derived from the Web of Science dataset and serve as input parameters. More specifically, the model starts from Step~\textbf{0}, which initiates the citation network. Then, it iteratively adds one paper at a time, by executing Steps~\textbf{1} to \textbf{3} for each paper:
\begin{itemize}
\item Step~\textbf{0}: Initiate the citation network by $1,000,000$ isolated nodes (papers). For each paper, randomly sample the reference length $l$ and fitness $f$ from distributions $L$ and $F$, respectively. Designate the first $10,000$ papers without any reference as the initial citable papers for subsequent papers; 
\item Step~\textbf{1}: For each newly added paper, if the number of selected references does not exceed $l$, move to Step~\textbf{2}; Else, continue to the next paper;
\item Step~\textbf{2}: Randomly select a reference $r$ with ($\KI_r$,$F_r$,$A_r$) from all citable papers according to the overall selecting probability $p_{s,o} \propto F_rA_r$, as combined by fitness and aging effect;
\item Step~\textbf{3}: Assign the probability $p_c \propto 1-\frac{\KI_r+1}{2}$, which determines the likelihood of co-citing any reference of reference $r$. When co-citing, randomly select a reference $r'$ with $\KI_{r'}$ from the reference list of $r$ according to the local selecting probability $p_{s,c} \propto -\KI_{r'}$, and update $r=r'$, $\KI_r=\KI_{r'}$, and return to Step~\textbf{3}; Else, return to Step~\textbf{2}.
\end{itemize}
\end{itemize}

\textbf{Parallel Simulation Experiments}: To isolate the effects of these two mechanisms in Step 3, we designed a set of six parallel simulation experiments. The principal ``empirical'' experiment sets both $p_{s,c}$ and $p_c$ to be negatively correlated with $\KI$, mirroring our real-world observations (Fig.~\ref{FigS_model_prediction}A-C). In the other four counterfactual experiments, we systematically altered one of these behavioral rules at a time, setting the preference to be positive or orthogonal (uncorrelated) to $\KI$. And for the last (sixth) counterfactual experiment, we construct a neutral model, setting both the preferences to be orthogonal (uncorrelated) to $\KI$.

\textbf{Findings and Implications}: The results of this controlled analysis were conclusive. The ``empirical'' model successfully and accurately reproduced the relationships we observe in the real data: a negative correlation between a paper's $\KI$ and its eventual citation impact, and a positive correlation between its $\KI$ and its disruption score (Fig.~\ref{FigS_model_prediction}D). In contrast, in each of the four counterfactual experiments that alters one of the behavioral rules, the modification systematically changed the corresponding outcome (Fig.~\ref{FigS_model_prediction}E-H). For instance, when the ``path following'' mechanism ($p_c$) was changed, the relationship between $\KI$ and disruption was fundamentally altered, while the $\KI$-impact relationship remained intact. Moreover, in the neutral model exhibiting no rule with $\KI$, neither of the relationship with $\KI$ occurred (Fig.~\ref{FigS_model_prediction}I).

This series of controlled, generative experiments verify our explanatory frameworks. It demonstrates not only that our proposed mechanisms can reproduce the observed phenomena, but also that these specific mechanisms are necessary to do so. We can therefore conclude with high confidence that the negative relationship between $\KI$ and citation impact can be driven by authors' preference to cite well-established (low-$\KI$) works, while the positive relationship between $\KI$ and disruption can be driven by the tendency to cite independent (high-$\KI$) works without co-citing their predecessors.

\subsubsection{Conclusion}
In summary, our dual Monte Carlo approach provides a comprehensive validation. The rewiring method confirms that our empirical finding in $\KI$'s decline is non-random, signaling a genuine shift in scientists' behaviour. The generative method provides powerful validation for the explanatory sufficiency of the link between an author's preference to compose independent knowledge and the outcome of producing disruptive but less recognized work.

\begin{figure}[htbp]
	\begin{center}
		\includegraphics[width=1\textwidth]{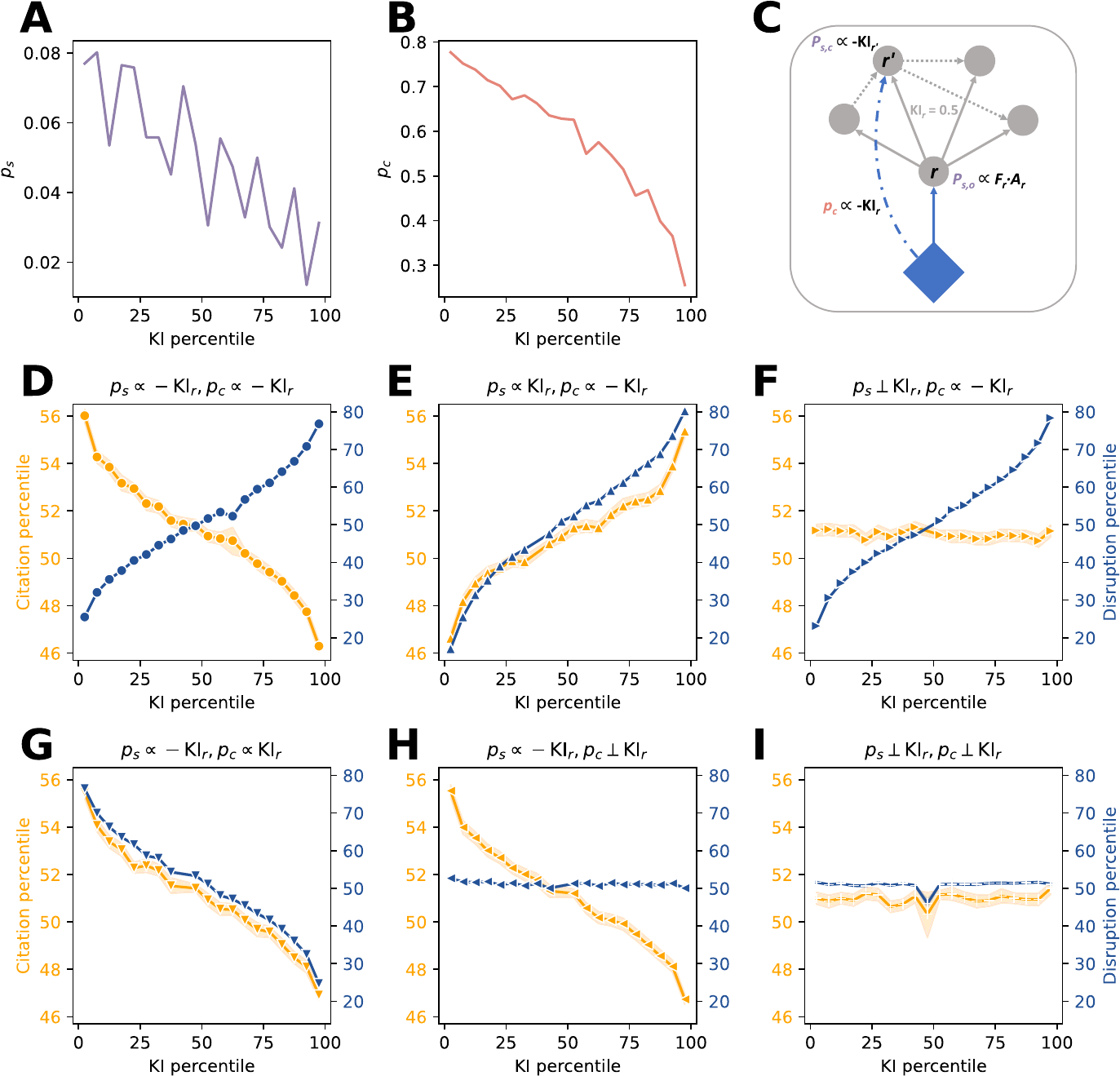}
		\caption{\textbf{$|$ Network generative Monte Carlo validation of $\KI$ metric.}
			(\textbf{A}) By going through all papers' reference lists, we obtain the distribution of the $\KI$ value among all the cited references, which represents the probability of a paper with $\KI$ being selected as a reference ($p_s$).
			(\textbf{B}) By going through all papers' reference lists, we identify the $dep$-type reference that is co-cited with its references by the same paper. We calculate the ratio of $dep$-type references among all references within the same group of $\KI$, which represents the probability of a paper with $\KI$ being co-cited with its references ($p_c$).
			(\textbf{C}) $\KI$ model: when a focal paper (blue diamond) selects references from all citable papers, the selecting probability $p_{s,o} \propto F_rA_r$ maintains the preferential attachment mechanism. Meanwhile, for a selected reference $r$, the co-citing probability $p_c \propto -\KI_r$ determines the likelihood that the focal paper simultaneously co-cites any reference of $r$. When co-citing, the local selecting probability $p_{s,c} \propto -\KI_{r'}$ determines the likelihood that a reference $r'$ of $r$ is selected.
			(\textbf{D}) The ``empirical'' $\KI$ model successfully reproduces the observed relationship between $\KI$ and citation impact, as well as disruption.
			(\textbf{E-H}) For each of the alternative $\KI$ model, the corresponding relationship with $\KI$ changed as the input mechanism altered.
			(\textbf{I}) For the neutral model, neither of the relationship with $\KI$ occurred as both the mechanisms altered.
			Bootstrapped 95\% confidence intervals are shown as gray zones.
			\label{FigS_model_prediction}}
	\end{center}
\end{figure}}

\clearpage
\subsection{Validating the robustness of analyses by Coarsened Exact Matching (CEM)}\label{SI_CEM}
{\fontsize{12pt}{14pt}\selectfont % {fontsize}{linespace}
To ensure that our causal estimates are not artifacts of arbitrary coarsening choices, we conducted a rigorous sensitivity analysis by varying the granularity of the CEM binning strategy. While the Coarsened Exact Matching (CEM) algorithm requires discretizing continuous variables, the choice of bin size involves a bias-variance trade-off: finer bins (more strata) reduce imbalance (bias) but may decrease the matched sample size (variance), whereas coarser bins preserve sample size at the cost of precision.

For categorical and numerical covariates with limited distinct values, like discipline and team freshness, we directly use their natural strata for binning. For other numerical covariates, we implemented a uniform binning strategy, systematically testing three distinct levels of granularity: 3-bin (coarse), 4-bin (moderate), and 5-bin (fine) specifications. Exceptionally, given the wide range of time period the dataset covers, we double the binning granularity for publication year. For each specification, we evaluated the matching performance using four key diagnostics: the number of valid strata, the matching rate, and both univariate and multivariate $L_1$ imbalance statistics. This multi-level approach allows us to verify the structural stability of the $\KI$-Disruption relationship across varying degrees of matching strictness.

\textbf{Matching Performance and Global Balance.} Fig.~\ref{FigOAS_CEM_coarsening_comparison} summarizes the diagnostic statistics across the three coarsening specifications. As expected, increasing the number of bins from 3 to 5 resulted in a significant increase in the number of generated strata (Panel A), reflecting a more precise matching structure. Crucially, due to the large scale of the OpenAlex dataset, the matching rate (Panel B) remained robustly high ($> 93\%$) even under the strictest 5-bin specification, indicating a strong common support between the treatment and control groups. Panels (C-D) illustrate the imbalance metrics. We observe a monotonic decrease in the multivariate $L_1$ statistic (Panel C) as the granularity increases, confirming that finer binning successfully removes higher-dimensional discrepancies. 
Complementing this, Panel (D) reveals that the mean univariate $L_1$ statistic remains consistently negligible across all specifications. Crucially, the values are on the order of $10^{-10}$, which is two orders of magnitude smaller than the multivariate imbalance. While minor nominal fluctuations are observed (e.g., the slight increase from 4 to 5 bins), these variations represent numerical noise rather than substantive imbalance. Taken together, these results indicate a coherent trade-off: refining the binning strategy effectively optimizes the global, multivariate balance (Panel C) while maintaining the univariate balance at an effectively perfect level (Panel D).

\textbf{Covariate Balance Assessment.} To visualize the quality of the match at a granular level, Fig.~\ref{FigOAS_CEM_SMD}A-C presents the Love Plots (Standardized Mean Differences, SMDs) for the 3-bin, 4-bin, and 5-bin specifications, respectively. Across all three panels, the post-matching SMDs (orange bars) are concentrated narrowly around zero, significantly outperforming the pre-matching imbalances (blue bars). Notably, even the coarse 3-bin strategy (Panel A) achieved SMDs well below the conventional threshold of 0.1. The fine 5-bin strategy (Panel C) yielded the most pristine balance, with almost all covariates exhibiting SMDs below 0.04, effectively approximating a randomized experiment. This confirms that our main findings are not driven by residual covariate imbalance.

\textbf{Robustness of Treatment Effects.} Fig.~\ref{FigOAS_CEM_SMD}D-F displays the estimated average treatment effect on the treated (ATT) matrices under the three binning strategies. 
The heatmaps exhibit a strikingly consistent morphological pattern across Panels (A) through (C). The magnitude of the ATT values and the gradient of the color spectrum remain stable regardless of whether 3, 4, or 5 bins are used for matching. The persistence of the positive gradient---where higher levels of $\KI$ consistently yield higher level of disruption---demonstrates that the estimated effect is robust to the choice of matching parameters.

% \subsubsection{Results based different coarsening schemes}
% Our CEM analysis robustly confirms the positive effect of increasing $\KI$ on scientific disruption, consistent with the findings obtained from PSM. Specifically, we observed that papers with higher $\KI$, when matched using CEM based on a wide range of covariates, consistently exhibit greater scientific disruption (Fig.~\ref{FigOAS_CEM_SMD}). For example, our CEM analyses show that increasing $\KI$ from the lowest to the highest decile resulted in an approximate increase of 20 percentile points in disruption ($p < 0.001$), which is highly consistent with the effects observed via PSM.

% This additional analysis using CEM reinforces the robustness of our central finding regarding the positive impact of $\KI$ on scientific disruption. It provides stronger assurances that our results are not overly sensitive to the specific causal inference method employed and that covariate balance is well-maintained.

\begin{figure}[ht]
    \begin{center}
        \includegraphics[width=.78\textwidth]{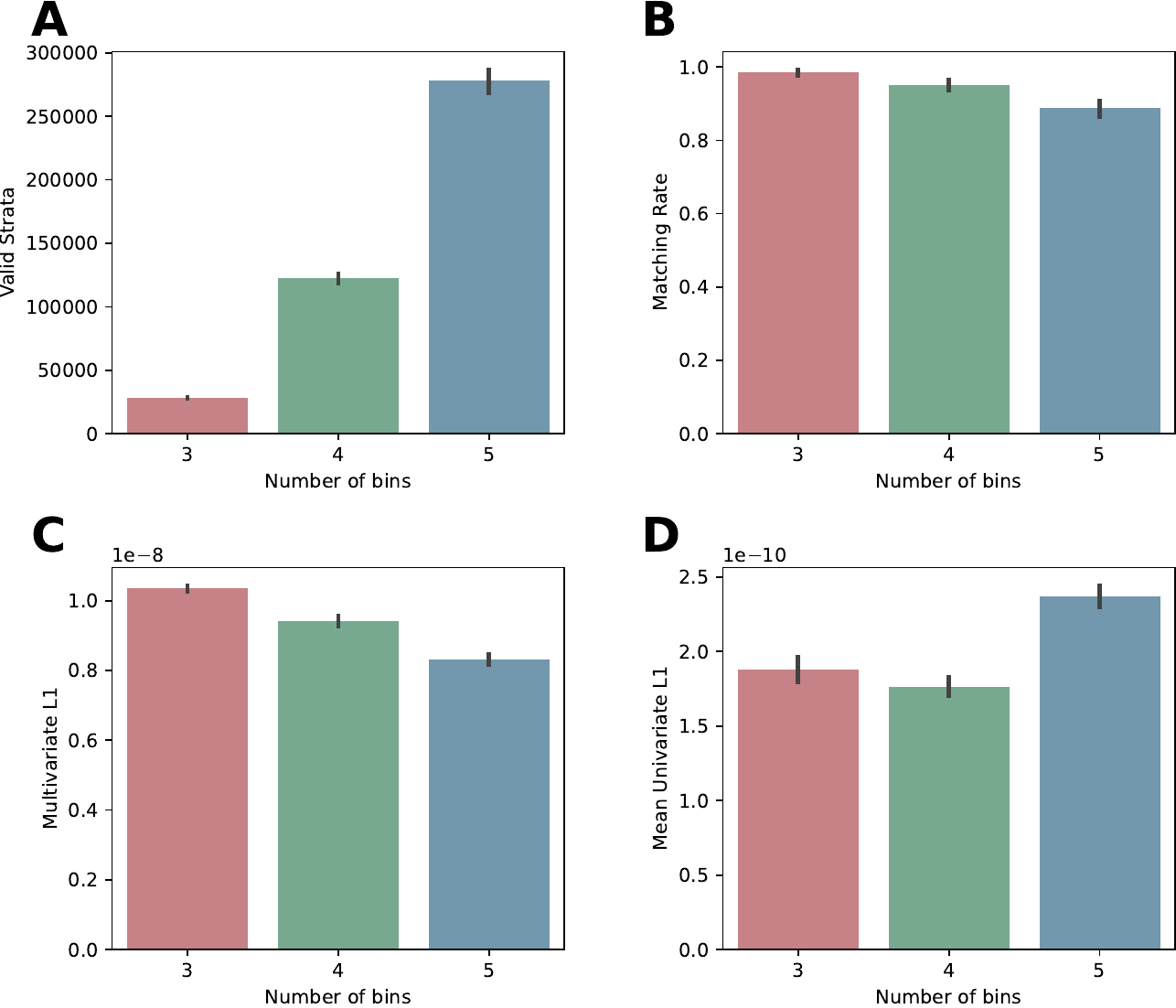}
        \caption{\textbf{$|$ Coarsening scheme comparison for CEM.}
        (\textbf{A}) The average number of valid strata (containing both treatment and control units) across bin configurations. %More bins create additional strata but do not necessary increase the proportion suitable for matched comparison analysis.
        (\textbf{B}) The average matching rate across different bin numbers. %Higher bin counts generally reduce matching rates due to increased stratification granularity.
        (\textbf{C}) The average multivariate $L_1$ imbalance, measuring overall covariate balance across all dimensions simultaneously. %Lower values indicate better balance.
        (\textbf{D}) The average univariate $L_1$ imbalance across individual covariates. This metric complements multivariate balance by revealing variable-specific matching quality.
        The error bars represent 95\% confidence intervals.
        \label{FigOAS_CEM_coarsening_comparison}}
    \end{center}
\end{figure}

\begin{figure}[ht]
    \begin{center}
        \includegraphics[width=1\textwidth]{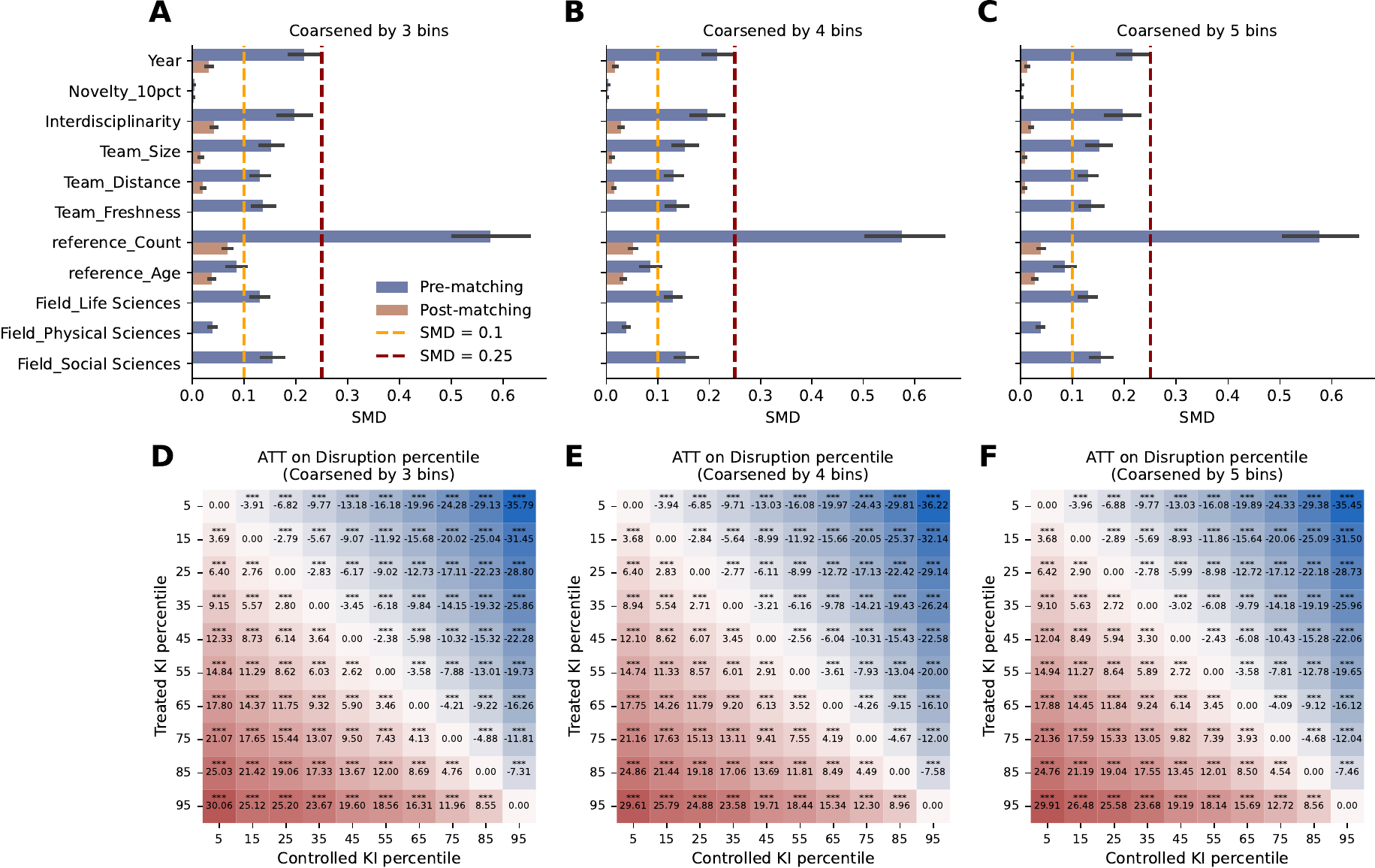}
        \caption{\textbf{$|$ Causal inference by CEM.}
        (\textbf{A-C}): The Love Plots (Standardized Mean Differences, SMDs) for the 3-bin, 4-bin, and 5-bin specifications, respectively. The error bars represent 95\% confidence intervals.
        (\textbf{D-F}): The average treatment effect on the treated (ATT) matrix of $\KI$ on disruption for the 3-bin, 4-bin, and 5-bin specifications, respectively.
        Each ATT is tested against the null hypothesis (ATT equals 0) using a two-sided \textit{t}-test. (*$p < 0.05$, **$p < 0.01$, ***$p < 0.001$).
        \label{FigOAS_CEM_SMD}}
    \end{center}
\end{figure}}

\clearpage
\subsection{Results based on propensity score matching (PSM)}\label{SI_PSM}
{\fontsize{12pt}{14pt}\selectfont % {fontsize}{linespace}
To further enhance the robustness of our findings by Coarsened Exact Matching (CEM), we have performed an additional causal inference analysis using Propensity Score Matching (PSM)~\cite{abadie2016matching}. PSM has been extensively employed to estimate causal treatment effects in the Science of Science~\cite{dong2023nobel,mutz2017effect}. It offers a robust approach to account for potential biases induced by covariates, particularly in scenarios where randomized experiments are impractical. We apply PSM to rigorously quantify the effect size of $\KI$ on scientific disruption. This method involves same procedures as CEM method, with a different matching and weighting strategies:

\begin{itemize}
    \item \textbf{Treatment and Control Groups}: We categorize papers into treated and control groups based on their $\KI$ values.
    \item \textbf{Covariates:} We utilize the set of ex-ante covariates: focal paper properties (discipline and publication year), reference properties (novelty, interdisciplinarity, reference count, and average reference age), and team properties (size, geographic distance, and collaboration freshness).
    \item \textbf{Propensity scoring and matching:} 
    We estimate the propensity score for each paper by fitting a generalized linear model considering all the covariates, where the treatment size is the dependent variable. For each paper in the control group, a counterpart from the treated group with a similar propensity score is matched. This matching ensures a balanced distribution of covariates across the two groups~(Fig.~\ref{FigOAS_PSM_SMD}A), effectively mimicking the conditions of a randomized controlled trial.
    \item \textbf{Weighting:} 
    For matched control and treated papers: Assigned a weight of 1. For unmatched (discarded) papers: Assigned a weight of 0.
    \item \textbf{Effect estimation:} To estimate the Average Treatment Effect on the Treated (ATT), we fit a Weighted Least Squares (WLS) regression model on the matched sample using the PSM-generated weights, which is specified as follows: 
    $$D = b + a_{ATT} \cdot T + \delta \cdot \rm Covariates + \epsilon$$
    Where: $D$ is the outcome variable of disruption. $T$ is the binary treatment indicator of $\KI$. Regarding doubly robust estimation, we control for the original Covariates to exploit residual imbalances. The regression is weighted by the PSM weights. The coefficient of interest, $a_{ATT}$, captures the marginal effect of the treatment on the outcome after balancing ex-ante characteristics.
    
    Our PSM analysis robustly confirms the positive effect of increasing $\KI$ on scientific disruption, consistent with the findings obtained from CEM. Specifically, we observed that papers with higher $\KI$, when matched using PSM based on a wide range of covariates, consistently exhibit greater scientific disruption~(Fig.~\ref{FigOAS_PSM_SMD}B). For example, our PSM analyses show that increasing $\KI$ from the lowest to the highest decile resulted in an approximate increase of 29.25 percentile points in disruption~($p < 0.001$), which is highly consistent with the effects observed via CEM.
    
    This additional analysis using PSM reinforces the robustness of our central finding regarding the positive impact of $\KI$ on scientific disruption. It provides stronger assurances that our results are not overly sensitive to the specific causal inference method employed.
\end{itemize}

\begin{figure}[htbp]
    \begin{center}
        \includegraphics[width=1\textwidth]{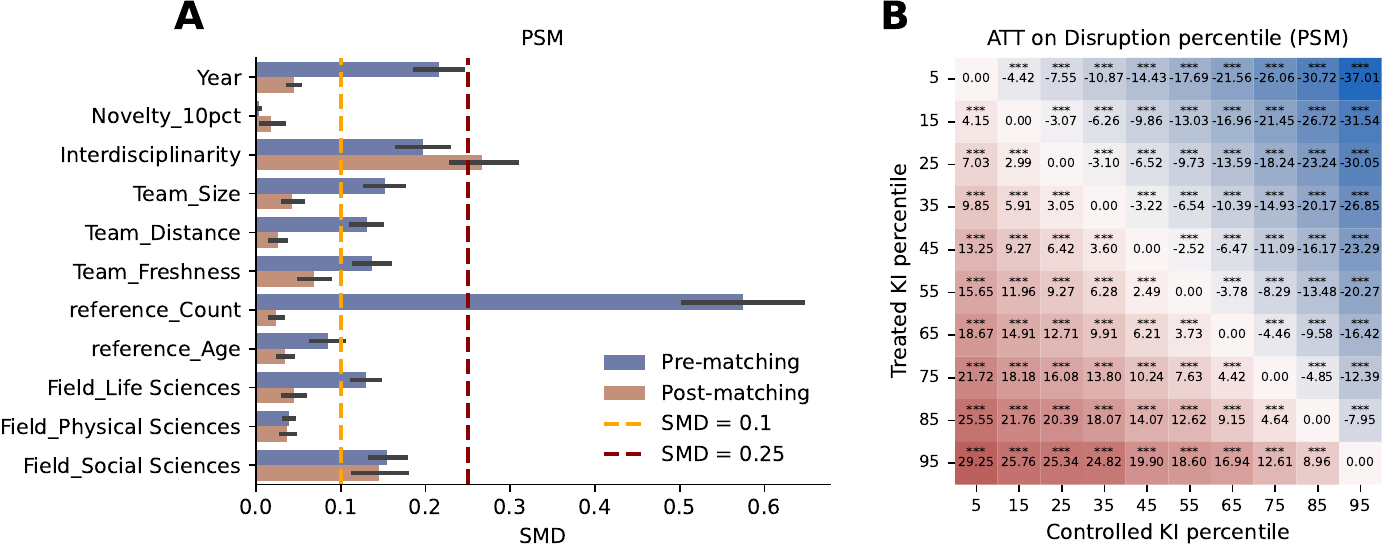}      
        \caption{\textbf{$|$ Causal inference by PSM.}
        \textbf{A}), Across all control-treatment comparisons, the Standardized Mean Difference~(SMD) plots consistently show a marked reduction in the length of orange bars~(post-matching) compared to blue bars~(pre-matching), indicating highly consistent balance improvement across various covariate dimensions. The error bars represent 95\% confidence intervals.
        \textbf{B}), 
        The ATT (average treatment effect on the treated) matrix of $\KI$ on disruption via PSM. Each ATT is tested against the null hypothesis (ATT equals 0) using a two-sided \textit{t}-test. (*$p < 0.05$, **$p < 0.01$, ***$p < 0.001$).
        \label{FigOAS_PSM_SMD}}
    \end{center}
\end{figure}}

\clearpage
\subsection{The robustness of regression, mediating, and confounding analyses based on alternative disruption metrics}\label{SI_ALTER_DISRUPTION}
% OLS, MEDIATING, AND CONFOUNDING ANALYSIS ON FOUR INDICATORS OF DISRUPTION

%%%%%%%%%%%%%%%%%%% OLS ANALYSIS %%%%%%%%%%%%%%%%%%%
\begin{table}[ht] % Do NOT use \begin{table*}
\centering
% Captions go above tables
\caption{\textbf{$|$ Normalized regression models of alternative disruption metrics.}}
\label{table_SI_ALL_OLS} % give each table a logical label name
\renewcommand\tabcolsep{0.16cm} % column spacing
\renewcommand{\arraystretch}{0.8} % line spacing
\begin{threeparttable}
    \fontsize{7.5}{14}\selectfont 

    \begin{tabular}{l cc p{0.6cm} cc p{0.6cm} cc p{0.6cm} cc}
        \toprule % head line for the table

        % sub-title setting
        & \multicolumn{2}{c}{Metric $D_0$} & 
        & \multicolumn{2}{c}{Metric $D_1$} & 
        & \multicolumn{2}{c}{Metric $D_2$} & 
        & \multicolumn{2}{c}{Metric $D_3$} \\
        
        % head line for the sub-title
        \cmidrule(lr){2-3} \cmidrule(lr){5-6} \cmidrule(lr){8-9} \cmidrule(lr){11-12}

        & Model & $R^2$\% & & Model & $R^2$\% & & Model & $R^2$\% & & Model & $R^2$\% \\ 
        
        \midrule % middle line
        
        $\KI$ & \textbf{0.4134}*** & \textbf{71.87} & & \textbf{0.1849}*** & \textbf{78.05} & & \textbf{0.3870}*** & \textbf{69.60} & & \textbf{0.1637}*** & \textbf{65.45} \\
        & (0.000) &  &  & (0.000) &  &  & (0.000) &  &  & (0.000) \\
        Novelty & 0.0002 & 0.00 & & 0.0042*** & 0.04 & & 0.0002 & 0.00 & & 0.0038*** & 0.04 \\
        & (0.000) &  &  & (0.000) &  &  & (0.000) &  &  & (0.000) \\
        Interdisciplinarity & -0.0061*** & 0.61 & & -0.0245*** & 0.84 & & -0.0139*** & 1.06 & & -0.0192*** & 1.33 \\
        & (0.000) &  &  & (0.000) &  &  & (0.000) &  &  & (0.000) \\
        Reference count & -0.0579*** & 10.82 & & 0.0081*** & 5.51 & & -0.0592*** & 8.94 & & 0.0243*** & 2.12 \\
        & (0.000) &  &  & (0.000) &  &  & (0.000) &  &  & (0.000) \\
        Reference age & -0.0148*** & 0.12 & & 0.0048*** & 0.11 & & -0.0082*** & 0.06 & & 0.0152*** & 1.45 \\
        & (0.000) &  &  & (0.000) &  &  & (0.000) &  &  & (0.000) \\
        Team size & -0.0159*** & 0.68 & & -0.0011*** & 0.50 & & -0.0158*** & 0.35 & & -0.0020*** & 0.15 \\
        & (0.000) &  &  & (0.000) &  &  & (0.000) &  &  & (0.000) \\
        Team distance & -0.0256*** & 1.21 & & -0.0028*** & 0.52 & & -0.0241*** & 0.92 & & -0.0043*** & 0.20 \\
        & (0.000) &  &  & (0.000) &  &  & (0.000) &  &  & (0.000) \\
        Team freshness & 0.0469*** & 2.44 & & 0.0133*** & 1.65 & & 0.0444*** & 2.40 & & 0.0153*** & 1.78 \\
        & (0.000) &  &  & (0.000) &  &  & (0.000) &  &  & (0.000) \\
        
        \midrule % middle line
        
        Discipline fixed effects & Yes & 10.60 & & Yes & 6.66 & & Yes & 11.22 & & Yes & 9.30 \\
        Year fixed effects & Yes & 1.64 & & Yes & 6.12 & & Yes & 5.44 & & Yes & 18.17 \\
        
        \midrule % middle line
        
        N & 33304580 &  &  & 33304580 &  &  & 31948768 &  &  & 31948768 \\
        $R^2$ & 0.231 &  &  & 0.039 &  &  & 0.199 &  &  & 0.031 \\
        
        \bottomrule % bottom line
    \end{tabular}
    \begin{tablenotes}
        \small
        \item[$\dagger$] We perform normalized ordinary-least-squares (OLS) regression analyses to examine the effect of various covariates on different disruption metrics (Disruption section in Methods). Each regression coefficient is tested against the null hypothesis (coefficient equals 0) using a two-sided \textit{t}-test. Standard errors are provided in parentheses for each coefficient. Note that we do not apply adjustments for multiple hypothesis testing in this analysis. (*$p < 0.05$, **$p < 0.01$, ***$p < 0.001$).
    \end{tablenotes}
\end{threeparttable}
\end{table}

%%%%%%%%%%%%%%%%%%% MEDIATING ANALYSIS %%%%%%%%%%%%%%%%%%%
\begin{table}[ht] % Do NOT use \begin{table*}
\centering
% Captions go above tables
\caption{\textbf{$|$ Normalized mediation regression on team size with alternative disruption metrics.}}
\label{table_SI_ALL_mediation_TS} % give each table a logical label name
\renewcommand\tabcolsep{0.16cm} % column spacing
\renewcommand{\arraystretch}{0.8} % line spacing
\begin{threeparttable}
    \fontsize{7.5}{14}\selectfont 

    \begin{tabular}{l cc p{0.3cm} cc p{0.3cm} cc p{0.3cm} cc}
        \toprule % head line for the table

        % sub-title setting
        & \multicolumn{2}{c}{Metric $D_0$} & 
        & \multicolumn{2}{c}{Metric $D_1$} & 
        & \multicolumn{2}{c}{Metric $D_2$} & 
        & \multicolumn{2}{c}{Metric $D_3$} \\
        
        % head line for the sub-title
        \cmidrule(lr){2-3} \cmidrule(lr){5-6} \cmidrule(lr){8-9} \cmidrule(lr){11-12}

        Mediator 
        & ACME ($\beta_1\theta_2$) & $|\frac{\rm ACME}{\rm TE}|$ & 
        & ACME ($\beta_1\theta_2$) & $|\frac{\rm ACME}{\rm TE}|$ & 
        & ACME ($\beta_1\theta_2$) & $|\frac{\rm ACME}{\rm TE}|$ & 
        & ACME ($\beta_1\theta_2$) & $|\frac{\rm ACME}{\rm TE}|$ \\ 
        
        \midrule % middle line
        
        $\KI$ & \textbf{-0.0196}*** & \textbf{0.5680} & & \textbf{-0.0079}*** & \textbf{0.9719} & & \textbf{-0.0177}*** & \textbf{0.5608} & & \textbf{-0.0067}*** & \textbf{0.8621} \\
        & (0.0011) &  &  & (0.0004) &  &  & (0.0010) &  &  & (0.0004) \\
        Novelty & -0.0000*** & 0.0000 & & 0.0000*** & 0.0002 & & -0.0000*** & 0.0000 & & 0.0000*** & 0.0002 \\
        & (0.0000) &  &  & (0.0000) &  &  & (0.0000) &  &  & (0.0000) \\
        Interdisciplinarity & -0.0020*** & 0.0575 & & -0.0001*** & 0.0155 & & -0.0017*** & 0.0553 & & -0.0003*** & 0.0352 \\
        & (0.0001) &  &  & (0.0000) &  &  & (0.0001) &  &  & (0.0000) \\
        Reference count & -0.0049*** & 0.1413 & & -0.0014*** & 0.1686 & & -0.0042*** & 0.1344 & & -0.0008*** & 0.1005 \\
        & (0.0003) &  &  & (0.0001) &  &  & (0.0002) &  &  & (0.0000) \\
        Reference age & -0.0011*** & 0.0330 & & -0.0012*** & 0.1493 & & -0.0013*** & 0.0427 & & -0.0020*** & 0.2620 \\
        & (0.0000) &  &  & (0.0001) &  &  & (0.0001) &  &  & (0.0001) \\
        % Team size & -0.0159*** & 0.68 & & -0.0011*** & 0.50 & & -0.0158*** & 0.35 & & -0.0020*** & 0.15 \\
        % & (0.0000) &  &  & (0.0000) &  &  & (0.0000) &  &  & (0.0000) \\
        Team distance & -0.0092*** & 0.2678 & & -0.0024*** & 0.2891 & & -0.0086*** & 0.2734 & & -0.0021*** & 0.2692 \\
        & (0.0005) &  &  & (0.0001) &  &  & (0.0004) &  &  & (0.0001) \\
        Team freshness & 0.0082*** & 0.2368 & & 0.0028*** & 0.3473 & & 0.0078*** & 0.2461 & & 0.0027*** & 0.3539 \\
        & (0.0005) &  &  & (0.0002) &  &  & (0.0004) &  &  & (0.0002) \\
        
        \bottomrule % bottom line
    \end{tabular}
    \begin{tablenotes}
        \small
        \item[$\dagger$] We perform normalized mediation analysis to assess the effect of team size on different disruption metrics (Disruption section in Methods). The candidate mediating variables include: $\KI$, novelty, interdisciplinarity, reference count, reference age, team distance, and team freshness, with discipline and publication year controlled. ACME refers to average causal mediation effects driven from the mediating variable and TE is defined as the total effect. Given that the TE is identical to all candidate mediators in each model, we opt to not report this value but the ratio of ADE among TE to highlight the relative mediating effect. Each regression coefficient is tested against the null hypothesis (coefficient equals 0) using a two-sided \textit{t}-test. Standard errors are provided in parentheses for each coefficient. Note that we do not apply adjustments for multiple hypothesis testing in this analysis. (*$p < 0.05$, **$p < 0.01$, ***$p < 0.001$).
    \end{tablenotes}
\end{threeparttable}
\end{table}

\begin{table}[ht] % Do NOT use \begin{table*}
\centering
% Captions go above tables
\caption{\textbf{$|$ Normalized mediation regression on team distance with alternative disruption metrics.}}
\label{table_SI_ALL_mediation_TD} % give each table a logical label name
\renewcommand\tabcolsep{0.16cm} % column spacing
\renewcommand{\arraystretch}{0.8} % line spacing
\begin{threeparttable}
    \fontsize{7.5}{14}\selectfont 

    \begin{tabular}{l cc p{0.3cm} cc p{0.3cm} cc p{0.3cm} cc}
        \toprule % head line for the table

        % sub-title setting
        & \multicolumn{2}{c}{Metric $D_0$} & 
        & \multicolumn{2}{c}{Metric $D_1$} & 
        & \multicolumn{2}{c}{Metric $D_2$} & 
        & \multicolumn{2}{c}{Metric $D_3$} \\
        
        % head line for the sub-title
        \cmidrule(lr){2-3} \cmidrule(lr){5-6} \cmidrule(lr){8-9} \cmidrule(lr){11-12}

        Mediator 
        & ACME ($\beta_1\theta_2$) & $|\frac{\rm ACME}{\rm TE}|$ & 
        & ACME ($\beta_1\theta_2$) & $|\frac{\rm ACME}{\rm TE}|$ & 
        & ACME ($\beta_1\theta_2$) & $|\frac{\rm ACME}{\rm TE}|$ & 
        & ACME ($\beta_1\theta_2$) & $|\frac{\rm ACME}{\rm TE}|$ \\ 
        
        \midrule % middle line
        
        $\KI$ & \textbf{-0.0330}*** & \textbf{0.5151} & & \textbf{-0.0134}*** & \textbf{0.8232} & & \textbf{-0.0306}*** & \textbf{0.5129} & & \textbf{-0.0116}*** & \textbf{0.8016} \\
        & (0.0001) &  &  & (0.0001) &  &  & (0.0001) &  &  & (0.0001) \\
        Novelty & 0.0000*** & 0.0000 & & -0.0000*** & 0.0000 & & 0.0000*** & 0.0000 & & -0.0000*** & 0.0000 \\
        & (0.0000) &  &  & (0.0000) &  &  & (0.0000) &  &  & (0.0000) \\
        Interdisciplinarity & 0.0008*** & 0.0121 & & 0.0001*** & 0.0033 & & 0.0007*** & 0.0113 & & 0.0001*** & 0.0074 \\
        & (0.0000) &  &  & (0.0000) &  &  & (0.0000) &  &  & (0.0000) \\
        Reference count & -0.0181*** & 0.2824 & & -0.0051*** & 0.3145 & & -0.0169*** & 0.2836 & & -0.0031*** & 0.2132 \\
        & (0.0001) &  &  & (0.0000) &  &  & (0.0001) &  &  & (0.0000) \\
        Reference age & -0.0002*** & 0.0036 & & -0.0002*** & 0.0139 & & -0.0003*** & 0.0042 & & -0.0004*** & 0.0245 \\
        & (0.0000) &  &  & (0.0000) &  &  & (0.0000) &  &  & (0.0000) \\
        Team size & -0.0039*** & 0.0605 & & -0.0009*** & 0.0548 & & -0.0035*** & 0.0593 & & -0.0009*** & 0.0604 \\
        & (0.0002) &  &  & (0.0001) &  &  & (0.0002) &  &  & (0.0001) \\
        % Team distance & -0.0092*** & 0.2678 & & -0.0024*** & 0.2891 & & -0.0086*** & 0.2734 & & -0.0021*** & 0.2692 \\
        % & (0.0005) &  &  & (0.0001) &  &  & (0.0004) &  &  & (0.0001) \\
        Team freshness & 0.0005*** & 0.0079 & & 0.0002*** & 0.0109 & & 0.0006*** & 0.0102 & & 0.0002*** & 0.0149 \\
        & (0.0000) &  &  & (0.0000) &  &  & (0.0000) &  &  & (0.0000) \\
        
        \bottomrule % bottom line
    \end{tabular}
    \begin{tablenotes}
        \small
        \item[$\dagger$] We perform normalized mediation analysis to assess the effect of team distance on different disruption metrics (Disruption section in Methods). The candidate mediating variables include: $\KI$, novelty, interdisciplinarity, reference count, reference age, team size, and team freshness, with discipline and publication year controlled. ACME refers to average causal mediation effects driven from the mediating variable and TE is defined as the total effect. Given that the TE is identical to all candidate mediators in each model, we opt to not report this value but the ratio of ADE among TE to highlight the relative mediating effect. Each regression coefficient is tested against the null hypothesis (coefficient equals 0) using a two-sided \textit{t}-test. Standard errors are provided in parentheses for each coefficient. Note that we do not apply adjustments for multiple hypothesis testing in this analysis. (*$p < 0.05$, **$p < 0.01$, ***$p < 0.001$).
    \end{tablenotes}
\end{threeparttable}
\end{table}

\begin{table}[ht] % Do NOT use \begin{table*}
\centering
% Captions go above tables
\caption{\textbf{$|$ Normalized mediation regression on team freshness with alternative disruption metrics.}}
\label{table_SI_ALL_mediation_TF} % give each table a logical label name
\renewcommand\tabcolsep{0.16cm} % column spacing
\renewcommand{\arraystretch}{0.8} % line spacing
\begin{threeparttable}
    \fontsize{7.5}{14}\selectfont 

    \begin{tabular}{l cc p{0.3cm} cc p{0.3cm} cc p{0.3cm} cc}
        \toprule % head line for the table

        % sub-title setting
        & \multicolumn{2}{c}{Metric $D_0$} & 
        & \multicolumn{2}{c}{Metric $D_1$} & 
        & \multicolumn{2}{c}{Metric $D_2$} & 
        & \multicolumn{2}{c}{Metric $D_3$} \\
        
        % head line for the sub-title
        \cmidrule(lr){2-3} \cmidrule(lr){5-6} \cmidrule(lr){8-9} \cmidrule(lr){11-12}

        Mediator 
        & ACME ($\beta_1\theta_2$) & $|\frac{\rm ACME}{\rm TE}|$ & 
        & ACME ($\beta_1\theta_2$) & $|\frac{\rm ACME}{\rm TE}|$ & 
        & ACME ($\beta_1\theta_2$) & $|\frac{\rm ACME}{\rm TE}|$ & 
        & ACME ($\beta_1\theta_2$) & $|\frac{\rm ACME}{\rm TE}|$ \\ 
        
        \midrule % middle line
        
        $\KI$ & \textbf{0.0465}*** & \textbf{0.4922} & & \textbf{0.0189}*** & \textbf{0.5699} & & \textbf{0.0422}*** & \textbf{0.4818} & & \textbf{0.0159}*** & \textbf{0.5100} \\
        & (0.0002) &  &  & (0.0001) &  &  & (0.0001) &  &  & (0.0001) \\
        Novelty & -0.0000*** & 0.0000 & & 0.0000*** & 0.0000 & & -0.0000*** & 0.0000 & & 0.0000*** & 0.0000 \\
        & (0.0000) &  &  & (0.0000) &  &  & (0.0000) &  &  & (0.0000) \\
        Interdisciplinarity & 0.0001*** & 0.0000 & & 0.0002*** & 0.0033 & & 0.0001*** & 0.0010 & & 0.0000*** & 0.0005 \\
        & (0.0000) &  &  & (0.0000) &  &  & (0.0000) &  &  & (0.0000) \\
        Reference count & 0.0136*** & 0.1439 & & 0.0038*** & 0.1150 & & 0.0125*** & 0.1433 & & 0.0022*** & 0.0717 \\
        & (0.0001) &  &  & (0.0000) &  &  & (0.0001) &  &  & (0.0000) \\
        Reference age & 0.0000*** & 0.0002 & & 0.0000*** & 0.0005 & & -0.0001*** & 0.0006 & & -0.0001*** & 0.0023 \\
        & (0.0000) &  &  & (0.0000) &  &  & (0.0000) &  &  & (0.0000) \\
        Team size & -0.0035*** & 0.0375 & & -0.0009*** & 0.0276 & & -0.0034*** & 0.0384 & & -0.0009*** & 0.0287 \\
        & (0.0002) &  &  & (0.0001) &  &  & (0.0002) &  &  & (0.0001) \\
        Team distance & -0.0003*** & 0.0036 & & -0.0001*** & 0.0026 & & -0.0004*** & 0.0047 & & -0.0001*** & 0.0032 \\
        & (0.0000) &  &  & (0.0000) &  &  & (0.0000) &  &  & (0.0000) \\
        % Team freshness & 0.0005*** & 0.0079 & & 0.0002*** & 0.0109 & & 0.0006*** & 0.0102 & & 0.0002*** & 0.0149 \\
        % & (0.0000) &  &  & (0.0000) &  &  & (0.0000) &  &  & (0.0000) \\
        
        \bottomrule % bottom line
    \end{tabular}
    \begin{tablenotes}
        \small
        \item[$\dagger$] We perform normalized mediation analysis to assess the effect of team freshness on different disruption metrics (Disruption section in Methods). The candidate mediating variables include: $\KI$, novelty, interdisciplinarity, reference count, reference age, team size, and team distance, with discipline and publication year controlled. ACME refers to average causal mediation effects driven from the mediating variable and TE is defined as the total effect. Given that the TE is identical to all candidate mediators in each model, we opt to not report this value but the ratio of ADE among TE to highlight the relative mediating effect. Each regression coefficient is tested against the null hypothesis (coefficient equals 0) using a two-sided \textit{t}-test. Standard errors are provided in parentheses for each coefficient. Note that we do not apply adjustments for multiple hypothesis testing in this analysis. (*$p < 0.05$, **$p < 0.01$, ***$p < 0.001$).
    \end{tablenotes}
\end{threeparttable}
\end{table}

\begin{table}[ht] % Do NOT use \begin{table*}
\centering
% Captions go above tables
\caption{\textbf{$|$ Normalized confounding regression on paper impact with alternative disruption metrics.}}
\label{table_SI_ALL_confounding_impact} % give each table a logical label name
\renewcommand\tabcolsep{0.16cm} % column spacing
\renewcommand{\arraystretch}{0.8} % line spacing
\begin{threeparttable}
    \fontsize{7.5}{14}\selectfont 

    \begin{tabular}{l cc p{0.0cm} cc p{0.0cm} cc p{0.0cm} cc}
        \toprule % head line for the table

        % sub-title setting
        & \multicolumn{2}{c}{Metric $D_0$} & 
        & \multicolumn{2}{c}{Metric $D_1$} & 
        & \multicolumn{2}{c}{Metric $D_2$} & 
        & \multicolumn{2}{c}{Metric $D_3$} \\
        
        % head line for the sub-title
        \cmidrule(lr){2-3} \cmidrule(lr){5-6} \cmidrule(lr){8-9} \cmidrule(lr){11-12}

        Mediator 
        & Adjusted ($\delta_1$) & RCE ($|\frac{\alpha_1 - \delta_1}{\alpha_1}|$) & 
        & Adjusted ($\delta_1$) & RCE ($|\frac{\alpha_1 - \delta_1}{\alpha_1}|$) & 
        & Adjusted ($\delta_1$) & RCE ($|\frac{\alpha_1 - \delta_1}{\alpha_1}|$) & 
        & Adjusted ($\delta_1$) & RCE ($|\frac{\alpha_1 - \delta_1}{\alpha_1}|$) \\ 
        
        \midrule % middle line
        
        $\KI$ & \textbf{-0.0248}*** & \textbf{0.8164} & & \textbf{0.0189}*** & \textbf{1.6907} & & \textbf{-0.0471}*** & \textbf{0.6735} & & \textbf{-0.0352}*** & \textbf{0.4967} \\
        & (0.0004) &  &  & (0.0004) &  &  & (0.0004) &  &  & (0.0004) \\
        Novelty & -0.1349*** & 0.0001 & & -0.0273*** & 0.0001 & & -0.1441*** & 0.0001 & & -0.0700*** & 0.0001 \\
        & (0.0004) &  &  & (0.0004) &  &  & (0.0004) &  &  & (0.0004) \\
        Interdisciplinarity & -0.1355*** & 0.0044 & & -0.0274*** & 0.0027 & & -0.1446*** & 0.0031 & & -0.0701*** & 0.0016 \\
        & (0.0004) &  &  & (0.0004) &  &  & (0.0004) &  &  & (0.0004) \\
        Reference count & -0.0762*** & 0.4356 & & -0.0100*** & 0.6335 & & -0.0916*** & 0.3644 & & -0.0670*** & 0.0437 \\
        & (0.0005) &  &  & (0.0004) &  &  & (0.0005) &  &  & (0.0004) \\
        Reference age & -0.1355*** & 0.0045 & & -0.0258*** & 0.0545 & & -0.1445*** & 0.0025 & & -0.0682*** & 0.0267 \\
        & (0.0004) &  &  & (0.0004) &  &  & (0.0004) &  &  & (0.0004) \\
        Team size & -0.1328*** & 0.0157 & & -0.0268*** & 0.0197 & & -0.1424*** & 0.0119 & & -0.0700*** & 0.0005 \\
        & (0.0004) &  &  & (0.0004) &  &  & (0.0004) &  &  & (0.0004) \\
        Team distance & -0.1295*** & 0.0402 & & -0.0259*** & 0.0532 & & -0.1393*** & 0.0334 & & -0.0693*** & 0.0104 \\
        & (0.0004) &  &  & (0.0004) &  &  & (0.0004) &  &  & (0.0004) \\
        Team freshness & -0.1298*** & 0.0380 & & -0.0255*** & 0.0661 & & -0.1399*** & 0.0295 & & -0.0686*** & 0.0199 \\
        & (0.0004) &  &  & (0.0004) &  &  & (0.0004) &  &  & (0.0004) \\
        
        \bottomrule % bottom line
    \end{tabular}
    \begin{tablenotes}
        \small
        \item[$\dagger$] We perform normalized confounding analysis to assess the effect of paper impact on different disruption metrics (Disruption section in Methods). The candidate confounding variables include: $\KI$, novelty, interdisciplinarity, reference count, reference age, team size, team distance, and team freshness, with discipline and publication year controlled. $\alpha_1$ refers to the Unadjusted effect of paper impact and $\delta_1$ refers to the Adjusted effect isolated from the confounding variable. Given that the Unadjusted effect ($\alpha_1$) is identical to all candidate confounders in each model, we opt to not report this value but the relative change in estimate (RCE) to highlight the relative confounding effect. Each regression coefficient is tested against the null hypothesis (coefficient equals 0) using a two-sided \textit{t}-test. Standard errors are provided in parentheses for each coefficient. Note that we do not apply adjustments for multiple hypothesis testing in this analysis. (*$p < 0.05$, **$p < 0.01$, ***$p < 0.001$).
    \end{tablenotes}
\end{threeparttable}
\end{table}

\color{black}
%%%%%%%%%%%%%%%% SUPPLEMENTARY FIGURES %%%%%%%%%%%%%%%
\clearpage
\subsection{Reproduced main results based on SciSciNet publication dataset}\label{SI_SciSciNet_results}

\begin{figure}[h] % Do not use \begin{figure*}
	\centering
	\includegraphics[width=0.87\textwidth]{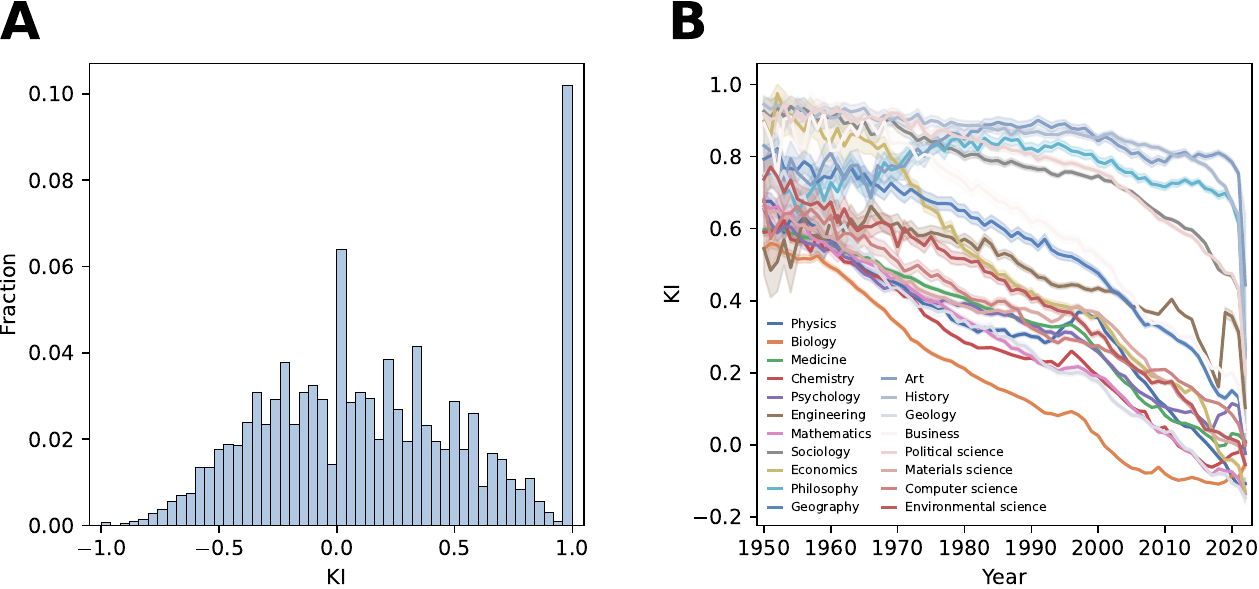}
    \caption{\textbf{$|$ The distribution and evolution pattern of knowledge independence in SciSciNet dataset.}
        (\textbf{A}) The $\KI$ distribution of 49,640,852 papers with at least two references published between 1950 and 2022 in the SciSciNet.
        (\textbf{B}) The pervasive downtrend of $\KI$ over time across disciplines. Bootstrapped 95\% confidence intervals are shown as shaded bands.
        \label{figSSN_1}}
\end{figure}

\clearpage
\begin{figure}[htbp] % Do not use \begin{figure*}
	\centering
	\includegraphics[width=1\textwidth]{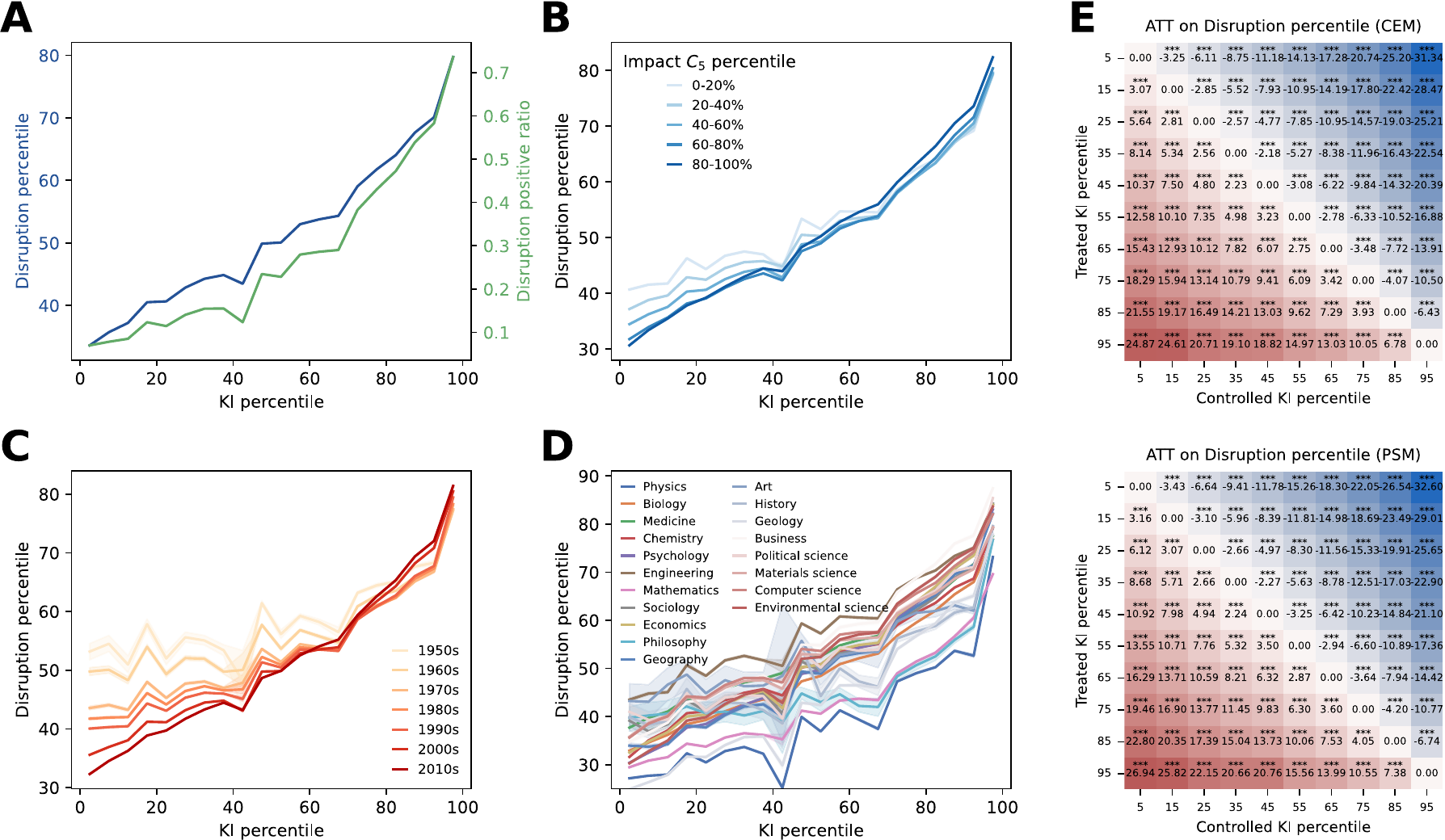}
    \caption{\textbf{$|$ Knowledge independence is associated with scientific disruption in SciSciNet dataset.}
        (\textbf{A}) For the 41,359,799 papers published before 2021 in the SciSciNet that are cited at least once, the average disruption percentile (blue curve, left y-axis) and the disruption positive ratio (green curve, right y-axis) increase with the $\KI$ percentile.
        (\textbf{B}) The association between $\KI$ and disruption persists, regardless of the impact $C_5$ of the focal paper.
        (\textbf{C}) The association between $\KI$ and disruption persists across decades.
        (\textbf{D}) The association between $\KI$ and disruption persists, regardless of discipline.
        Bootstrapped 95\% confidence intervals are shown as shaded bands in panels~\textbf{A-D}.
        (\textbf{E}) The ATT matrixs of $\KI$ on disruption via CEM (upper panel) and PSM (lower panel).
        Each controlled group is set as a baseline, and ATTs are calculated for comparisons between the baseline and each of the treated groups. Blue cells represent negative ATTs, while red ones represent positive ATTs, with color intensity proportional to the absolute value.
        Each ATT is tested against the null hypothesis (ATT equals 0) using a two-sided \textit{t}-test. (*$p < 0.05$, **$p < 0.01$, ***$p < 0.001$).
        \label{figSSN_2}}
\end{figure}

\clearpage
\begin{figure}[htbp] % Do not use \begin{figure*}
	\centering
	\includegraphics[width=1.0\textwidth]{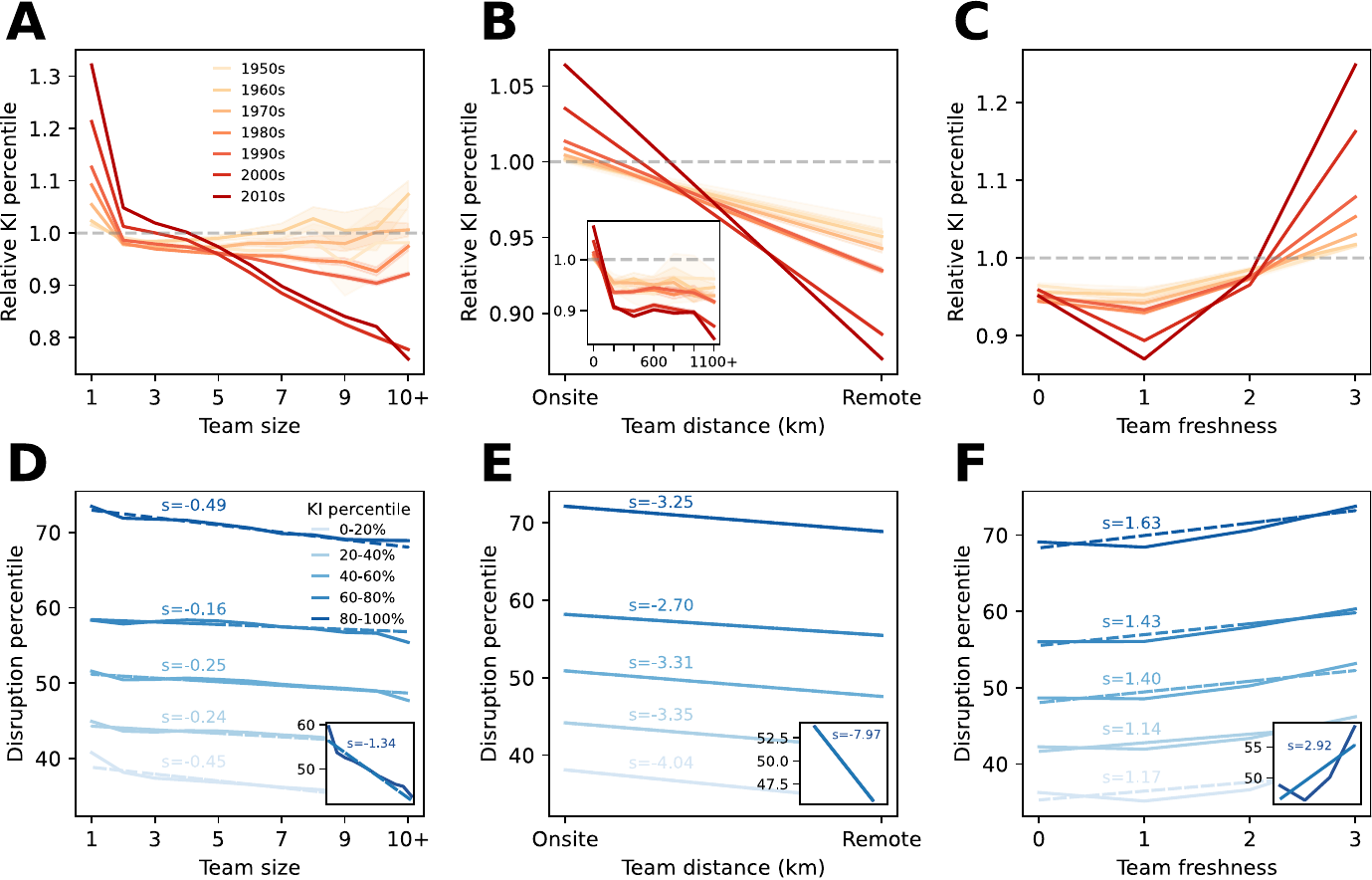}
    \caption{\textbf{$|$ Team's preference for knowledge independence in SciSciNet dataset.}
        (\textbf{A-C}) The $\KI$ values in each bin are rescaled by the average value of the respective time period to highlight the trends. Over time, the relative $\KI$ percentile decreases with team size (panel~\textbf{A}) and team distance (panel~\textbf{B}), and increases with team freshness (panel~\textbf{C}).
        Here we calculate the geographic team distance by the coordinate information of affiliations, and categorize teams into two types---onsite ($\leq100$ km) and remote ($>100$ km) teams (panel~\textbf{B}). The relationship between $\KI$ and average collaboration distance is displayed in inset of panel~\textbf{B}.
        (\textbf{D-F}) Solid lines depict the relationships between disruption and team properties with fixed $\KI$ values, alongside the uncontrolled cases (insets). Dashed lines represent linear fitted curves. Teams with higher $\KI$ are consistently more disruptive. Moreover, the slopes of the fitted curves with fixed $\KI$ are flatter compared to the uncontrolled curves (insets).
        Bootstrapped 95\% confidence intervals are shown as shaded bands.
        \label{figSSN_4}}
\end{figure}

\clearpage
\begin{figure}[htbp] % Do not use \begin{figure*}
	\centering
	\includegraphics[width=1.0\textwidth]{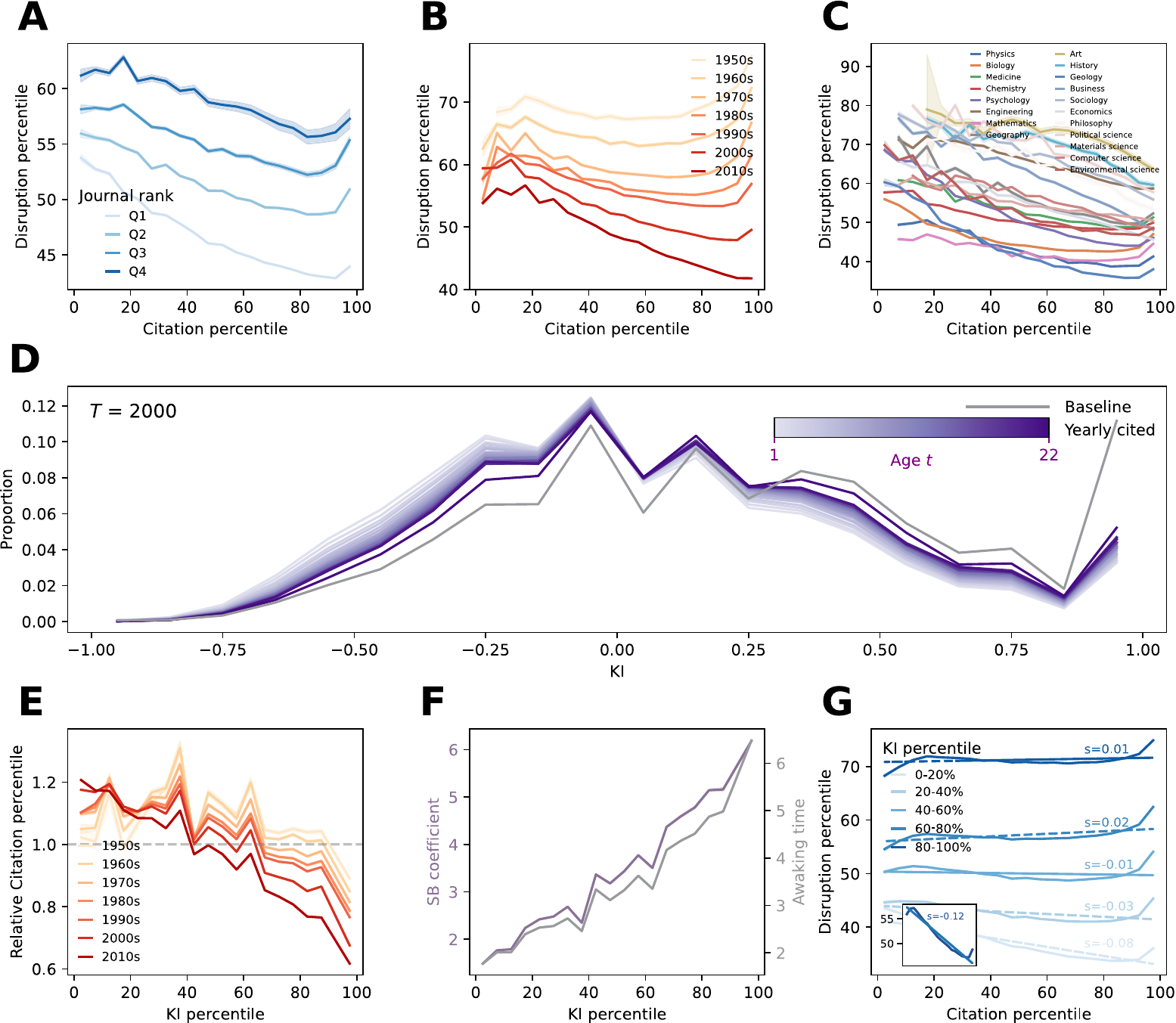}
    \caption{\textbf{$|$ Knowledge independence explains the negative relationship between disruption and impact in SciSciNet dataset.}
            We observe a universal negative relationship between papers' disruption and impact, regardless of the journal rank (panel~\textbf{A}), time period (panel~\textbf{B}), and discipline (panel~\textbf{C}). Here, a paper's impact is measured by the percentile of its total citations within the same publication year and discipline, and the journal rank is based on Scimago Journal Rank.
            (\textbf{D}) For papers published in the year $T=2000$, we measure the $\KI$ distribution among these papers as the baseline distribution (gray curve). Then, for all papers published in a subsequent year, $T+t$ ($t\geq 1$), we examine their bibliographies, extract the references that were published in $T$, and generate the $\KI$ distribution of those references (purple curve). Compared to the baseline, papers with negative $\KI$ are overrepresented while those with positive $\KI$ are underrepresented. Moreover, as the cited age $t$ increases, the corresponding distribution converges towards the baseline (see figure~\ref{figSSN_PS} for different values of $T$).
            (\textbf{E}) The citation percentiles in each bin are rescaled by the average value of the respective time period to highlight the trends. Over time, the relative citation percentile decreases with $\KI$ percentile.
            (\textbf{F}) A paper with a higher $\KI$ is more likely to become a Sleeping Beauty (SB coefficient, blue curve, left y-axis) and takes a longer silence to usher in the citation burst (Awaking time, green curve, right y-axis).
            (\textbf{G}) Solid lines depict the relationships between disruption and impact with fixed $\KI$ values, alongside the uncontrolled case (inset). Dashed lines represent linear fitted curves. For papers of comparable impact, those with higher $\KI$ are consistently more disruptive. Moreover, the slopes of the fitted curves with fixed $\KI$ are significantly flatter compared to the uncontrolled curve (inset).
            Bootstrapped 95\% confidence intervals are shown as shaded bands.
        \label{figSSN_3}}
\end{figure}

% \clearpage
% \begin{figure}[htbp] % Do not use \begin{figure*}
% 	\centering
% 	\includegraphics[width=0.6\textwidth]{figures_si/FigSSN_Reference_count_distribution.pdf}
%     \caption{\textbf{$|$ Distribution of reference count in SciSciNet dataset.}
%         The distribution of reference count for articles recorded in SciSciNet follows a stretched exponential pattern, with a large number of papers containing relatively short reference lists.
%         \label{figSSN_RC_Dis}}
% \end{figure}

\clearpage
\begin{figure}[htbp] % Do not use \begin{figure*}
	\centering
	\includegraphics[width=1.0\textwidth]{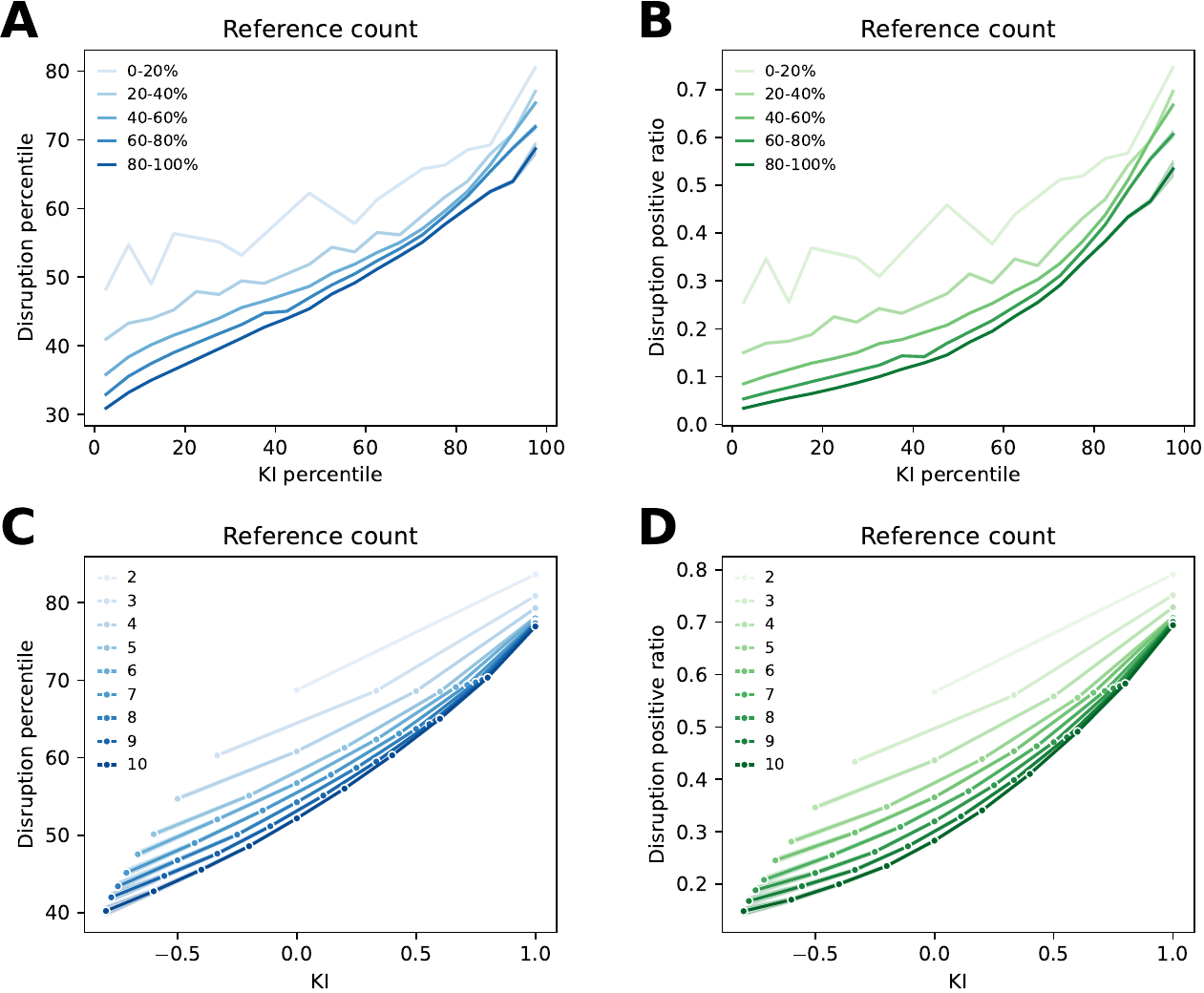}
    \caption{\textbf{$|$ The association between disruption and $\KI$ is robust across different reference counts in SciSciNet dataset.}
        (\textbf{A-B}) When controlling for the percentile of reference count, both the percentile and positive ratio of disruption continue to increase with $\KI$ across all levels of reference count.
        (\textbf{C-D}) When controlling for the reference count within the bottom group ($<=10$), where each reference count induces a limited range of $\KI$ values, both the percentile and positive ratio of disruption continue to increase with $\KI$ across all groups of reference count.
        \label{figSSN_RC}}
\end{figure}

\clearpage
\begin{figure}[htbp] % Do not use \begin{figure*}
	\centering
	\includegraphics[width=1.0\textwidth]{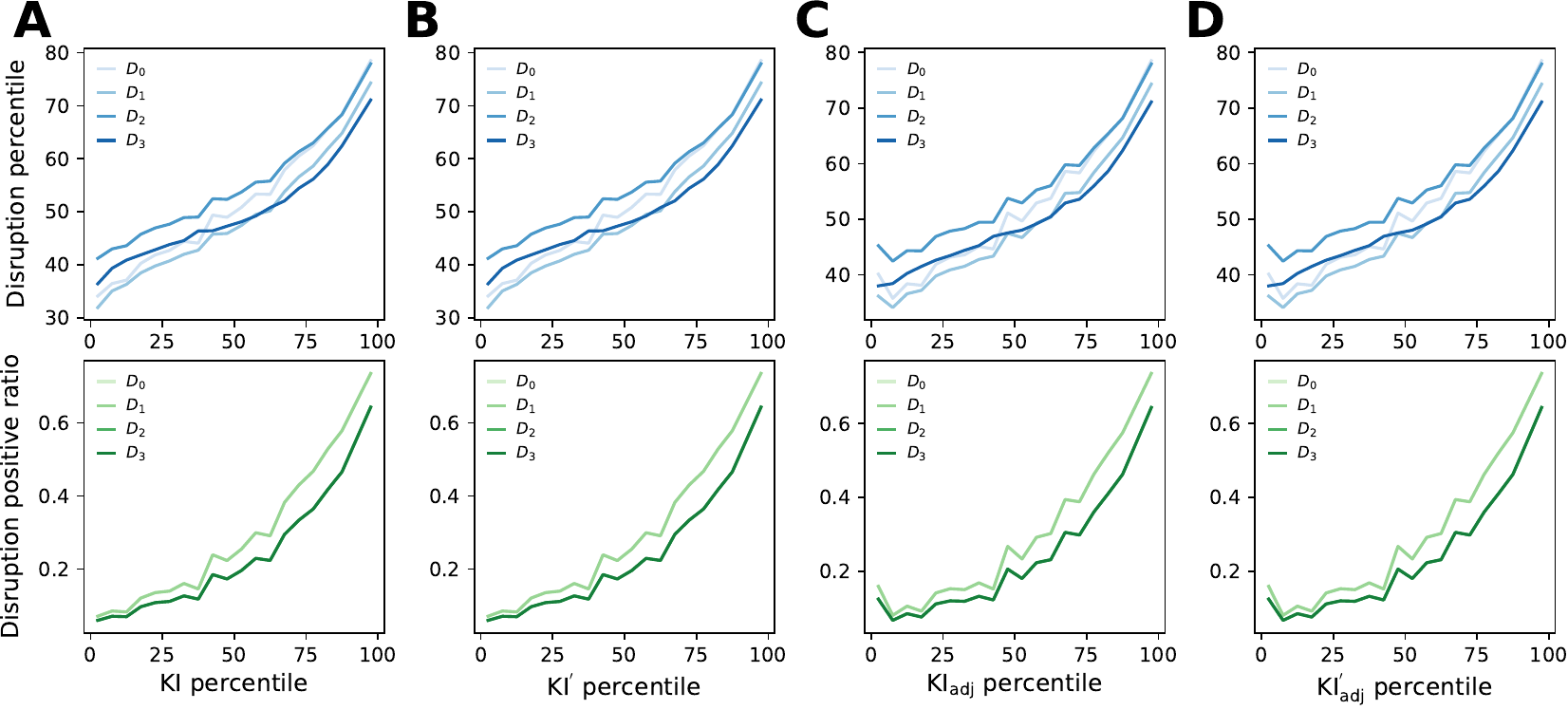}
    \caption{\textbf{$|$ The association between disruption and $\KI$ is robust across alternative measures in SciSciNet dataset.}
        (\textbf{A-D})
        \textbf{Upper panel}: from the perspective of percentile, all disruption measures show a consistent increasing trend with all $\KI$ measures.
        \textbf{Lower panel}: from the perspective of positive ratio, all disruption measures similarly increase with all $\KI$ measures. The positive ratio represents the likelihood of having a higher proportion of $ind$-type references compared to $dep$-type references. Consequently, the variation curves for $D_1$ and $D_3$ align with those for $D_0$ and $D_2$, respectively.
        \label{figSSN_AI}}
\end{figure}

\clearpage
\begin{figure}[htbp] % Do not use \begin{figure*}
	\centering
	\includegraphics[width=1.0\textwidth]{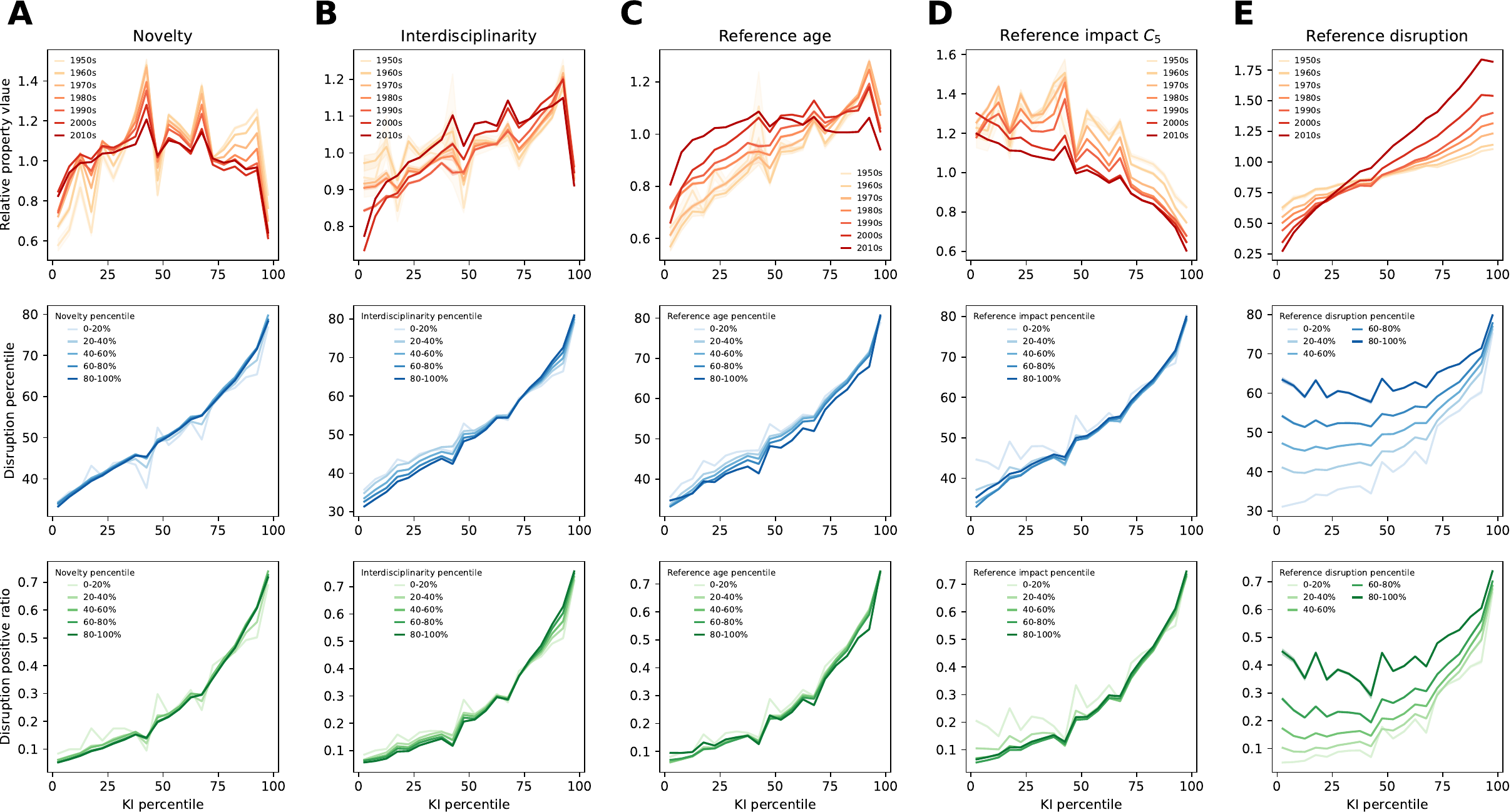}
    \caption{\textbf{$|$ The association between disruption and $\KI$ is robust when controlling for focal paper's properties in SciSciNet dataset.}
        (\textbf{A-E}) When controlling for focal paper's novelty (panel~\textbf{A}), interdisciplinarity (panel~\textbf{B}), reference age (panel~\textbf{C}), reference impact $C_5$ (panel~\textbf{D}), and reference disruption (panel~\textbf{E}), respectively, both the percentile (blue curves) and positive ratio (green curves) of disruption continue to increase with $\KI$ across all levels of above covariates.
        \label{figSSN_RP}}
\end{figure}

\clearpage
\begin{figure}[htbp] % Do not use \begin{figure*}
	\centering
	\includegraphics[width=1.0\textwidth]{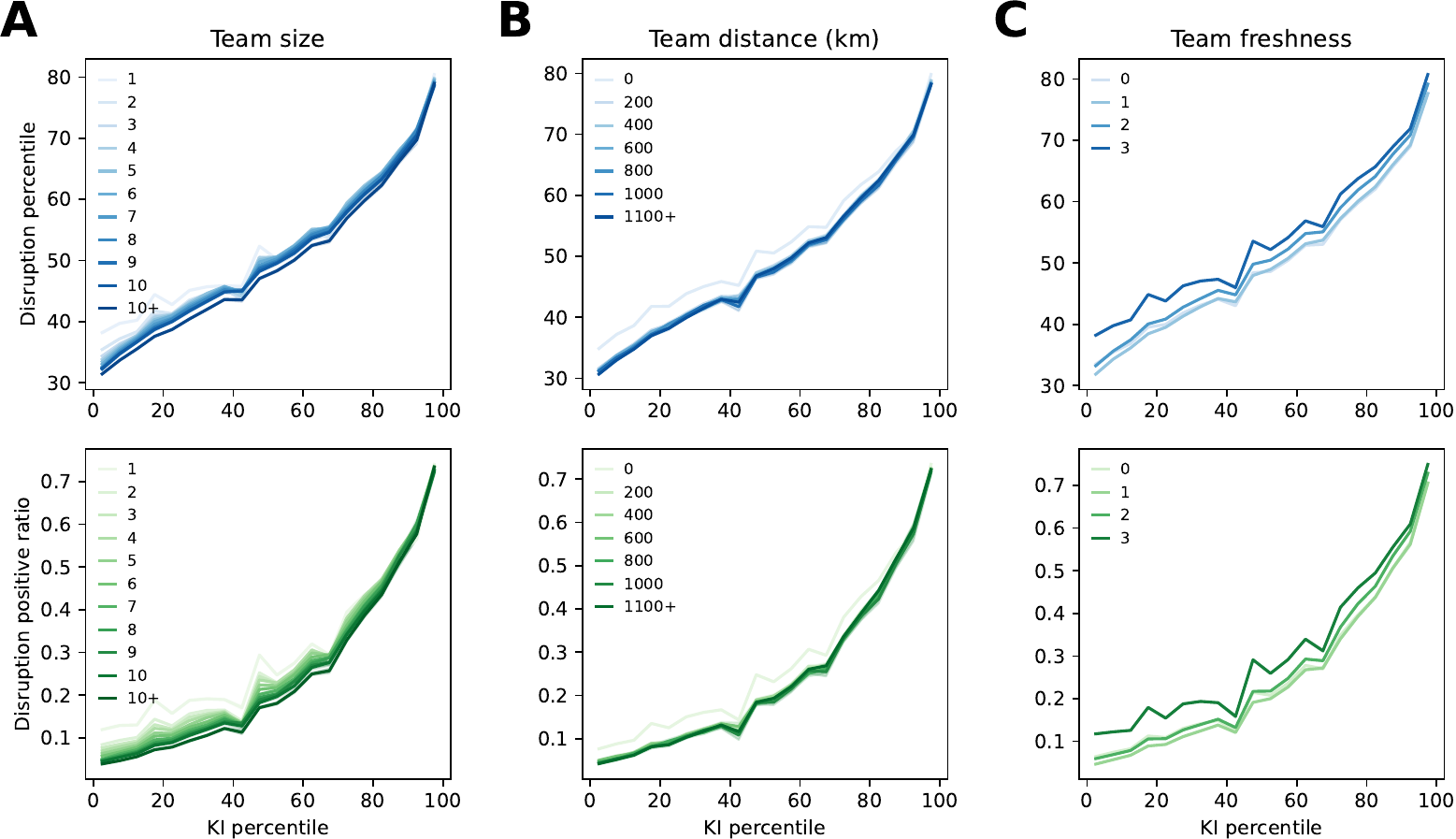}
    \caption{\textbf{$|$ The association between disruption and $\KI$ is robust when controlling for team compositions in SciSciNet dataset.}
        (\textbf{A-C}) When controlling for team size (panel~\textbf{A}), geographic distance (panel~\textbf{B}), and collaboration freshness (panel~\textbf{C}) separately, both the percentile (blue curves) and positive ratio (green curves) of disruption continue to increase with $\KI$ across all levels of above team compositions. Additionally, papers produced by small, onsite, and fresh teams tend to exhibit slightly higher disruption.
        \label{figSSN_TC}}
\end{figure}

\clearpage
\begin{figure}[htbp] % Do not use \begin{figure*}
	\centering
	\includegraphics[width=1.0\textwidth]{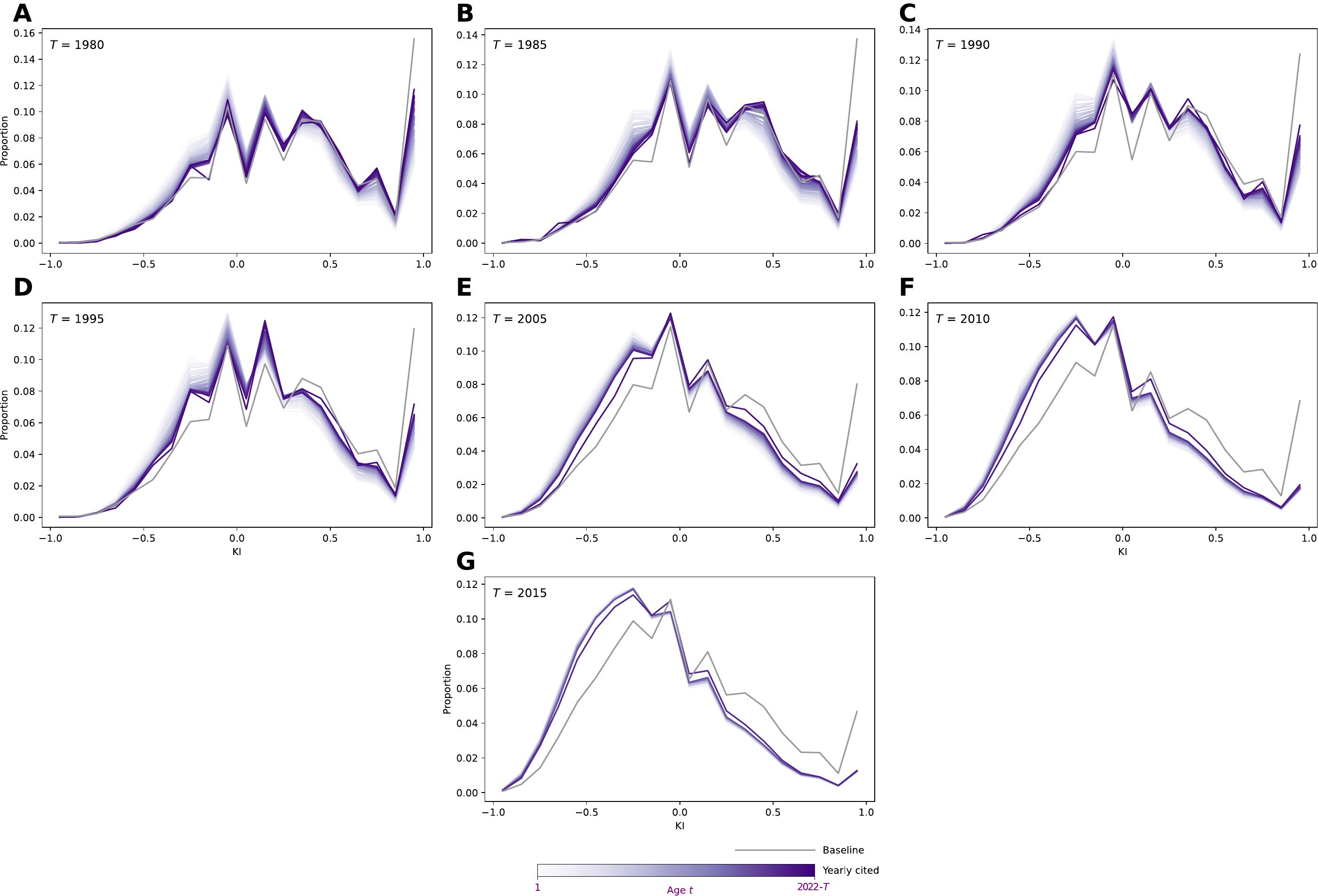}
    \caption{\textbf{$|$ Papers with higher $\KI$ are underrepresented among cited references and receiving delayed citations in SciSciNet dataset.}
        (\textbf{A-G})
        For papers published in the year $T$, we measure the $\KI$ distribution among these papers as the baseline distribution (gray curve). Then, for all papers published in a subsequent year, $T+t$ ($t\geq 1$), we examine their bibliographies, extract the references that were published in $T$, and generate the $\KI$ distribution of those references (purple curve). Compared to the baseline, papers with negative $\KI$ are overrepresented while those with positive $\KI$ are underrepresented. Moreover, as the cited age $t$ increases, the corresponding distribution converges towards the baseline.
        \label{figSSN_PS}}
\end{figure}
\newpage

%%%%%%%%%%%%%%%% SUPPLEMENTARY TABLES %%%%%%%%%%%%%%%
\begin{table} % Do not use \begin{table*}
    \centering
    % Captions go above tables
    \caption{\textbf{$|$ Normalized regression models of disruption and the influencing factors in SciSciNet dataset.}}
    \label{tableSSN_OLS} % give each table a logical label name
    \renewcommand\tabcolsep{0.16cm} % column spacing
    \renewcommand{\arraystretch}{0.8} % line spacing
    \begin{threeparttable}
    \fontsize{7.5}{14}\selectfont % 设置表格字体大小为 8.75pt，行距为 12pt
    \begin{tabular}{lcccccccccc}
        \hline  % top line
        & Model 1 & Model 2 & Model 3 & Model 4 & Model 5 & Model 6 & Model 7 & Model 8 & Model 9 & $R^2$\% \\ 
        \hline  % middle line 1
        $\KI$ & \textbf{0.3964}*** & \textbf{0.3954}*** & \textbf{0.4001}*** & \textbf{0.3770}*** & \textbf{0.3986}*** & \textbf{0.3962}*** & \textbf{0.3952}*** & \textbf{0.3926}*** & \textbf{0.3773}*** & \textbf{67.32} \\
         & (0.000) & (0.000) & (0.000) & (0.000) & (0.000) & (0.000) & (0.000) & (0.000) & (0.000) \\
        % Log $C_5$ &  & -0.0416*** &  &  &  &  &  &  &  &  &  & -0.0322*** & 3.91 \\
         % &  & (0.000) &  &  &  &  &  &  &  &  &  & (0.000) \\
        Novelty &  & -0.0248*** &  &  &  &  &  &  & -0.0183*** & 0.43 \\
         &  & (0.000) &  &  &  &  &  &  & (0.000) \\
        Interdisciplinarity &  &  & -0.0269*** &  &  &  &  &  & -0.0189*** & 0.19 \\
         &  &  & (0.000) &  &  &  &  &  & (0.000) \\
        Reference count &  &  &  & -0.0619*** &  &  &  &  & -0.0566*** & 10.68 \\
         &  &  &  & (0.000) &  &  &  &  & (0.000) \\
        Reference age &  &  &  &  & -0.0147*** &  &  &  & -0.0136*** & 0.19 \\
         &  &  &  &  & (0.000) &  &  &  & (0.000) \\
        % Reference $C_5$ &  &  &  &  &  &  & -0.0197*** &  &  &  &  & -0.0168*** & 0.39 \\
        %  &  &  &  &  &  &  & (0.000) &  &  &  &  & (0.000) \\
        % Reference disruption &  &  &  &  &  &  &  & 0.2293*** &  &  &  & 0.2231*** & 37.17 \\
        %  &  &  &  &  &  &  &  & (0.000) &  &  &  & (0.000) \\
        Team size &  &  &  &  &  & -0.0093*** &  &  & -0.0108*** & 0.37 \\
         &  &  &  &  &  & (0.000) &  &  & (0.000) \\
        Team distance &  &  &  &  &  &  & -0.0233*** &  & -0.0202*** & 1.05 \\
         &  &  &  &  &  &  & (0.000) &  & (0.000) \\
        Team freshness &  &  &  &  &  &  &  & 0.0386*** & 0.0384*** & 2.13 \\
         &  &  &  &  &  &  &  & (0.000) & (0.000) \\
        \hline % middle line 2
        Discipline fixed effects  & Yes & Yes & Yes & Yes & Yes & Yes & Yes & Yes & Yes & 13.48 \\
        Year fixed effects & Yes & Yes & Yes & Yes & Yes & Yes & Yes & Yes & Yes & 4.16 \\
        \hline % middle line 3
        N & 19934738 & 19934738 & 19934738 & 19934738 & 19934738 & 19934738 & 19934738 & 19934738 & 19934738 \\
        $R^2$ & 0.199 & 0.200 & 0.200 & 0.202 & 0.200 & 0.199 & 0.200 & 0.201  & 0.205 \\
        \hline  % bottom line
    \end{tabular}
    \begin{tablenotes}
        \small
        \item[$\dagger$] We perform normalized ordinary-least-squares (OLS) regression analyses to examine the effect of $\KI$ on disruption in SciSciNet dataset, while controlling for various covariates. Each regression coefficient is tested against the null hypothesis (coefficient equals 0) using a two-sided \textit{t}-test. Standard errors are provided in parentheses for each coefficient. Note that we do not apply adjustments for multiple hypothesis testing in this analysis. (*$p < 0.05$, **$p < 0.01$, ***$p < 0.001$).
    \end{tablenotes}
    \end{threeparttable}
\end{table}

\clearpage
\begin{table} % Do not use \begin{table*}
\centering
% Captions go above tables
\caption{\textbf{$|$ Normalized mediation regression on team size in SciSciNet dataset.}}
\label{tableSSN_mediation_TS} % give each table a logical label name
\renewcommand\tabcolsep{0.2cm} % column spacing
\renewcommand{\arraystretch}{1.5} % line spacing
\begin{threeparttable}
\fontsize{12}{10}\selectfont % 设置表格字体大小为 8.75pt，行距为 12pt
\begin{tabular}{llcccc} % four columns, alignment for each
    \hline
    Independent variable & Mediator & ACME ($\beta_1\theta_2$) & ADE ($\theta_1$) & TE ($\theta_1 + \beta_1\theta_2$) & $|$ACME / TE$|$ \\
    \hline
    Team size & $\KI$ & \textbf{-0.0118}***  & \textbf{-0.0099}*** & -0.0217*** & \textbf{0.5442} \\
     &  & (0.0025) & (0.0021) & (0.0046) \\\\
    % Team size & \sout{Log $C_5$} & -0.0121***  & -0.0082*** & -0.0217*** \\
    %  &  & (0.0026) & (0.0018) & (0.0046) \\\\
    Team size & Novelty & -0.0004***  & -0.0212*** & -0.0217*** & 0.0204 \\
     &  & (0.0001) & (0.0045) & (0.0046) \\\\
    Team size & Interdisciplinarity & -0.0003***  & -0.0214*** & -0.0217*** & 0.0123 \\
     &  & (0.0001) & (0.0046) & (0.0046) \\\\
    Team size & Reference count & -0.0006***  & -0.0210*** & -0.0217*** & 0.0289 \\
     &  & (0.0001) & (0.0045) & (0.0046) \\\\
    Team size & Reference age & -0.0033***  & -0.0184*** & -0.0217*** & 0.1510 \\
     &  & (0.0007) & (0.0040) & (0.0046) \\\\
    % Team size & Reference $C_5$ & -0.0005***  & -0.0210*** & -0.0217*** & 0.0233 \\
    %  &  & (0.0001) & (0.0046) & (0.0046) \\\\
    % Team size & \sout{Reference disruption} & -0.0083***  & -0.0120*** & -0.0217*** \\
    %  &  & (0.0019) & (0.0026) & (0.0046) \\\\
    Team size & Team distance & -0.0059***  & -0.0158*** & -0.0217*** & 0.2733 \\
     &  & (0.0012) & (0.0035) & (0.0046) \\\\
    Team size & Team freshness & 0.0051***  & -0.0268*** & -0.0217*** & 0.2352 \\
     &  & (0.0011) & (0.0057) & (0.0046) \\
    \hline
\end{tabular}
\begin{tablenotes}
    \small
    \item[$\dagger$] We perform normalized mediation analysis to assess the effect of team size on disruption in SciSciNet dataset. The candidate mediating variables include: $\KI$, novelty, interdisciplinarity, reference count, reference age, team distance, and team freshness, with discipline and publication year controlled. ACME refers to average causal mediation effects driven from the mediating variable, ADE stands for average direct effects caused by team size, and TE is defined as the sum of the ACME and ADE. Each regression coefficient is tested against the null hypothesis (coefficient equals 0) using a two-sided \textit{t}-test. Standard errors are provided in parentheses for each coefficient. Note that we do not apply adjustments for multiple hypothesis testing in this analysis. (*$p < 0.05$, **$p < 0.01$, ***$p < 0.001$).
\end{tablenotes}
\end{threeparttable}
\end{table}

\clearpage
\begin{table} % Do not use \begin{table*}
\centering
% Captions go above tables
\caption{\textbf{$|$ Normalized mediation regression on team distance in SciSciNet dataset.}}
\label{tableSSN_mediation_TD} % give each table a logical label name
\renewcommand\tabcolsep{0.2cm} % column spacing
\renewcommand{\arraystretch}{1.5} % line spacing
\begin{threeparttable}
\fontsize{12}{10}\selectfont % 设置表格字体大小为 8.75pt，行距为 12pt
\begin{tabular}{llcccc} % four columns, alignment for each
    \hline
    Independent variable & Mediator & ACME ($\beta_1\theta_2$) & ADE ($\theta_1$) & TE ($\theta_1 + \beta_1\theta_2$) & $|$ACME / TE$|$ \\
    \hline
    Team distance & $\KI$ & \textbf{-0.0203}***  & \textbf{-0.0236}*** & -0.0440*** & \textbf{0.4630} \\
     &  & (0.0002) & (0.0004) & (0.0004) \\\\
    % Team distance & \sout{Log $C_5$} & -0.0106***  & -0.0323*** & -0.0440*** \\
    %  &  & (0.0001) & (0.0004) & (0.0004) \\\\
    Team distance & Novelty & -0.0003***  & -0.0437*** & -0.0440*** & 0.0068 \\
     &  & (0.0000) & (0.0004) & (0.0004) \\\\
    Team distance & Interdisciplinarity & 0.0007***  & -0.0446*** & -0.0440*** & 0.0152 \\
     &  & (0.0000) & (0.0004) & (0.0004) \\\\
    Team distance & Reference count & -0.0111***  & -0.0329*** & -0.0440*** & 0.2524 \\
     &  & (0.0001) & (0.0004) & (0.0004) \\\\
    Team distance & Reference age & -0.0009***  & -0.0431*** & -0.0440*** & 0.0203 \\
     &  & (0.0000) & (0.0004) & (0.0004) \\\\
    % Team distance & Reference $C_5$ & -0.0002***  & -0.0436*** & -0.0440*** & 0.0051 \\
    %  &  & (0.0000) & (0.0004) & (0.0004) \\\\
    % Team distance & \sout{Reference disruption} & -0.0179***  & -0.0249*** & -0.0440*** \\
    %  &  & (0.0004) & (0.0006) & (0.0004) \\\\
    Team distance & Team size & -0.0022***  & -0.0417*** & -0.0440*** & 0.0508 \\
     &  & (0.0005) & (0.0006) & (0.0004) \\\\
    Team distance & Team freshness & 0.0020***  & -0.0459*** & -0.0440*** & 0.0452 \\
     &  & (0.0000) & (0.0004) & (0.0004) \\
    \hline
\end{tabular}
\begin{tablenotes}
    \small
    \item[$\dagger$] We perform normalized mediation analysis to assess the effect of team distance on disruption in SciSciNet dataset. The candidate mediating variables include: $\KI$, novelty, interdisciplinarity, reference count, reference age, team size, and team freshness, with discipline and publication year controlled. ACME refers to average causal mediation effects driven from the mediating variable, ADE stands for average direct effects caused by team distance, and TE is defined as the sum of the ACME and ADE. Each regression coefficient is tested against the null hypothesis (coefficient equals 0) using a two-sided \textit{t}-test. Standard errors are provided in parentheses for each coefficient. Note that we do not apply adjustments for multiple hypothesis testing in this analysis. (*$p < 0.05$, **$p < 0.01$, ***$p < 0.001$).
\end{tablenotes}
\end{threeparttable}
\end{table}

\clearpage
\begin{table} % Do not use \begin{table*}
\centering
% Captions go above tables
\caption{\textbf{$|$ Normalized mediation regression on team freshness in SciSciNet dataset.}}
\label{tableSSN_mediation_TF} % give each table a logical label name
\renewcommand\tabcolsep{0.2cm} % column spacing
\renewcommand{\arraystretch}{1.5} % line spacing
\begin{threeparttable}
\fontsize{12}{10}\selectfont % 设置表格字体大小为 8.75pt，行距为 12pt
\begin{tabular}{llcccc} % four columns, alignment for each
    \hline
    Independent variable & Mediator & ACME ($\beta_1\theta_2$) & ADE ($\theta_1$) & TE ($\theta_1 + \beta_1\theta_2$) & $|$ACME / TE$|$ \\
    \hline
    Team freshness & $\KI$ & \textbf{0.0366}***  & \textbf{0.0393}*** & 0.0758*** & \textbf{0.4821} \\
     &  & (0.0002) & (0.0004) & (0.0004) \\\\
    % Team freshness & \sout{Log $C_5$} & 0.0108***  & 0.0616*** & 0.0758*** & 0. \\
    %  &  & (0.0001) & (0.0004) & (0.0004) \\\\
    Team freshness & Novelty & 0.0002***  & 0.0757*** & 0.0758*** & 0.0020 \\
     &  & (0.0000) & (0.0005) & (0.0004) \\\\
    Team freshness & Interdisciplinarity & 0.0002***  & 0.0757*** & 0.0758*** & 0.0020 \\
     &  & (0.0000) & (0.0004) & (0.0004) \\\\
    Team freshness & Reference count & 0.0111***  & 0.0648*** & 0.0758*** & 0.1461 \\
     &  & (0.0001) & (0.0004) & (0.0004) \\\\
    Team freshness & Reference age & 0.0011***  & 0.0747*** & 0.0758*** & 0.0150 \\
     &  & (0.0000) & (0.0004) & (0.0004) \\\\
    % Team freshness & Reference $C_5$ & 0.0000***  & 0.0752*** & 0.0758*** & 0.0001 \\
    %  &  & (0.0000) & (0.0005) & (0.0004) \\\\
    % Team freshness & \sout{Reference disruption} & 0.0255***  & 0.0470*** & 0.0758*** & 0. \\
    %  &  & (0.0002) & (0.0004) & (0.0004) \\\\
    Team freshness & Team size & -0.0018***  & 0.0776*** & 0.0758*** & 0.0232 \\
     &  & (0.0004) & (0.0006) & (0.0004) \\\\
    Team freshness & Team distance & -0.0012***  & 0.0770*** & 0.0758*** & 0.0156 \\
     &  & (0.0000) & (0.0004) & (0.0004) \\
    \hline
\end{tabular}
\begin{tablenotes}
    \small
    \item[$\dagger$] We perform normalized mediation analysis to assess the effect of team freshness on disruption in SciSciNet dataset. The candidate mediating variables include: $\KI$, novelty, interdisciplinarity, reference count, reference age, team size, and team distance, with discipline and publication year controlled. ACME refers to average causal mediation effects driven from the mediating variable, ADE stands for average direct effects caused by team freshness, and TE is defined as the sum of the ACME and ADE. Each regression coefficient is tested against the null hypothesis (coefficient equals 0) using a two-sided \textit{t}-test. Standard errors are provided in parentheses for each coefficient. Note that we do not apply adjustments for multiple hypothesis testing in this analysis. (*$p < 0.05$, **$p < 0.01$, ***$p < 0.001$).
\end{tablenotes}
\end{threeparttable}
\end{table}

\clearpage
\begin{table} % Do not use \begin{table*}
\centering
% Captions go above tables
\caption{\textbf{$|$ Normalized confounding regression on paper impact in SciSciNet dataset.}}
\label{tableSSN_mediation_impact} % give each table a logical label name
\renewcommand\tabcolsep{0.2cm} % column spacing
\renewcommand{\arraystretch}{1.5} % line spacing
\begin{threeparttable}
\fontsize{12}{10}\selectfont % 设置表格字体大小为 8.75pt，行距为 12pt
\begin{tabular}{llccc} % four columns, alignment for each
    \hline
    Independent variable & Confounder & Unadjusted ($\alpha_1$) & Adjusted ($\delta_1$) & RCE ($|\alpha_1 - \delta_1|/|\alpha_1|$) \\
    \hline
    Citation percentile & $\KI$ & -0.1094***  & \textbf{-0.0184}***  & \textbf{0.8316} \\
     &  & (0.0005) & (0.0005) \\\\
    Citation percentile & Novelty & -0.1094***  & -0.1074*** & 0.0181 \\
     &  & (0.0005) & (0.0005) \\\\
    Citation percentile & Interdisciplinarity & -0.1094***  & -0.1103*** & 0.0085 \\
     &  & (0.0005) & (0.0005) \\\\
    Citation percentile & Reference count & -0.1094***  & -0.0623*** & 0.4309 \\
     &  & (0.0005) & (0.0005) \\\\
    Citation percentile & Reference age & -0.1094***  & -0.1059*** & 0.0321 \\
     &  & (0.0005) & (0.0005) \\\\
    % Citation percentile & Reference $C_5$ & -0.1094***  & -0.1087*** & 0.0031 \\
    %  &  & (0.0005) & (0.0005) \\\\
    % Citation percentile & \sout{Reference disruption} & -0.1094***  & -0.0534*** \\
    %  &  & (0.0005) & (0.0006) \\\\
    Citation percentile & Team size & -0.1094***  & -0.1083*** & 0.0097 \\
     &  & (0.0005) & (0.0005) \\\\
    Citation percentile & Team distance & -0.1094***  & -0.1064*** & 0.0270 \\
     &  & (0.0005) & (0.0005) \\\\
    Citation percentile & Team freshness & -0.1094***  & -0.1053*** & 0.0379 \\
     &  & (0.0005) & (0.0005) \\
    \hline
\end{tabular}
\begin{tablenotes}
    \small
    \item[$\dagger$] We perform normalized confounding analysis to assess the effect of paper impact on disruption in SciSciNet dataset. The candidate confounding variables include: $\KI$, novelty, interdisciplinarity, reference count, reference age, team size, team distance, and team freshness, with discipline and publication year controlled. The relative change in estimate (RCE) refers to the relative change of effect after controlling for the confounder. Each regression coefficient is tested against the null hypothesis (coefficient equals 0) using a two-sided \textit{t}-test. Standard errors are provided in parentheses for each coefficient. Note that we do not apply adjustments for multiple hypothesis testing in this analysis. (*$p < 0.05$, **$p < 0.01$, ***$p < 0.001$).
\end{tablenotes}
\end{threeparttable}
\end{table}

\clearpage
\subsection{Reproduced main results based on Web of Science publication dataset}\label{SI_WebofScience_results}
\begin{figure}[h] % Do not use \begin{figure*}
	\centering
	\includegraphics[width=0.87\textwidth]{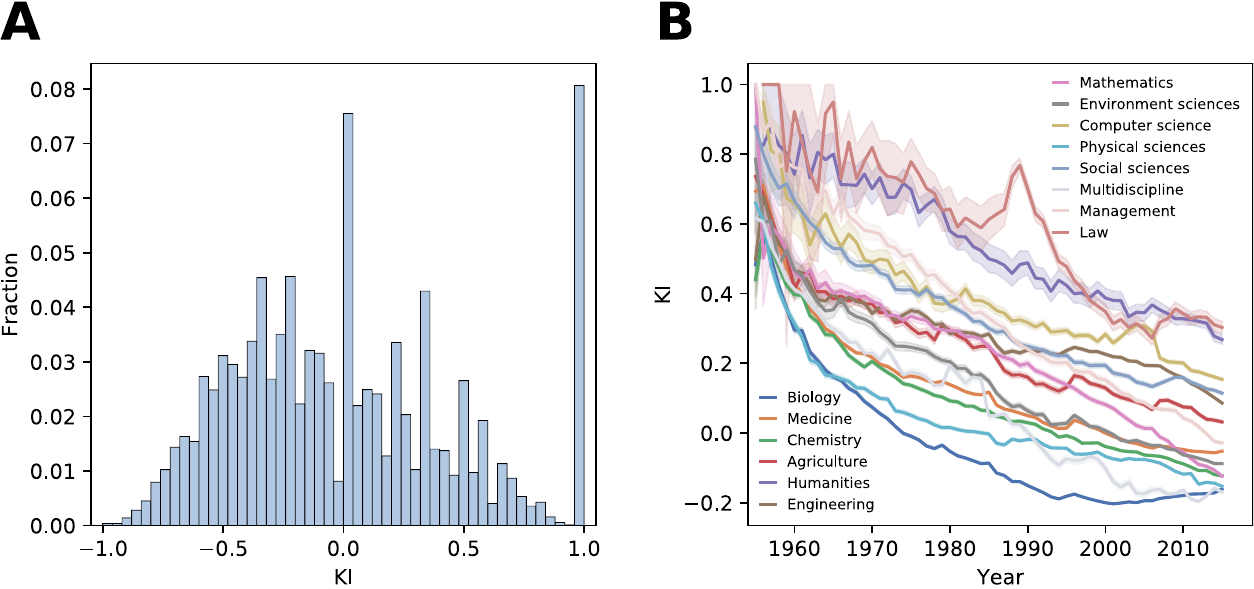}
    \caption{\textbf{$|$ The distribution and evolution pattern of knowledge independence in Web of Science dataset.}
        (\textbf{B}) The $\KI$ distribution of 30,560,343 papers with at least two references published between 1955 and 2017 in the Web of Science.
        (\textbf{C}) The evolving downtrend of $\KI$ over time across disciplines. Bootstrapped 95\% confidence intervals are shown as shaded bands.
        \label{FigWOS_1_KI_Stats}}
\end{figure}

\clearpage
\begin{figure}[htbp] % Do not use \begin{figure*}
	\centering
	\includegraphics[width=1\textwidth]{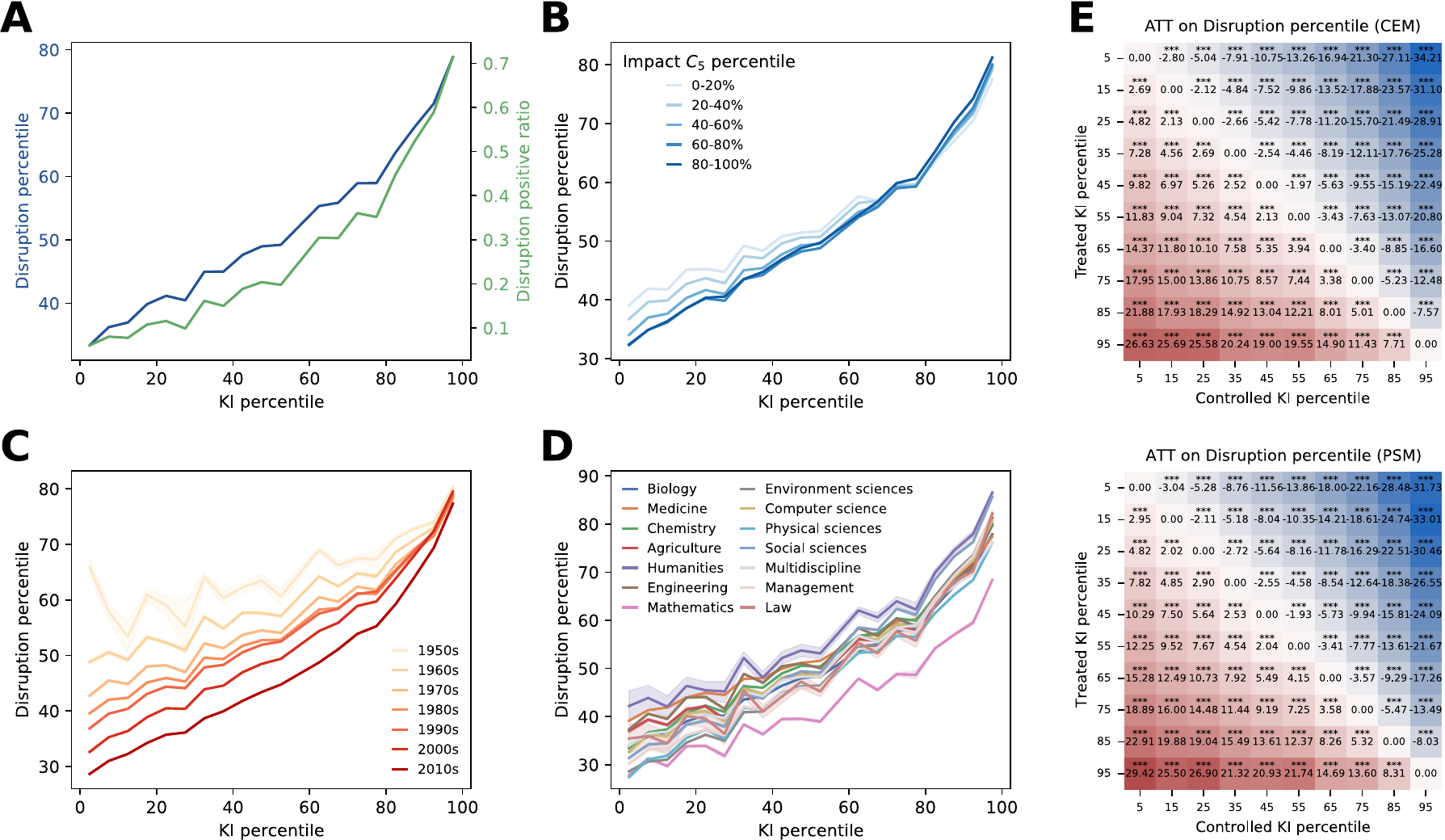}
    \caption{\textbf{$|$ Knowledge independence is associated with scientific disruption in Web of Science dataset.}
        (\textbf{A}) For the 27,769,300 papers published before 2016 in the Web of Science that are cited at least once, the average disruption percentile (blue curve, left y-axis) and the disruption positive ratio (green curve, right y-axis) increase with the $\KI$ percentile.
        (\textbf{B}) The association between $\KI$ and disruption persists, regardless of the impact of the focal paper. Here, impact is measured as the number of citations received within the first five years after publication, denoted by $C_5$.
        (\textbf{C}) The association between $\KI$ and disruption persists across decades.
        (\textbf{D}) The association between $\KI$ and disruption persists, regardless of discipline.
        Bootstrapped 95\% confidence intervals are shown as shaded bands in panels~\textbf{A-D}.
        (\textbf{E}) The ATT matrixs of $\KI$ on disruption via CEM (upper panel) and PSM (lower panel).
        Each controlled group is set as a baseline, and ATTs are calculated for comparisons between the baseline and each of the treated groups. Blue cells represent negative ATTs, while red ones represent positive ATTs, with color intensity proportional to the absolute value.
        Each ATT is tested against the null hypothesis (ATT equals 0) using a two-sided \textit{t}-test. (*$p < 0.05$, **$p < 0.01$, ***$p < 0.001$).
        \label{FigWOS_2_KI_D}}
\end{figure}

\clearpage
\begin{figure}[htbp] % Do not use \begin{figure*}
	\centering
	\includegraphics[width=1.0\textwidth]{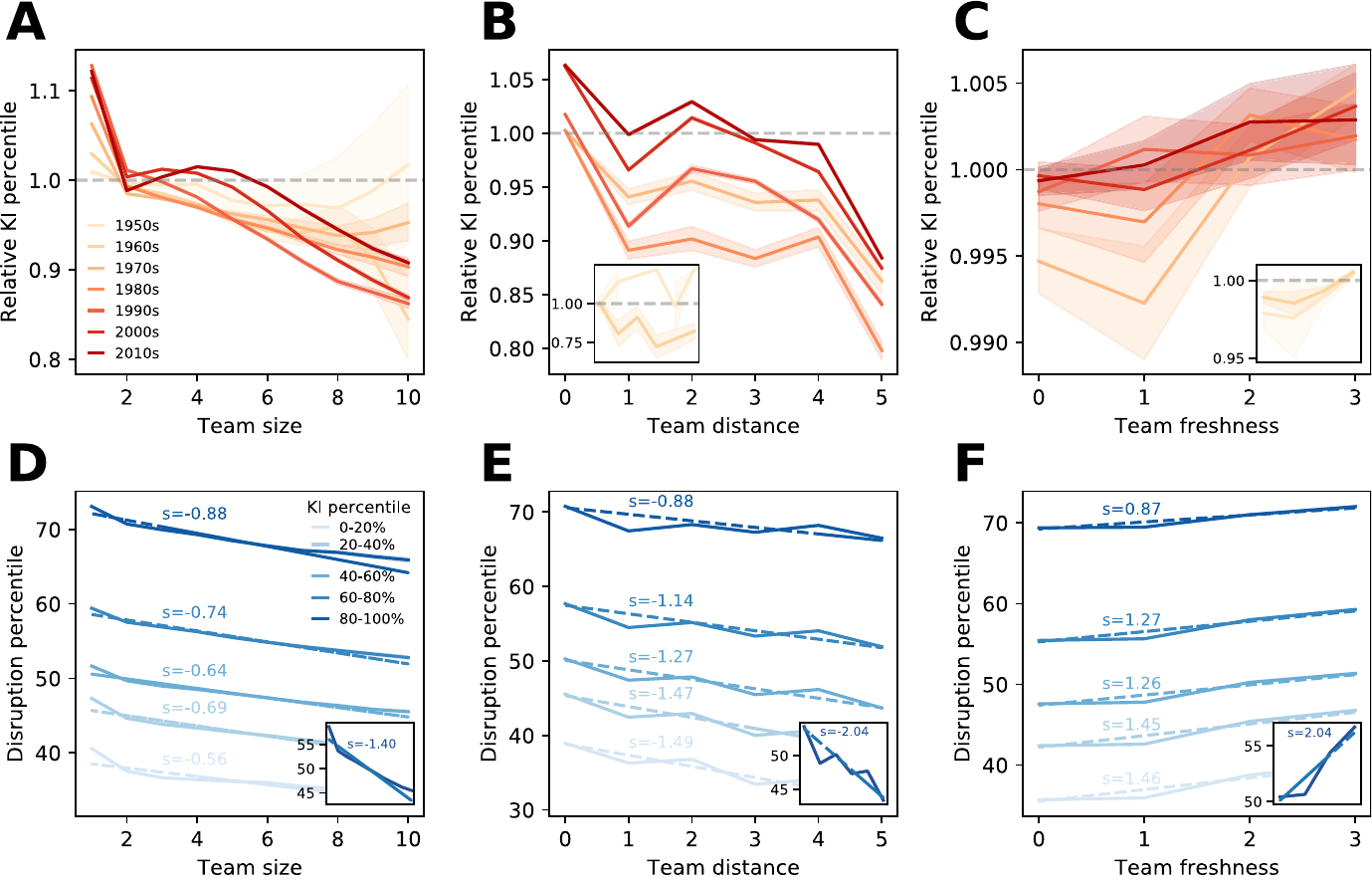}
    \caption{\textbf{$|$ Team's preference for knowledge independence in Web of Science dataset.}
        (\textbf{A-C}) The $\KI$ percentiles in each bin are rescaled by the average value of the respective time period to highlight the trends. Curves with larger bounds are displayed in the inset to improve visualization. Over time, the relative $\KI$ percentile decreases with team size (panel~\textbf{A}) and team distance (panel~\textbf{B}), and increases with team freshness (panel~\textbf{C}); the only exception is observed for the 1950s in the team distance analysis (inset of panel~\textbf{B}).
        (\textbf{D-F}) Solid lines depict the relationships between disruption and team properties with fixed $\KI$ values, alongside the uncontrolled cases (insets). Dashed lines represent linear fitted curves. Teams with higher $\KI$ are consistently more disruptive. Moreover, the slopes of the fitted curves with fixed $\KI$ are flatter compared to the uncontrolled curves (insets).
        Bootstrapped 95\% confidence intervals are shown as shaded bands. 
        Considering the lack of coordinate information in Web of Science dataset, we categorize the team distance of a given focal paper by hierarchical categorization of the affiliations: $0$ (Sub-organization / Department level): $1$: Organization / University level; $2$: City level; $3$: State / Province level; $4$: National level; $5$: International level. This methodology is also applied in Fig.~\ref{FigWOS_TC}.
        \label{FigWOS_4_team_preference}}
\end{figure}

\clearpage
\begin{figure}[htbp] % Do not use \begin{figure*}
	\centering
	\includegraphics[width=1.0\textwidth]{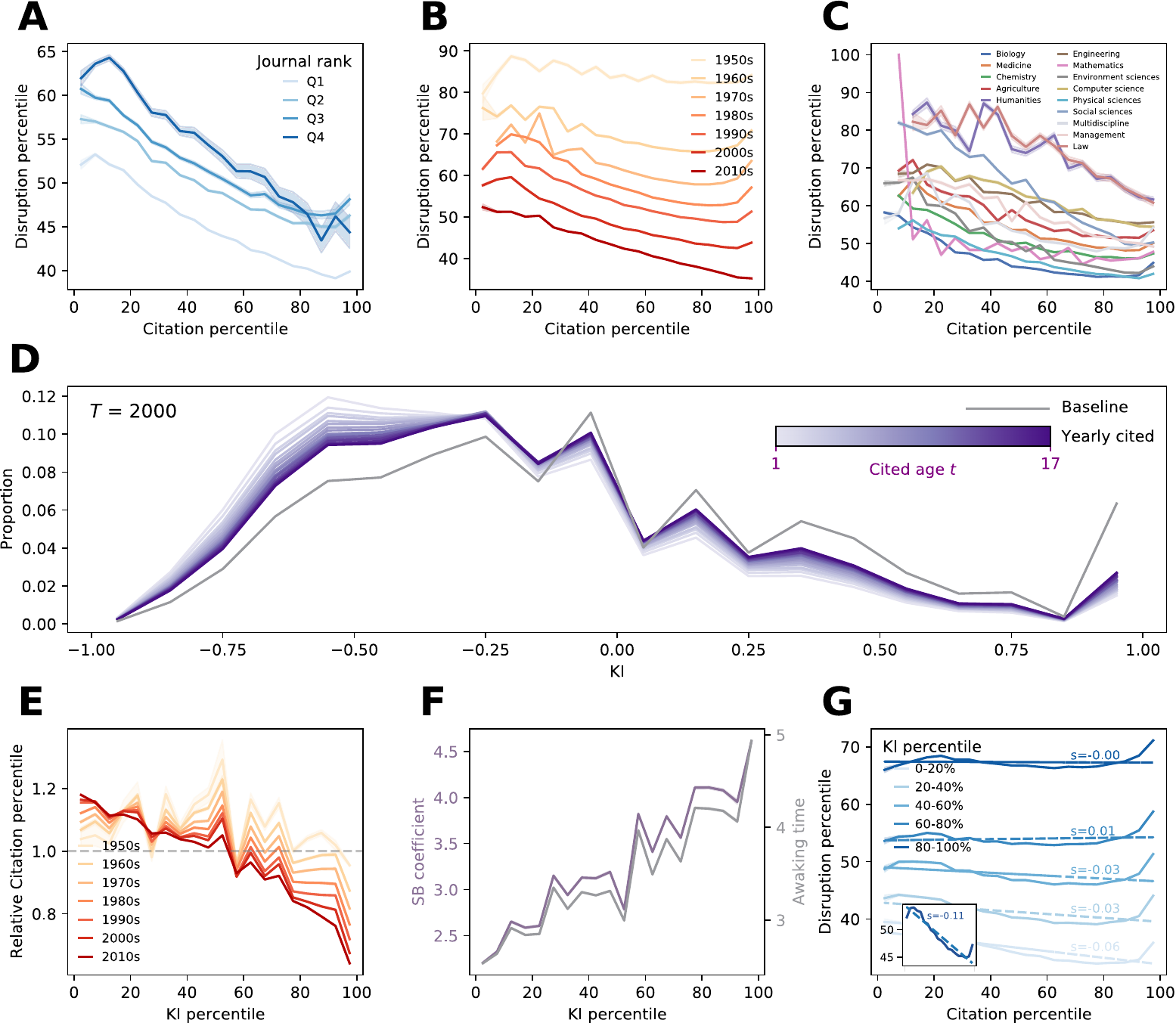}
    \caption{\textbf{$|$ Knowledge independence explains the negative relationship between disruption and impact in Web of Science dataset.}
        We observe a universal negative relationship between papers' disruption and impact, regardless of the journal rank (panel~\textbf{A}), time period (panel~\textbf{B}), and discipline (panel~\textbf{C}). Here, a paper's impact is measured by the percentile of its total citations within the same publication year and discipline, and the journal rank is based on Scimago Journal Rank.
        (\textbf{D}) For papers published in the year $T=2000$, we measure the $\KI$ distribution among these papers as the baseline distribution (gray curve). Then, for all papers published in a subsequent year, $T+t$ ($t\geq 1$), we examine their bibliographies, extract the references that were published in $T$, and generate the $\KI$ distribution of those references (purple curve). Compared to the baseline, papers with negative $\KI$ are overrepresented while those with positive $\KI$ are underrepresented. Moreover, as the cited age $t$ increases, the corresponding distribution converges towards the baseline (see figure~\ref{FigWOS_PS} for different values of $T$).
        (\textbf{E}) The citation percentiles in each bin are rescaled by the average value of the respective time period to highlight the trends. Over time, the relative citation percentile decreases with $\KI$ percentile.
        (\textbf{F}) A paper with a higher $\KI$ is more likely to become a Sleeping Beauty (SB coefficient, blue curve, left y-axis) and takes a longer silence to usher in the citation burst (Awaking time, green curve, right y-axis).
        (\textbf{G}) Solid lines depict the relationships between disruption and impact with fixed $\KI$ values, alongside the uncontrolled case (inset). Dashed lines represent linear fitted curves. For papers of comparable impact, those with higher $\KI$ are consistently more disruptive. Moreover, the slopes of the fitted curves with fixed $\KI$ are significantly flatter compared to the uncontrolled curve (inset).
        Bootstrapped 95\% confidence intervals are shown as shaded bands.
        \label{FigWOS_3}}
\end{figure}

% \clearpage
% \begin{figure}[htbp] % Do not use \begin{figure*}
% 	\centering
% 	\includegraphics[width=0.6\textwidth]{figures_si/FigWOS_Reference_count_distribution.pdf}
%     \caption{\textbf{$|$ Distribution of reference count in Web of Science dataset.}
%         The distribution of reference count for articles recorded in Web of Science follows a stretched exponential pattern, with a large number of papers containing relatively short reference lists.
%         \label{FigWOS_RC_Dis}}
% \end{figure}

\clearpage
\begin{figure}[htbp] % Do not use \begin{figure*}
	\centering
	\includegraphics[width=1.0\textwidth]{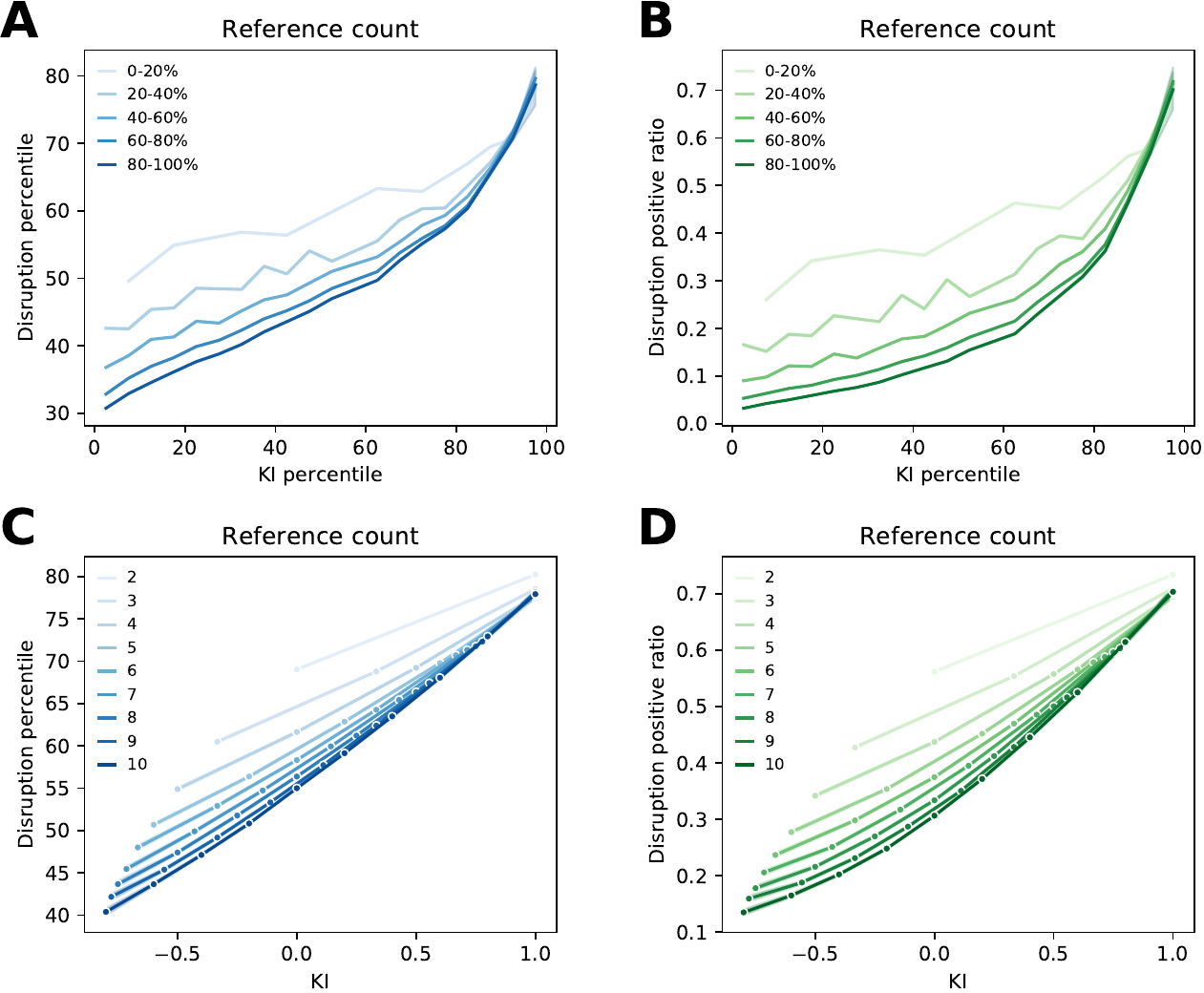}
    \caption{\textbf{$|$ The association between disruption and $\KI$ is robust across different reference counts in Web of Science dataset.}
        (\textbf{A-B}) When controlling for the percentile of reference count, both the percentile and positive ratio of disruption continue to increase with $\KI$ across all levels of reference count.
        (\textbf{C-D}) When controlling for the reference count within the bottom group ($<=10$), where each reference count induces a limited range of $\KI$ values, both the percentile and positive ratio of disruption continue to increase with $\KI$ across all groups of reference count.
        \label{FigWOS_RC}}
\end{figure}

\clearpage
\begin{figure}[htbp] % Do not use \begin{figure*}
	\centering
	\includegraphics[width=1.0\textwidth]{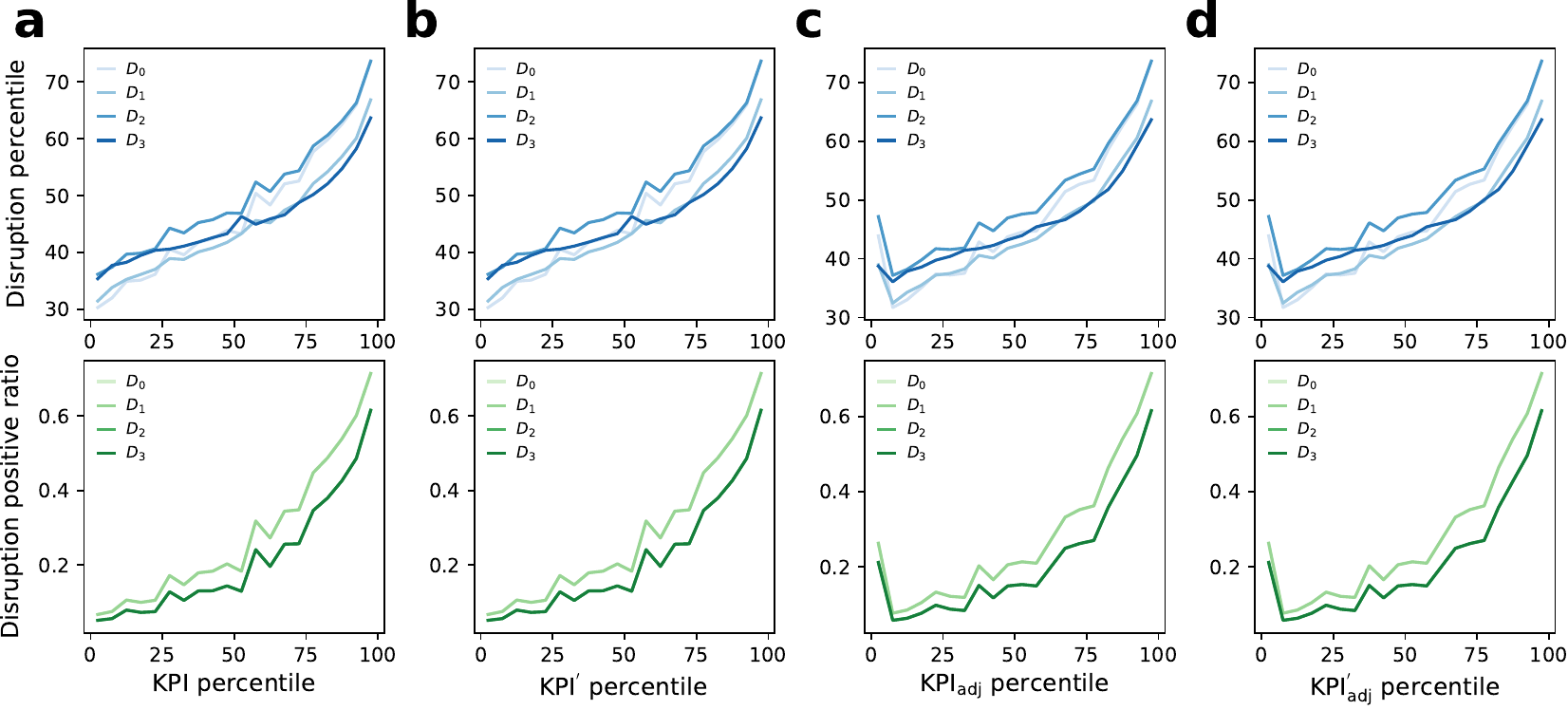}
    \caption{\textbf{$|$ The association between disruption and $\KI$ is robust across alternative measures in Web of Science dataset.}
        (\textbf{A-D})
        \textbf{Upper panel}: from the perspective of percentile, all disruption measures show a consistent increasing trend with all $\KI$ measures.
        \textbf{Lower panel}: from the perspective of positive ratio, all disruption measures similarly increase with all $\KI$ measures. The positive ratio represents the likelihood of having a higher proportion of $ind$-type references compared to $dep$-type references. Consequently, the variation curves for $D_1$ and $D_3$ align with those for $D_0$ and $D_2$, respectively.
        \label{FigWOS_AI}}
\end{figure}

\clearpage
\begin{figure}[htbp] % Do not use \begin{figure*}
	\centering
    \includegraphics[width=1.0\textwidth]{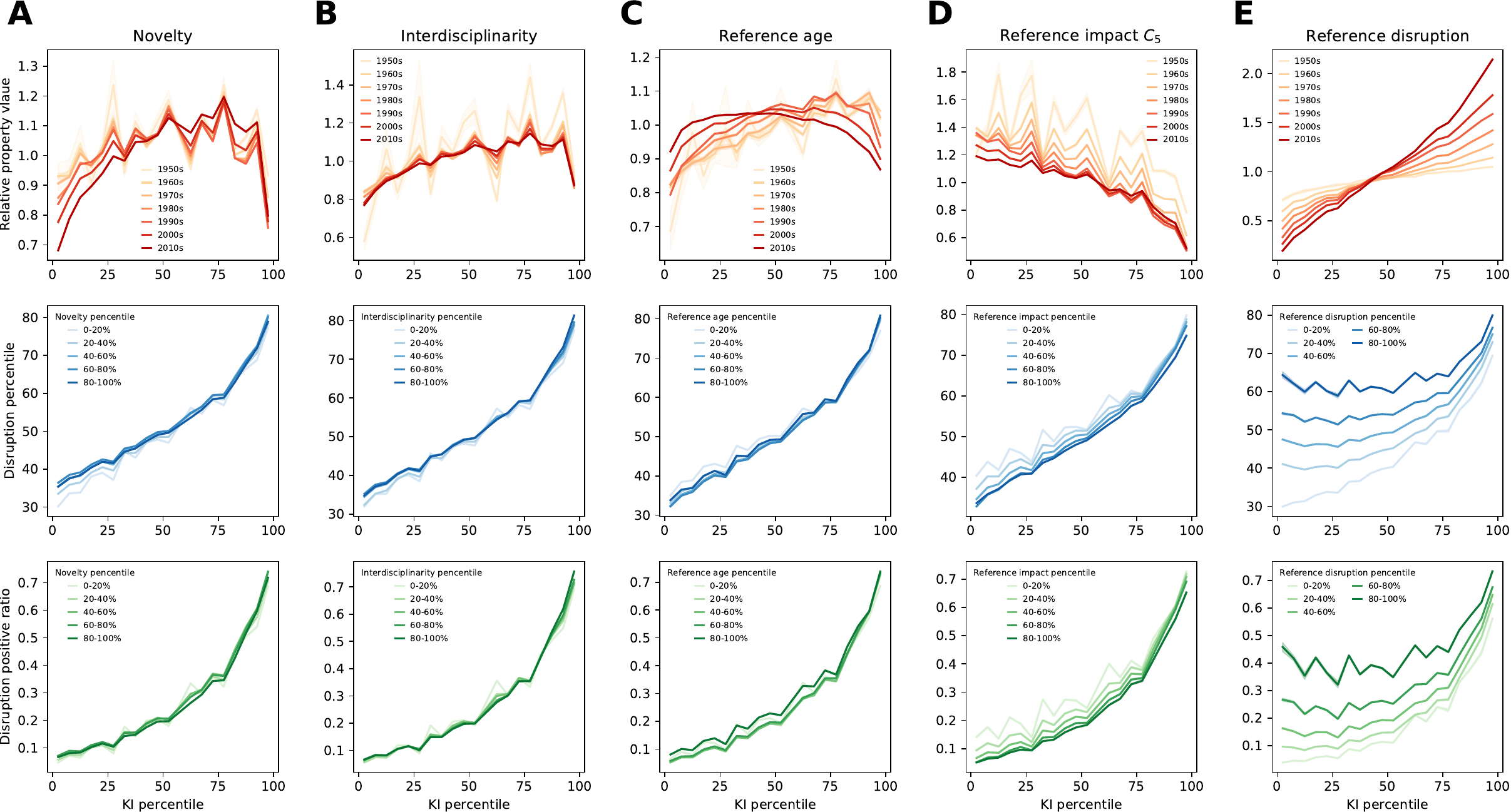}
    \caption{\textbf{$|$ The association between disruption and $\KI$ is robust when controlling for focal paper's properties in Web of Science dataset.}
        (\textbf{A-E}) When controlling for focal paper's novelty (panel~\textbf{A}), interdisciplinarity (panel~\textbf{B}), reference age (panel~\textbf{C}), reference impact $C_5$ (panel~\textbf{D}), and reference disruption (panel~\textbf{E}), respectively, both the percentile (blue curves) and positive ratio (green curves) of disruption continue to increase with $\KI$ across all levels of above covariates.
        \label{FigWOS_RP}}
\end{figure}

\clearpage
\begin{figure}[htbp] % Do not use \begin{figure*}
	\centering
	\includegraphics[width=1.0\textwidth]{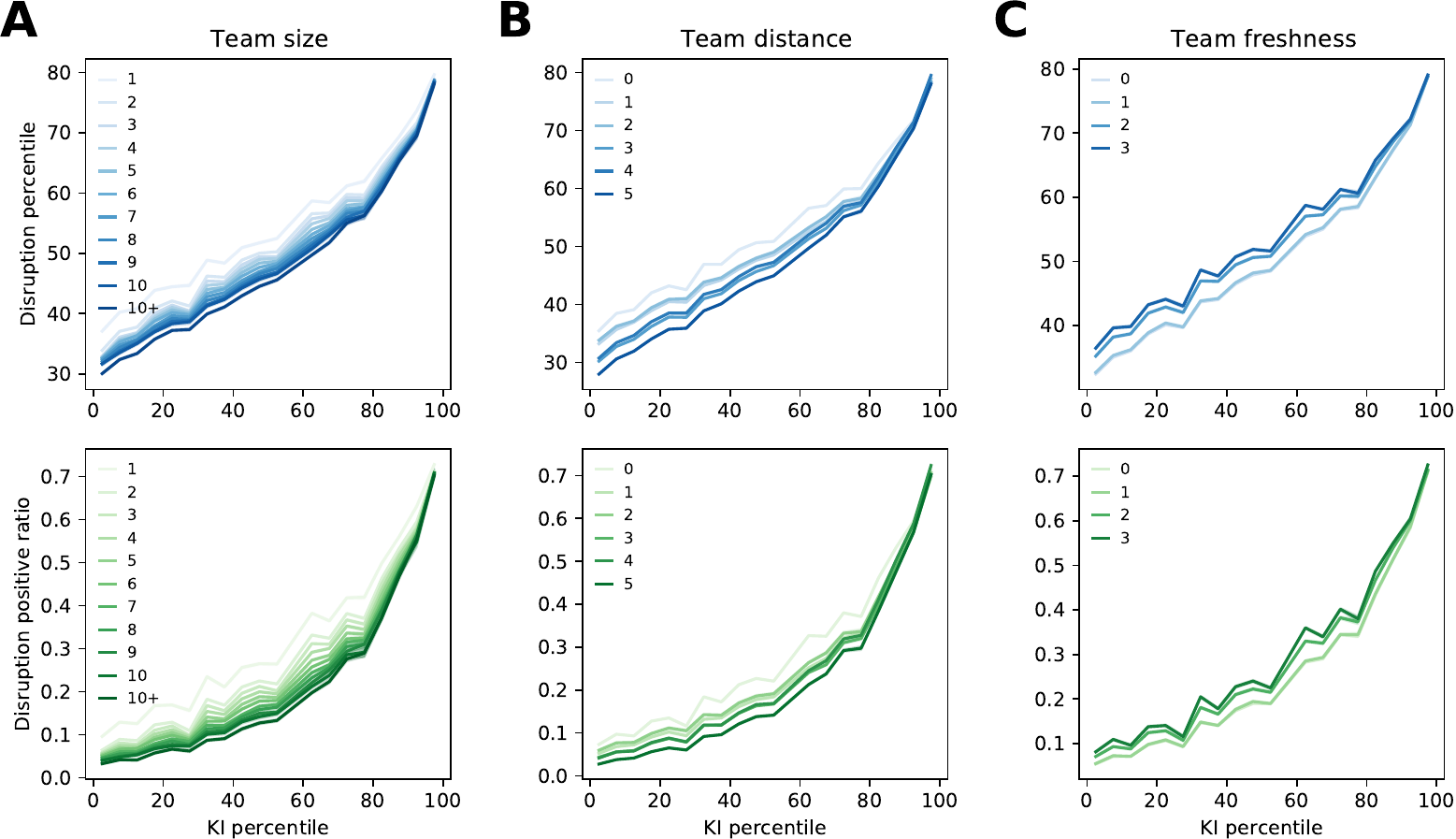}
    \caption{\textbf{$|$ The association between disruption and $\KI$ is robust when controlling for team compositions in Web of Science dataset.}
        (\textbf{A-C}) When controlling for team size (panel~\textbf{A}), geographic distance (panel~\textbf{B}), and collaboration freshness (panel~\textbf{C}) separately, both the percentile (blue curves) and positive ratio (green curves) of disruption continue to increase with $\KI$ across all levels of above team compositions. Additionally, papers produced by small, onsite, and fresh teams tend to exhibit slightly higher disruption.
        Considering the lack of coordinate information in Web of Science dataset, we categorize the team distance of a given focal paper by hierarchical categorization of the affiliations: $0$ (Sub-organization / Department level): $1$: Organization / University level; $2$: City level; $3$: State / Province level; $4$: National level; $5$: International level.
        \label{FigWOS_TC}}
\end{figure}

\clearpage
\begin{figure}[htbp] % Do not use \begin{figure*}
	\centering
	\includegraphics[width=1.0\textwidth]{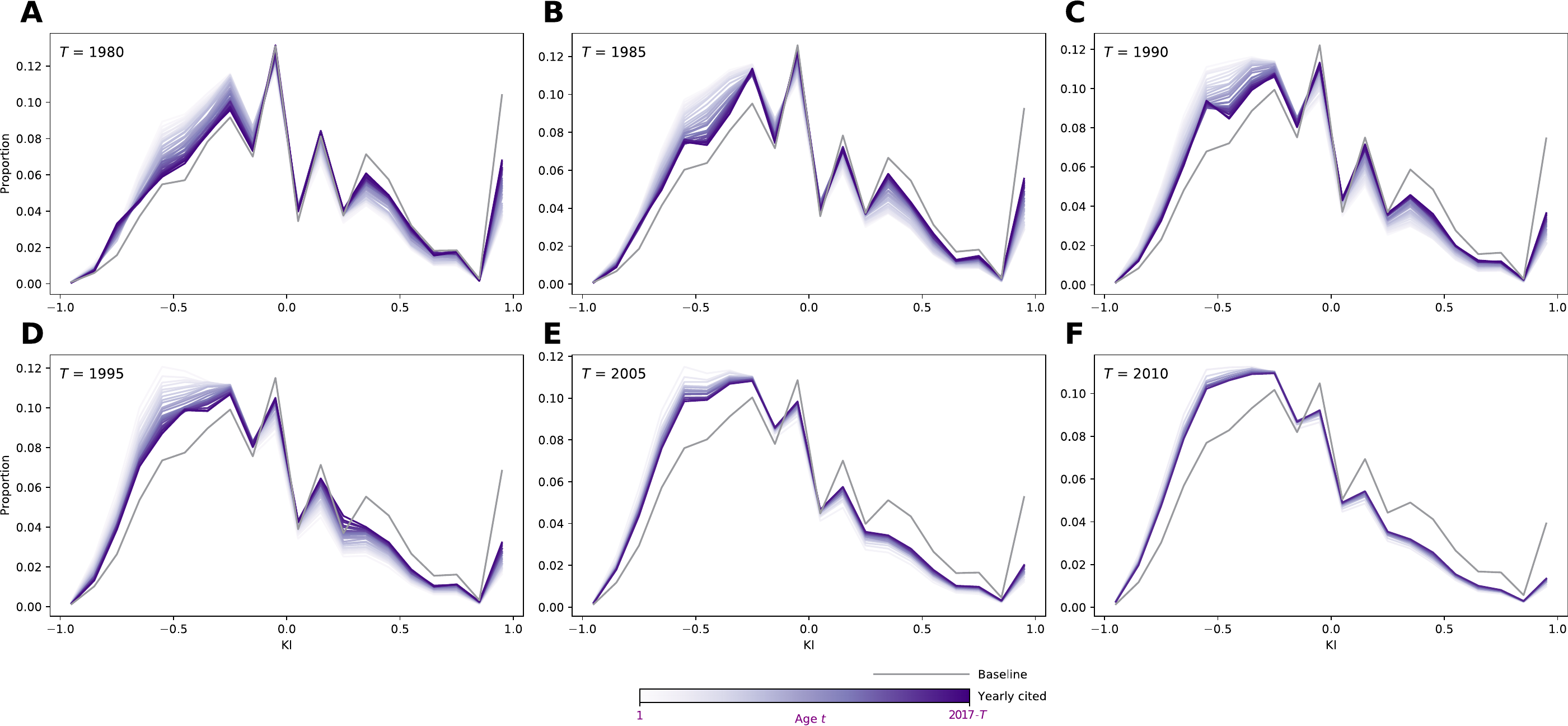}
    \caption{\textbf{$|$ Papers with higher $\KI$ are underrepresented among cited references and receiving delayed citations in Web of Science dataset.}
        (\textbf{A-F}) 
        For papers published in the year $T$, we measure the $\KI$ distribution among these papers as the baseline distribution (gray curve). Then, for all papers published in a subsequent year, $T+t$ ($t\geq 1$), we examine their bibliographies, extract the references that were published in $T$, and generate the $\KI$ distribution of those references (purple curve). Compared to the baseline, papers with negative $\KI$ are overrepresented while those with positive $\KI$ are underrepresented. Moreover, as the cited age $t$ increases, the corresponding distribution converges towards the baseline.
        \label{FigWOS_PS}}
\end{figure}
\newpage

%%%%%%%%%%%%%%%% SUPPLEMENTARY TABLES %%%%%%%%%%%%%%%

\begin{table} % Do NOT use \begin{table*}
    \centering
    % Captions go above tables
    \caption{\textbf{$|$ Normalized regression models of disruption and the influencing factors in Web of Science dataset.}}
    \label{tableWOS_OLS} % give each table a logical label name
    \renewcommand\tabcolsep{0.16cm} % column spacing
    \renewcommand{\arraystretch}{0.8} % line spacing
    \begin{threeparttable}
    \fontsize{7.5}{14}\selectfont % 设置表格字体大小为 8.75pt，行距为 12pt
    \begin{tabular}{lcccccccccc}
        \hline  % top line
        & Model 1 & Model 2 & Model 3 & Model 4 & Model 5 & Model 6 & Model 7 & Model 8 & Model 9 & $R^2$\% \\ 
        \hline  % middle line 1
        $\KI$ & \textbf{0.4204}*** & \textbf{0.4203}*** & \textbf{0.4189}*** & \textbf{0.4081}*** & \textbf{0.4184}*** & \textbf{0.4203}*** & \textbf{0.4186}*** & \textbf{0.4204}*** & \textbf{0.4036}*** & \textbf{71.16} \\
         & (0.000) & (0.000) & (0.000) & (0.000) & (0.000) & (0.000) & (0.000) & (0.000) & (0.000) \\
        % Log $C_5$ &  & -0.0416*** &  &  &  &  &  &  &  &  &  & -0.0322*** & 3.91 \\
         % &  & (0.000) &  &  &  &  &  &  &  &  &  & (0.000) \\
        Novelty &  & 0.0063*** &  &  &  &  &  &  & 0.0020*** & 0.04 \\
         &  & (0.000) &  &  &  &  &  &  & (0.000) \\
        Interdisciplinarity &  &  & 0.0164*** &  &  &  &  &  & 0.0168*** & 0.74 \\
         &  &  & (0.000) &  &  &  &  &  & (0.000) \\
        Reference count &  &  &  & -0.0364*** &  &  &  &  & -0.0342*** & 9.56 \\
         &  &  &  & (0.000) &  &  &  &  & (0.000) \\
        Reference age &  &  &  &  & 0.0538*** &  &  &  & 0.0508*** & 0.85 \\
         &  &  &  &  & (0.000) &  &  &  & (0.000) \\
        % Reference $C_5$ &  &  &  &  &  &  & -0.0197*** &  &  &  &  & -0.0168*** & 0.39 \\
        %  &  &  &  &  &  &  & (0.000) &  &  &  &  & (0.000) \\
        % Reference disruption &  &  &  &  &  &  &  & 0.2293*** &  &  &  & 0.2231*** & 37.17 \\
        %  &  &  &  &  &  &  &  & (0.000) &  &  &  & (0.000) \\
        Team size &  &  &  &  &  & -0.0061*** &  &  & -0.0034*** & 0.07 \\
         &  &  &  &  &  & (0.000) &  &  & (0.000) \\
        Team distance &  &  &  &  &  &  & -0.0330*** &  & -0.0308*** & 2.77 \\
         &  &  &  &  &  &  & (0.000) &  & (0.000) \\
        Team freshness &  &  &  &  &  &  &  & 0.0002 & 0.0002 & 0.35 \\
         &  &  &  &  &  &  &  & (0.000) & (0.000) \\
        \hline % middle line 2
        Discipline fixed effects  & Yes & Yes & Yes & Yes & Yes & Yes & Yes & Yes & Yes & 7.30 \\
        Year fixed effects & Yes & Yes & Yes & Yes & Yes & Yes & Yes & Yes & Yes & 7.16 \\
        \hline % middle line 3
        N & 12145196 & 12145196 & 12145196 & 12145196 & 12145196 & 12145196 & 12145196 & 12145196 & 12145196 \\
        $R^2$ & 0.221 & 0.221 & 0.221 & 0.222 & 0.224 & 0.221 & 0.222 & 0.221  & 0.226 \\
        \hline  % bottom line
    \end{tabular}
    \begin{tablenotes}
        \small
        \item[$\dagger$] We perform normalized ordinary-least-squares (OLS) regression analyses to examine the effect of $\KI$ on disruption in Web of Science dataset, while controlling for various covariates. Each regression coefficient is tested against the null hypothesis (coefficient equals 0) using a two-sided \textit{t}-test. Standard errors are provided in parentheses for each coefficient. Note that we do not apply adjustments for multiple hypothesis testing in this analysis. (*$p < 0.05$, **$p < 0.01$, ***$p < 0.001$).
    \end{tablenotes}
    \end{threeparttable}
\end{table}

\clearpage
\begin{table} % Do not use \begin{table*}
\centering
% Captions go above tables
\caption{\textbf{$|$ Normalized mediation regression on team size in Web of Science dataset.}}
\label{tableWOS_mediation_TS} % give each table a logical label name
\renewcommand\tabcolsep{0.2cm} % column spacing
\renewcommand{\arraystretch}{1.5} % line spacing
\begin{threeparttable}
\fontsize{12}{10}\selectfont % 设置表格字体大小为 8.75pt，行距为 12pt
\begin{tabular}{llcccc} % four columns, alignment for each
    \hline
    Independent variable & Mediator & ACME ($\beta_1\theta_2$) & ADE ($\theta_1$) & TE ($\theta_1 + \beta_1\theta_2$) & $|$ACME / TE$|$ \\
    \hline
    Team size & $\KI$ & \textbf{-0.0043}***  & \textbf{-0.0064}*** & -0.0107*** & \textbf{0.4029} \\
     &  & (0.0003) & (0.0006) & (0.0009) \\\\
    % Team size & \sout{Log $C_5$} & -0.0065***  & -0.0058*** & -0.0107*** \\
    %  &  & (0.0006) & (0.0007) & (0.0009) \\\\
    Team size & Novelty & -0.0001***  & -0.0106*** & -0.0107*** & 0.0139 \\
     &  & (0.0000) & (0.0009) & (0.0009) \\\\
    Team size & Interdisciplinarity & 0.0002***  & -0.0109*** & -0.0107*** & 0.0149 \\
     &  & (0.0000) & (0.0009) & (0.0009) \\\\
    Team size & Reference count & -0.0006***  & -0.0101*** & -0.0107*** & 0.0577 \\
     &  & (0.0002) & (0.0009) & (0.0009) \\\\
    Team size & Reference age & -0.0015***  & -0.0093*** & -0.0107*** & 0.1359 \\
     &  & (0.0001) & (0.0008) & (0.0009) \\\\
    % Team size & Reference $C_5$ & -0.0004***  & -0.0120*** & -0.0107*** & 0.0359 \\
    %  &  & (0.0000) & (0.0009) & (0.0009) \\\\
    % Team size & \sout{Reference disruption} & -0.0064***  & -0.0059*** & -0.0107*** \\
    %  &  & (0.0007) & (0.0006) & (0.0009) \\\\
    Team size & Team distance & -0.0037***  & -0.0070*** & -0.0107*** & 0.3443 \\
     &  & (0.0003) & (0.0007) & (0.0009) \\\\
    Team size & Team freshness & -0.0000***  & -0.0107*** & -0.0107*** & 0.0000 \\
     &  & (0.0000) & (0.0009) & (0.0009) \\
    \hline
\end{tabular}
\begin{tablenotes}
    \small
    \item[$\dagger$] We perform normalized mediation analysis to assess the effect of team size on disruption in Web of Science dataset. The candidate mediating variables include: $\KI$, novelty, interdisciplinarity, reference count, reference age, team distance, and team freshness, with discipline and publication year controlled. ACME refers to average causal mediation effects driven from the mediating variable, ADE stands for average direct effects caused by team size, and TE is defined as the sum of the ACME and ADE. Each regression coefficient is tested against the null hypothesis (coefficient equals 0) using a two-sided \textit{t}-test. Standard errors are provided in parentheses for each coefficient. Note that we do not apply adjustments for multiple hypothesis testing in this analysis. (*$p < 0.05$, **$p < 0.01$, ***$p < 0.001$).
\end{tablenotes}
\end{threeparttable}
\end{table}

\clearpage
\begin{table} % Do not use \begin{table*}
\centering
% Captions go above tables
\caption{\textbf{$|$ Normalized mediation regression on team distance in Web of Science dataset.}}
\label{tableWOS_mediation_TD} % give each table a logical label name
\renewcommand\tabcolsep{0.2cm} % column spacing
\renewcommand{\arraystretch}{1.5} % line spacing
\begin{threeparttable}
\fontsize{12}{10}\selectfont % 设置表格字体大小为 8.75pt，行距为 12pt
\begin{tabular}{llcccc} % four columns, alignment for each
    \hline
    Independent variable & Mediator & ACME ($\beta_1\theta_2$) & ADE ($\theta_1$) & TE ($\theta_1 + \beta_1\theta_2$) & $|$ACME / TE$|$ \\
    \hline
    Team distance & $\KI$ & \textbf{-0.0254}***  & \textbf{-0.0298}*** & -0.0553*** & \textbf{0.4602} \\
     &  & (0.0002) & (0.0005) & (0.0005) \\\\
    % Team distance & \sout{Log $C_5$} & -0.0174***  & -0.0399*** & -0.0553*** \\
    %  &  & (0.0001) & (0.0005) & (0.0005) \\\\
    Team distance & Novelty & -0.0000***  & -0.0552*** & -0.0553*** & 0.0008 \\
     &  & (0.0000) & (0.0005) & (0.0005) \\\\
    Team distance & Interdisciplinarity & 0.0023***  & -0.0576*** & -0.0553*** & 0.0422 \\
     &  & (0.0000) & (0.0005) & (0.0005) \\\\
    Team distance & Reference count & -0.0085***  & -0.0468*** & -0.0553*** & 0.1536 \\
     &  & (0.0002) & (0.0005) & (0.0005) \\\\
    Team distance & Reference age & -0.0021***  & -0.0532*** & -0.0553*** & 0.0379 \\
     &  & (0.0000) & (0.0005) & (0.0005) \\\\
    % Team distance & Reference $C_5$ & -0.0003***  & -0.0547*** & -0.0553*** & 0.0056 \\
    %  &  & (0.0000) & (0.0005) & (0.0005) \\\\
    % Team distance & \sout{Reference disruption} & -0.0244***  & -0.0330*** & -0.0553*** \\
    %  &  & (0.0003) & (0.0005) & (0.0005) \\\\
    Team distance & Team size & -0.0005***  & -0.0548*** & -0.0553*** & 0.0086 \\
     &  & (0.0000) & (0.0005) & (0.0005) \\\\
    Team distance & Team freshness & -0.0000***  & -0.0553*** & -0.0553*** & 0.0000 \\
     &  & (0.0000) & (0.0005) & (0.0005) \\
    \hline
\end{tabular}
\begin{tablenotes}
    \small
    \item[$\dagger$] We perform normalized mediation analysis to assess the effect of team distance on disruption in Web of Science dataset. The candidate mediating variables include: $\KI$, novelty, interdisciplinarity, reference count, reference age, team size, and team freshness, with discipline and publication year controlled. ACME refers to average causal mediation effects driven from the mediating variable, ADE stands for average direct effects caused by team distance, and TE is defined as the sum of the ACME and ADE. Each regression coefficient is tested against the null hypothesis (coefficient equals 0) using a two-sided \textit{t}-test. Standard errors are provided in parentheses for each coefficient. Note that we do not apply adjustments for multiple hypothesis testing in this analysis. (*$p < 0.05$, **$p < 0.01$, ***$p < 0.001$).
\end{tablenotes}
\end{threeparttable}
\end{table}

\clearpage
\begin{table} % Do not use \begin{table*}
\centering
% Captions go above tables
\caption{\textbf{$|$ Normalized mediation regression on team freshness in Web of Science dataset.}}
\label{tableWOS_mediation_TF} % give each table a logical label name
\renewcommand\tabcolsep{0.2cm} % column spacing
\renewcommand{\arraystretch}{1.5} % line spacing
\begin{threeparttable}
\fontsize{12}{10}\selectfont % 设置表格字体大小为 8.75pt，行距为 12pt
\begin{tabular}{llcccc} % four columns, alignment for each
    \hline
    Independent variable & Mediator & ACME ($\beta_1\theta_2$) & ADE ($\theta_1$) & TE ($\theta_1 + \beta_1\theta_2$) & $|$ACME / TE$|$ \\
    \hline
    Team freshness & $\KI$ & \textbf{0.0003}***  & \textbf{0.0006}*** & 0.0008*** & \textbf{0.3044} \\
     &  & (0.0002) & (0.0005) & (0.0006) \\\\
    % Team freshness & \sout{Log $C_5$} & -0.0001***  & -0.0003*** & 0.0008*** \\
    %  &  & (0.0001) & (0.0006) & (0.0006) \\\\
    Team freshness & Novelty & 0.0000***  & 0.0008*** & 0.0008*** & 0.0014 \\
     &  & (0.0000) & (0.0006) & (0.0006) \\\\
    Team freshness & Interdisciplinarity & -0.0000***  & 0.0009*** & 0.0008*** & 0.0401 \\
     &  & (0.0000) & (0.0006) & (0.0006) \\\\
    Team freshness & Reference count & 0.0002***  & 0.0006*** & 0.0008*** & 0.2572 \\
     &  & (0.0001) & (0.0006) & (0.0006) \\\\
    Team freshness & Reference age & -0.0000  & 0.0008*** & 0.0008*** & 0.0010 \\
     &  & (0.0000) & (0.0006) & (0.0006) \\\\
    % Team freshness & Reference $C_5$ & 0.0000***  & 0.0008*** & 0.0008*** & 0.0118 \\
    %  &  & (0.0000) & (0.0006) & (0.0006) \\\\
    % Team freshness & \sout{Reference disruption} & 0.0003***  & -0.0006*** & 0.0008*** \\
    %  &  & (0.0003) & (0.0006) & (0.0006) \\\\
    Team freshness & Team size & 0.0000***  & 0.0008*** & 0.0008*** & 0.0016 \\
     &  & (0.0000) & (0.0006) & (0.0006) \\\\
    Team freshness & Team distance & 0.0001***  & 0.0007*** & 0.0008*** & 0.1398 \\
     &  & (0.0000) & (0.0006) & (0.0006) \\
    \hline
\end{tabular}
\begin{tablenotes}
    \small
    \item[$\dagger$] We perform normalized mediation analysis to assess the effect of team freshness on disruption in Web of Science dataset. The candidate mediating variables include: $\KI$, novelty, interdisciplinarity, reference count, reference age, team size, and team distance, with discipline and publication year controlled. ACME refers to average causal mediation effects driven from the mediating variable, ADE stands for average direct effects caused by team freshness, and TE is defined as the sum of the ACME and ADE. Each regression coefficient is tested against the null hypothesis (coefficient equals 0) using a two-sided \textit{t}-test. Standard errors are provided in parentheses for each coefficient. Note that we do not apply adjustments for multiple hypothesis testing in this analysis. (*$p < 0.05$, **$p < 0.01$, ***$p < 0.001$).
\end{tablenotes}
\end{threeparttable}
\end{table}

\clearpage
\begin{table} % Do not use \begin{table*}
\centering
% Captions go above tables
\caption{\textbf{$|$ Normalized confounding regression on paper impact in Web of Science dataset.}}
\label{tableWOS_confounding_impact} % give each table a logical label name
\renewcommand\tabcolsep{0.2cm} % column spacing
\renewcommand{\arraystretch}{1.5} % line spacing
\begin{threeparttable}
\fontsize{12}{10}\selectfont % 设置表格字体大小为 8.75pt，行距为 12pt
\begin{tabular}{llccc} % four columns, alignment for each
    \hline
    Independent variable & Confounder & Unadjusted ($\alpha_1$) & Adjusted ($\delta_1$) & RCE ($|\alpha_1 - \delta_1|/|\alpha_1|$) \\
    \hline
    Citation percentile & $\KI$ & -0.1302***  & \textbf{-0.0276}***  & \textbf{0.7878} \\
     &  & (0.0006) & (0.0006) \\\\
    Citation percentile & Novelty & -0.1302***  & -0.1305*** & 0.0022 \\
     &  & (0.0006) & (0.0006) \\\\
    Citation percentile & Interdisciplinarity & -0.1302***  & -0.1328*** & 0.0202 \\
     &  & (0.0006) & (0.0006) \\\\
    Citation percentile & Reference count & -0.1302***  & -0.0872*** & 0.3306 \\
     &  & (0.0006) & (0.0009) \\\\
    Citation percentile & Reference age & -0.1302***  & -0.1217*** & 0.0649 \\
     &  & (0.0006) & (0.0006) \\\\
    % Citation percentile & Reference $C_5$ & -0.1302***  & -0.1001*** & 0.0073 \\
    %  &  & (0.0006) & (0.0006) \\\\
    % Citation percentile & \sout{Reference disruption} & -0.1302***  & -0.0534*** \\
    %  &  & (0.0006) & (0.0006) \\\\
    Citation percentile & Team size & -0.1302***  & -0.1300*** & 0.0016 \\
     &  & (0.0006) & (0.0006) \\\\
    Citation percentile & Team distance & -0.1302***  & -0.1257*** & 0.0345 \\
     &  & (0.0006) & (0.0006) \\\\
    Citation percentile & Team freshness & -0.1302***  & -0.1302*** & 0.0000 \\
     &  & (0.0006) & (0.0006) \\
    \hline
\end{tabular}
\begin{tablenotes}
    \small
    \item[$\dagger$] We perform normalized confounding analysis to assess the effect of paper impact on disruption in Web of Science dataset. The candidate confounding variables include: $\KI$, novelty, interdisciplinarity, reference count, reference age, team size, team distance, and team freshness, with discipline and publication year controlled. The relative change in estimate (RCE) refers to the relative change of effect after controlling for the confounder. Each regression coefficient is tested against the null hypothesis (coefficient equals 0) using a two-sided \textit{t}-test. Standard errors are provided in parentheses for each coefficient. Note that we do not apply adjustments for multiple hypothesis testing in this analysis. (*$p < 0.05$, **$p < 0.01$, ***$p < 0.001$).
\end{tablenotes}
\end{threeparttable}
\end{table}

\clearpage
\subsection{Extended analyses to OECD patent dataset}\label{SI_OECD_results}
{\fontsize{12pt}{14pt}\selectfont % {fontsize}{linespace}
We have replicated our core analyses using the OECD-Patent data\footnote{https://www.oecd.org/en/data/datasets/intellectual-property-statistics.html}, with the latest data version of Feb-2025. This dataset, known for its extensive coverage and larger volume compared to other patent databases like PatentView, allows for a more comprehensive analysis of global technological innovation trends.

Our analysis of the patent data revealed a distinctive citation landscape. We found that the distribution of $\KI$ in patents is highly skewed, with over half of the patents having a maximum $\KI$ score of 1.0 (Fig.~\ref{FigOECD_1_KI_Stats}A). The remaining $\KI$ scores are also systematically biased towards positive values. This unique distribution reflects a key difference between scientific and technological knowledge systems, which we attribute to the inherent review mechanisms of patent applications. Patent examiners systematically assess the relevance of cited prior art; patents with overly similar or dependent citations are often rejected due to a lack of novelty and inventiveness, which, by design, leads to a systemic preference for citation independence.

We also analyzed the temporal evolution of patent $\KI$, which supports the influence of this institutional mechanism~(Fig.~\ref{FigOECD_1_KI_Stats}B). From 1978 to 2010, the average $\KI$ of patents showed a clear downward trend, consistent with the decline cycle of the patent disruption displayed in Figure~2 by Park et al.~(2023)~\cite{park2023papers}. However, this decline slowed around 2000, coinciding with the initial internationalization of patent examination driven by the cooperation between the European Patent Office (EPO) and the U.S. Patent and Trademark Office (USPTO). Notably, with the formal launch of the Cooperative Patent Classification (CPC) system by the EPO and the USPTO in 2013\footnote{European Patent Office, ``Highlights 2012: Cooperative Patent Classification'', EPO Annual Report 2012.}, which signaled a greater convergence of global patent examination standards, $\KI$ has since shown an overall upward trend.
\begin{figure}[h] % Do not use \begin{figure*}
	\centering
	\includegraphics[width=0.85\textwidth]{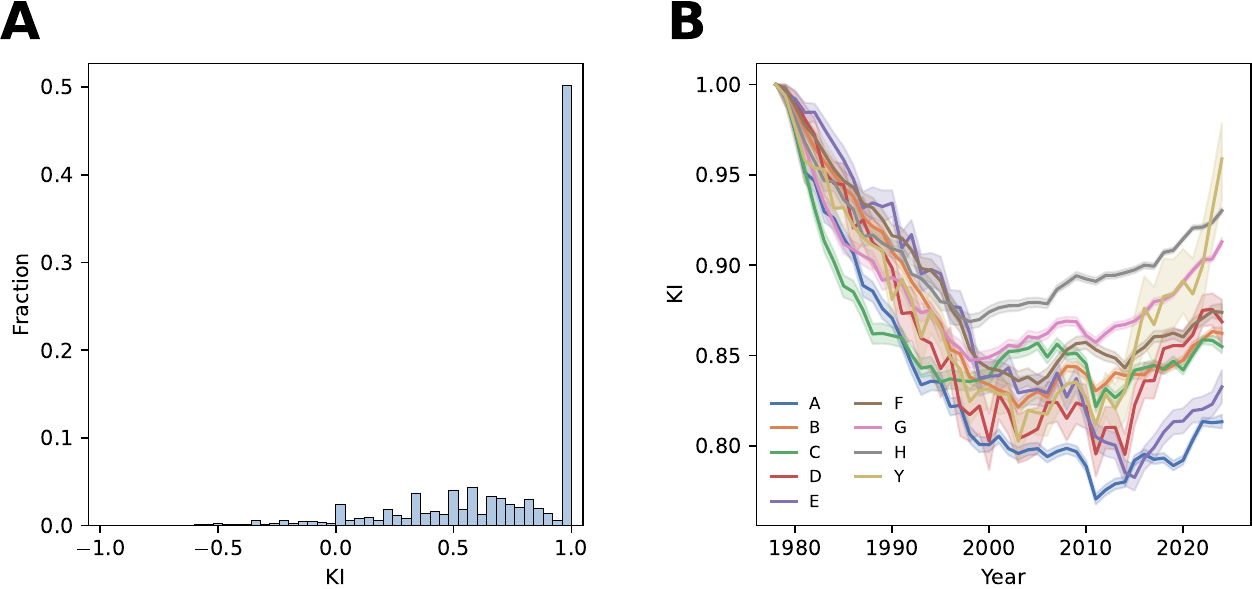}
    \caption{\textbf{$|$ The distribution and evolution pattern of knowledge independence in the OECD patent dataset.}
        (\textbf{A}) The $\KI$ distribution of 16,966,216 patents with at least two references published between 1978 and 2024 in the OECD.
        (\textbf{B}) Patents' $\KI$ exhibit a pervasive downtrend from 1978 to 2010 and upward trend since around 2010 across technical domains (the domain category mapping can be found via https://www.uspto.gov/web/patents/classification/cpc/html/cpc.html).
        % (A:~human necessities, B:~performing operations; transporting, C:~chemistry; metallurgy, D:~textiles; paper, E:~fixed constructions, F:~mechanical engineering; lighting; heating; weapons; blasting engines or pumps, G:~physics, H:~electricity, Y:~general tagging of new technological developments; general tagging of cross-sectional technologies spanning over several sections of the IPC; technical subjects covered by former USPC cross-reference art collections [XRACs] and digests). 
        Bootstrapped 95\% confidence intervals are shown as shaded bands.
        \label{FigOECD_1_KI_Stats}}
\end{figure}

Despite these institutional differences, we found compelling evidence that $\KI$'s relationship with technological innovation mirrors its role in scientific disruption. Specifically, our analysis confirms a significant positive association between $\KI$ and patent disruption~(Fig.~\ref{FigOECD_2_control_merge}A). This finding is robust across various statistical controls~(Fig.~\ref{FigOECD_2_control_merge}B-L) and is further supported by causal inference analyses~(Fig.~\ref{FigOECD_2_PSM_CEM}, Table~\ref{table_OECD_OLS}). Given the highly skewed distribution of $\KI$, with approximately half of the observations having a value of 1, we conducted a robustness check by binarizing the variable. We created a binary variable representing ``fully independence'' ($\KI=1$) versus ``non-fully independence'' ($\KI<1$). Despite this simplification, our analysis continued to find a robust positive association between patent $\KI$ and disruption~(Fig.~\ref{FigOECD_2_control_merge_binary}-\ref{FigOECD_2_PSM_CEM_binary}, Table~\ref{table_OECD_OLS_binary}), suggesting that our core finding is not sensitive to this distributional property. These results consistently show that a higher $\KI$, regardless of the institutional context, is a key driver of technological disruption. The consistency of these results across the scientific and technological domains strongly reinforces the generalizability of our conceptual framework and the predictive power of the $\KI$ metric on disruption.

Regarding the relationship between $\KI$ and patent impact, though it's a compelling area, we have opted to defer its comprehensive analysis to future work. This decision is based on two key considerations: first, patent impact is arguably more closely linked to commercial value than to citation counts; and second, the central focus of our current study is the citation-based recognition system. A systematic investigation into the connection between $\KI$ and patent impact will be a crucial component of our future research agenda.

\begin{figure}[ht] % Do not use \begin{figure*}
	\centering
	\includegraphics[width=1\textwidth]{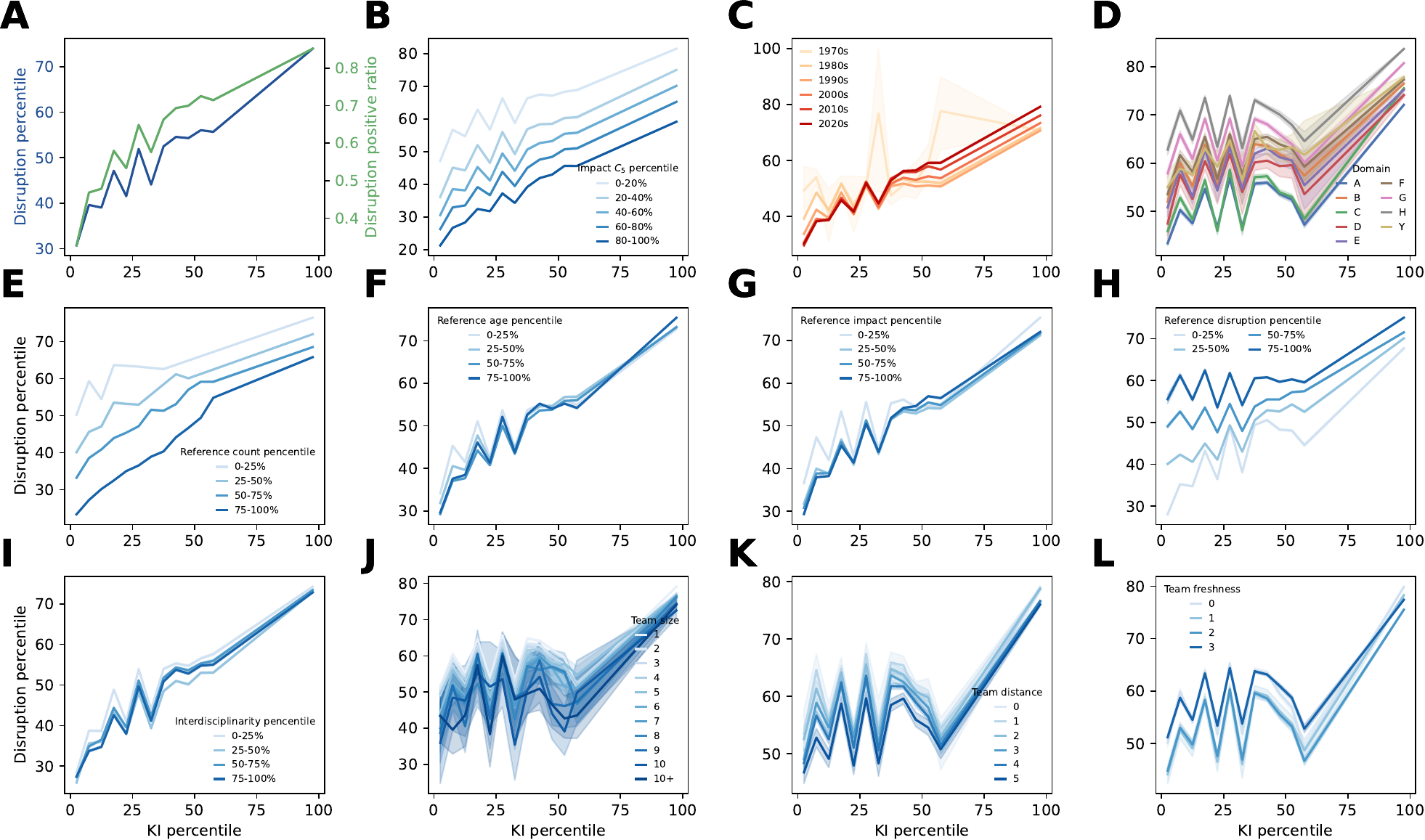}
    \caption{\textbf{$|$ Knowledge independence is associated with technical disruption in the OECD patent dataset.}
        (\textbf{A}) For the 11,349,527 patents published before 2023 in the OECD patent dataset that are cited at least once, the average disruption percentile (blue curve, left y-axis) and the disruption positive ratio (green curve, right y-axis) increase with the $\KI$ percentile.
        The association between $\KI$ and disruption persists, regardless of the impact $C_5$~(panel~\textbf{B}), publication decade~(panel~\textbf{C}), technical domain~(panel~\textbf{D}), reference count~(panel~\textbf{E}), reference age~(panel~\textbf{F}), reference impact $C_5$~(panel~\textbf{G}), reference disruption~(panel~\textbf{H}), interdisciplinarity~(panel~\textbf{I}), team size~(panel~\textbf{J}), geographic distance~(panel~\textbf{K}), and collaboration freshness~(panel~\textbf{L}).
        Bootstrapped 95\% confidence intervals are shown as shaded bands.
        \label{FigOECD_2_control_merge}}
\end{figure}

\begin{figure}[ht] % Do not use \begin{figure*}
	\centering
	\includegraphics[width=0.87\textwidth]{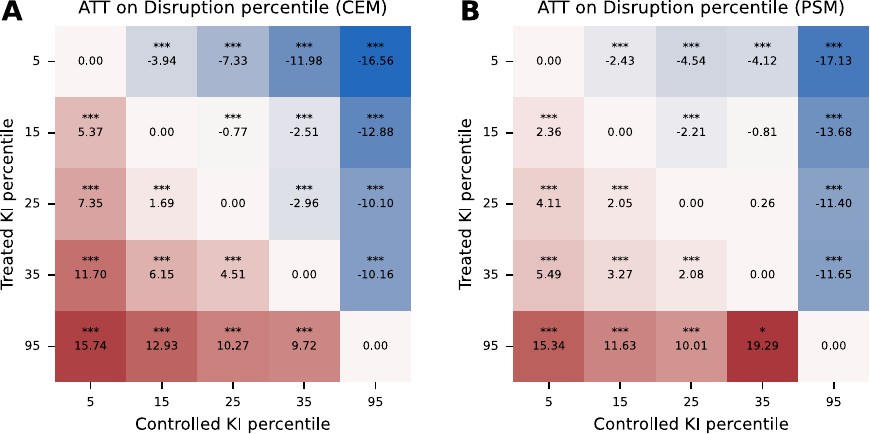}
    \caption{\textbf{$|$ Causal inference by CEM and PSM in the OECD patent dataset.}
        The ATT matrixs of $\KI$ on disruption via CEM (\textbf{A}) and PSM (\textbf{B}).
        Each controlled group is set as a baseline, and ATTs are calculated for comparisons between the baseline and each of the treated groups. Blue cells represent negative ATTs, while red ones represent positive ATTs, with color intensity proportional to the absolute value.
        Each ATT is tested against the null hypothesis (ATT equals 0) using a two-sided \textit{t}-test. (*$p < 0.05$, **$p < 0.01$, ***$p < 0.001$).
        \label{FigOECD_2_PSM_CEM}}
\end{figure}

\begin{table}[ht] % Do NOT use \begin{table*}
    \centering
    % Captions go above tables
    \caption{\textbf{$|$ Normalized regression models of disruption and the influencing factors in the OECD dataset.}}
    \label{table_OECD_OLS} % give each table a logical label name
    \renewcommand\tabcolsep{0.16cm} % column spacing
    \renewcommand{\arraystretch}{0.8} % line spacing
    \begin{threeparttable}
    \fontsize{7.5}{14}\selectfont % 设置表格字体大小为 8.75pt，行距为 12pt
    \begin{tabular}{lcccccccccc}
        \hline  % top line
        & Model 1 & Model 2 & Model 3 & Model 4 & Model 5 & Model 6 & Model 7 & Model 8 & $R^2$\% \\ 
        \hline  % middle line 1
        $\KI$ & \textbf{0.1712}*** & \textbf{0.1711}*** & \textbf{0.1644}*** & \textbf{0.1713}*** & \textbf{0.1707}*** & \textbf{0.1705}*** & \textbf{0.1713}*** & \textbf{0.1638}*** & \textbf{40.65} \\
         & (0.001) & (0.001) & (0.001) & (0.001) & (0.001) & (0.001) & (0.001) & (0.001) \\
        % Log $C_5$ &  & -0.0416*** &  &  &  &  &  &  &  &  &  & -0.0322*** & 3.91 \\
         % &  & (0.0000) &  &  &  &  &  &  &  &  &  & (0.0000) \\
        % Novelty &  & 0.0063*** &  &  &  &  &  &  & 0.0020*** & 0.04 \\
        %  &  & (0.0000) &  &  &  &  &  &  & (0.0000) \\
        Interdisciplinarity &  & -0.0045*** &  &  &  &  &  & -0.0028** & 0.07 \\
         &  & (0.001) &  &  &  &  &  & (0.001) \\
        Reference count &  &  & -0.1021*** &  &  &  &  & -0.1012*** & 19.52 \\
         &  &  & (0.001) &  &  &  &  & (0.001) \\
        Reference age &  &  &  & -0.0092*** &  &  &  & 0.0025* & 0.76 \\
         &  &  &  & (0.001) &  &  &  & (0.001) \\
        % Reference $C_5$ &  &  &  &  &  &  & -0.0197*** &  &  &  &  & -0.0168*** & 0.39 \\
        %  &  &  &  &  &  &  & (0.0000) &  &  &  &  & (0.0000) \\
        % Reference disruption &  &  &  &  &  &  &  & 0.2293*** &  &  &  & 0.2231*** & 37.17 \\
        %  &  &  &  &  &  &  &  & (0.0000) &  &  &  & (0.0000) \\
        Team size &  &  &  &  & -0.0222*** &  &  & -0.0037** & 0.54 \\
         &  &  &  &  & (0.001) &  &  & (0.001) \\
        Team distance &  &  &  &  &  & -0.0276*** &  & -0.0215*** & 0.74 \\
         &  &  &  &  &  & (0.001) &  & (0.001) \\
        Team freshness &  &  &  &  &  &  & -0.0129*** & -0.0101*** & 0.10 \\
         &  &  &  &  &  &  & (0.001) & (0.001) \\
        \hline % middle line 2
        Discipline fixed effects  & Yes & Yes & Yes & Yes & Yes & Yes & Yes & Yes & 28.82 \\
        Year fixed effects & Yes & Yes & Yes & Yes & Yes & Yes & Yes & Yes & 8.79 \\
        \hline % middle line 3
        N & 927551 & 927551 & 927551 & 927551 & 927551 & 927551 & 927551 & 927551 \\
        $R^2$ & 0.065 & 0.065 & 0.075 & 0.065 & 0.066 & 0.066 & 0.065  & 0.076 \\
        \hline  % bottom line
    \end{tabular}
    \begin{tablenotes}
        \small
        \item[$\dagger$] We perform normalized ordinary-least-squares (OLS) regression analyses to examine the effect of $\KI$ on disruption in the OECD patent dataset, while controlling for various covariates. Each regression coefficient is tested against the null hypothesis (coefficient equals 0) using a two-sided \textit{t}-test. Standard errors are provided in parentheses for each coefficient. Note that we do not apply adjustments for multiple hypothesis testing in this analysis. (*$p < 0.05$, **$p < 0.01$, ***$p < 0.001$).
    \end{tablenotes}
    \end{threeparttable}
\end{table}

\clearpage
\begin{figure}[ht] % Do not use \begin{figure*}
	\centering
	\includegraphics[width=1\textwidth]{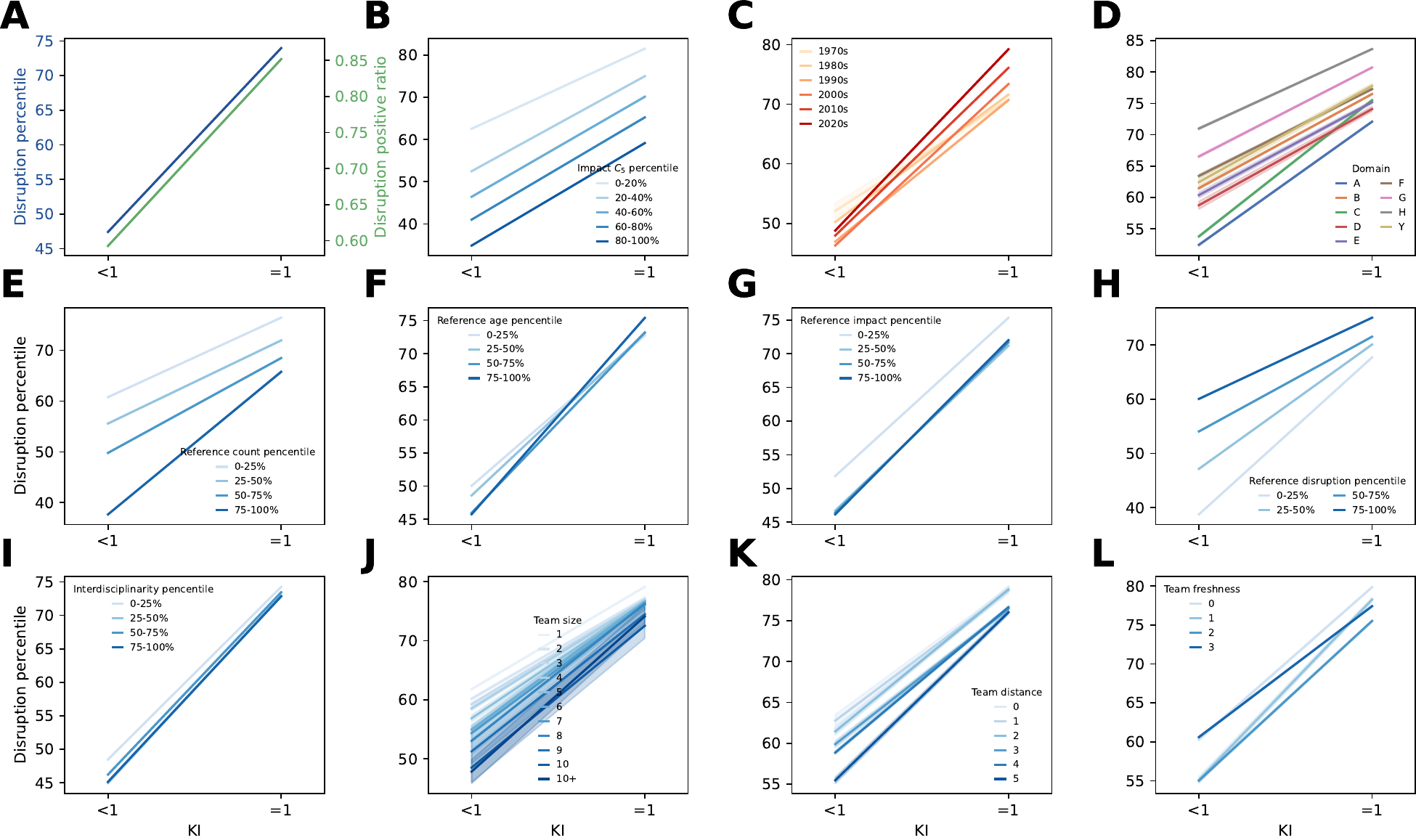}
    \caption{\textbf{$|$ Binary knowledge independence is associated with technical disruption in the OECD patent dataset.}
        (\textbf{A}) For the 11,349,527 patents published before 2023 in the OECD patent dataset that are cited at least once, the average disruption percentile (blue curve, left y-axis) and the disruption positive ratio (green curve, right y-axis) increase with the $\KI$ percentile.
        The association between $\KI$ and disruption persists, regardless of the impact $C_5$~(panel~\textbf{B}), publication decade~(panel~\textbf{C}), technical domain~(panel~\textbf{D}), reference count~(panel~\textbf{E}), reference age~(panel~\textbf{F}), reference impact $C_5$~(panel~\textbf{G}), reference disruption~(panel~\textbf{H}), interdisciplinarity~(panel~\textbf{I}), team size~(panel~\textbf{J}), geographic distance~(panel~\textbf{K}), and collaboration freshness~(panel~\textbf{L}).
        Bootstrapped 95\% confidence intervals are shown as shaded bands.
        \label{FigOECD_2_control_merge_binary}}
\end{figure}

\begin{figure}[ht] % Do not use \begin{figure*}
	\centering
	\includegraphics[width=0.87\textwidth]{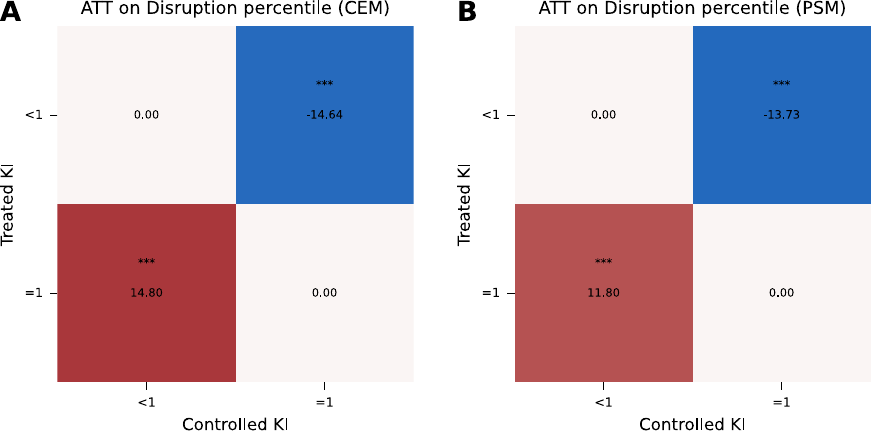}
    \caption{\textbf{$|$ Causal inference by CEM and PSM in the OECD patent dataset.}
        The ATT matrixs of binary $\KI$ on disruption via CEM (\textbf{A}) and PSM (\textbf{B}).
        Each controlled group is set as a baseline, and ATTs are calculated for comparisons between the baseline and each of the treated groups. Blue cells represent negative ATTs, while red ones represent positive ATTs, with color intensity proportional to the absolute value.
        Each ATT is tested against the null hypothesis (ATT equals 0) using a two-sided \textit{t}-test. (*$p < 0.05$, **$p < 0.01$, ***$p < 0.001$).
        \label{FigOECD_2_PSM_CEM_binary}}
\end{figure}

\begin{table}[ht] % Do NOT use \begin{table*}
    \centering
    % Captions go above tables
    \caption{\textbf{$|$ Normalized regression models of disruption and the influencing factors in the OECD dataset.}}
    \label{table_OECD_OLS_binary} % give each table a logical label name
    \renewcommand\tabcolsep{0.16cm} % column spacing
    \renewcommand{\arraystretch}{0.8} % line spacing
    \begin{threeparttable}
    \fontsize{7.5}{14}\selectfont % 设置表格字体大小为 8.75pt，行距为 12pt
    \begin{tabular}{lcccccccccc}
        \hline  % top line
        & Model 1 & Model 2 & Model 3 & Model 4 & Model 5 & Model 6 & Model 7 & Model 8 & $R^2$\% \\ 
        \hline  % middle line 1
        $\KI$ & \textbf{0.1884}*** & \textbf{0.1885}*** & \textbf{0.1742}*** & \textbf{0.1884}*** & \textbf{0.1877}*** & \textbf{0.1874}*** & \textbf{0.1883}*** & \textbf{0.1739}*** & \textbf{46.59} \\
         & (0.001) & (0.001) & (0.001) & (0.001) & (0.001) & (0.001) & (0.001) & (0.001) \\
        % Log $C_5$ &  & -0.0416*** &  &  &  &  &  &  &  &  &  & -0.0322*** & 3.91 \\
         % &  & (0.0000) &  &  &  &  &  &  &  &  &  & (0.0000) \\
        % Novelty &  & 0.0063*** &  &  &  &  &  &  & 0.0020*** & 0.04 \\
        %  &  & (0.0000) &  &  &  &  &  &  & (0.0000) \\
        Interdisciplinarity &  & 0.0027* &  &  &  &  &  & 0.0032** & 0.07 \\
         &  & (0.001) &  &  &  &  &  & (0.001) \\
        Reference count &  &  & -0.0831*** &  &  &  &  & -0.0831*** & 16.40 \\
         &  &  & (0.001) &  &  &  &  & (0.001) \\
        Reference age &  &  &  & 0.0017 &  &  &  & 0.0106*** & 0.64 \\
         &  &  &  & (0.001) &  &  &  & (0.001) \\
        % Reference $C_5$ &  &  &  &  &  &  & -0.0197*** &  &  &  &  & -0.0168*** & 0.39 \\
        %  &  &  &  &  &  &  & (0.0000) &  &  &  &  & (0.0000) \\
        % Reference disruption &  &  &  &  &  &  &  & 0.2293*** &  &  &  & 0.2231*** & 37.17 \\
        %  &  &  &  &  &  &  &  & (0.0000) &  &  &  & (0.0000) \\
        Team size &  &  &  &  & -0.0198*** &  &  & -0.0037** & 0.50 \\
         &  &  &  &  & (0.001) &  &  & (0.001) \\
        Team distance &  &  &  &  &  & -0.0244*** &  & -0.0201*** & 0.69 \\
         &  &  &  &  &  & (0.001) &  & (0.001) \\
        Team freshness &  &  &  &  &  &  & -0.0098*** & -0.0081*** & 0.08 \\
         &  &  &  &  &  &  & (0.001) & (0.001) \\
        \hline % middle line 2
        Discipline fixed effects  & Yes & Yes & Yes & Yes & Yes & Yes & Yes & Yes & 27.25 \\
        Year fixed effects & Yes & Yes & Yes & Yes & Yes & Yes & Yes & Yes & 7.78 \\
        \hline % middle line 3
        N & 927551 & 927551 & 927551 & 927551 & 927551 & 927551 & 927551 & 927551 \\
        $R^2$ & 0.071 & 0.071 & 0.077 & 0.071 & 0.071 & 0.071 & 0.071  & 0.078 \\
        \hline  % bottom line
    \end{tabular}
    \begin{tablenotes}
        \small
        \item[$\dagger$] We perform normalized ordinary-least-squares (OLS) regression analyses to examine the effect of binary $\KI$ on disruption in the OECD patent dataset, while controlling for various covariates. Each regression coefficient is tested against the null hypothesis (coefficient equals 0) using a two-sided \textit{t}-test. Standard errors are provided in parentheses for each coefficient. Note that we do not apply adjustments for multiple hypothesis testing in this analysis. (*$p < 0.05$, **$p < 0.01$, ***$p < 0.001$).
    \end{tablenotes}
    \end{threeparttable}
\end{table}}

\end{document}